\documentclass[12pt]{article}
\usepackage{natbib}
\usepackage{amsmath}
\usepackage{color}
\usepackage{rotating}
\usepackage{JASA_manu1}
\usepackage{lscape}
\usepackage{graphicx}
\usepackage{bm}
\usepackage{mathrsfs}
\usepackage{graphicx}
\usepackage{booktabs}
\usepackage{caption}
\usepackage{subfig}
\usepackage{amsmath}
\renewcommand{\theequation}{\thesection.\arabic{equation}}

\def\be{\begin{equation}}
\def\ee{\end{equation}}
\def\bea{\begin{eqnarray}}
\def\eea{\end{eqnarray}}
\def\bd{\begin{displaymath}}
\def\ed{\end{displaymath}}
\def\bda{\begin{eqnarray*}}
\def\eda{\end{eqnarray*}}
\def\bsm{\begin{small}}
\def\esm{\end{small}}

\def\nn{\nonumber}

\def\ha1{\hat \beta_1}

\def\bb0{\delta_\beta}

\def\bsc{\begin{scriptsize}}
\def\esc{\end{scriptsize}}

\input{epsf.sty}
\usepackage{epsfig}

\RequirePackage[OT1]{fontenc}
\RequirePackage{amsthm,amsmath}
\RequirePackage[colorlinks=true,citecolor=blue,urlcolor=blue]{hyperref}
\begin{document}

\title{Two-Sample  Tests for High Dimensional Means with Thresholding and Data Transformation\footnote{Emails: csx@gsm.pku.edu.cn, junli@math.kent.edu, pszhong@stt.msu.edu}}
\author{Song Xi Chen, Jun Li and Ping-Shou Zhong \\
Peking University and Iowa State University, Kent State University,\\
and Michigan State University }
\date{}
\maketitle

\begin{center}
\textbf{Abstract}
\end{center}
We consider testing for two-sample means of high dimensional populations by thresholding.  Two tests are investigated,  which are designed for better power performance when the two population mean vectors  differ only in sparsely populated coordinates.
 The first test is constructed %based on the original data and achieves a power improvement
 by carrying out thresholding to remove the non-signal bearing dimensions.  The second test combines data transformation via the precision matrix  with the  thresholding.
The benefits of the thresholding and the data transformations are showed  by a reduced variance of the test thresholding statistics, the improved power and a wider detection region of the tests.
%reducing the level of variance of the test statistics with thresholding.
%reducing the variance but also enhancing the signal strength. The asymptotic distributions of test statistics.
%It is shown that the second test is particularly powerful by incorporating the correlations %among the coordinates of the variables.
Simulation experiments and an empirical study are performed to confirm the theoretical findings and to demonstrate the practical implementations.

%\vspace*{.1in}


%\newpage
\setcounter{section}{1} \setcounter{equation}{0}

\section*{\large 1. Introduction}

Modern statistical data in biological and financial studies are increasingly high dimensional, but with relatively small sample sizes. This is the so-called ``large $p$, small $n$" phenomenon.
 If the dimension $p$ increases as the sample size $n$ increases, many classical approaches originally designed for fixed dimension problems (Hotelling's test and the likelihood ratio tests for the covariances) may no longer be feasible.  New methods are needed for the ``large $p$, small $n$" setting.

An important high dimensional inferential task is to test the equality of the mean vectors between two populations, which represent two treatments. Let $\bm{X}_{i1}\cdots,\bm{X}_{i n_i}$ be an independent and identically distributed sample drawn
from a $p$-dimensional distribution $F_i$,  for $i=1$ and 2 respectively.
The dimensionality $p$ can be  much larger than the two sample sizes $n_1$ and $n_2$ so that $p/n_i \to \infty$.
% each $X$ is $p$-dimensional random vector.
Let $\bm{\mu}_i$ and $\bm{\Sigma}_i$ be the means and the covariance of $F_i$.
%And $\bm{X}_{ij}=\bm{\mu}_i+\bm{W}_{ij}$, where $\bm{\mu}_i=(\mu_{i1},\cdots,\mu_{ip})$ is a %$p$%dimensional mean vector and $\mbox{Var}(\bm{W}_{ij})=\Sigma_i=(\sigma_{i,jk})$ is the covariance%matrix%of $F_i$.
The primary interest is testing
\be
H_{0} : \bm{\mu}_1 = \bm{\mu}_2  \quad  \mbox{versus} \quad H_{1} : \bm{\mu}_1
\ne \bm{\mu}_2. \label{eq:hypo1}
\ee
%Let $\bar{\bm{X}_i}=\frac{1}{n_i}\sum_{j=1}^{n_i}\bm{X}_{ij}$ be the sample means and %$\bm{S}_n=\frac{1}{n_1+n_2-2}\sum_{i=1}^2\sum_{j=1}^{n_i}(\bm{X}_{ij}-\bar{\bm{X}_i})(\bm{X}_{ij}-\bar{\bm{X}_i})^{\prime}$ be the pooled sample covariance matrix under %$\bm{\Sigma}_1=\bm{\Sigma}_2$.
 Hotelling's $T^2$ test %statistic $\frac{n_1n_2}{n_1+n_2}(\bar{\bm{X}}_1-\bar{\bm{X}}_2)^{\prime}\bm{S}_n^{-1}(\bar{\bm{X}}_1-\bar{\bm{X}}_2)$
has been the classical test for the above hypotheses for fixed dimension $p$ and is still applicable if $p \le n_1+n_2-2$.
%\footnote{Is it true ? I thought it is $p \le n_1 + n_2 -2$ ?(\bf{yes, I correct it})} [see Anderson (2003) for details].
However, as shown in Bai and Saranadasa (1996), Hotelling's test suffers from a significant power loss when $p/(n_1+n_2-2)$ approaches to 1 from below.
When  $p > n_1+n_2-2$, the test is not applicable as the pooled sample covariance matrix, say $\bm{S}_n$,  is no longer invertible.

There are proposals which modify Hotelling's $T^2$ statistic for high dimensional situations.  Bai and  Saranadasa (1996) proposed the following alteration
\begin{eqnarray}
M_n=(\bar{\bm{X}}_1-\bar{\bm{X}}_2)^{T}(\bar{\bm{X}}_1-\bar{\bm{X}}_2)-\mbox{tr}(\bm{S}_n)/n, \label{BS}
\end{eqnarray}
by removing the inverse of the sample covariance matrix $\bm{S}_n^{-1}$ from the Hotelling's statistic, where $n=n_1n_2/(n_1+n_2)$.
Chen and Qin (2010) considered a linear combination of U-statistics
\begin{eqnarray}
T_{n}=\frac{1}{n_1(n_1-1)}\sum_{i\ne j}^{n_1}\bm{X}_{1i}^{T}\bm{X}_{1j}+\frac{1}{n_2(n_2-1)}\sum_{i\ne j}^{n_2}\bm{X}_{2i}^{T}\bm{X}_{2j}
-\frac{2}{n_1n_2}\sum_{i}^{n_1}\sum_{j}^{n_2}\bm{X}_{1i}^{T}\bm{X}_{2j},
\label{eq1-0}
\end{eqnarray}
and showed that the corresponding test can operate under much relaxed regimes regarding the dimensionality and sample size constraint and
without assuming $\Sigma_1 =\Sigma_2$.
 Srivastava, Katayama and Kano (2013) proposed using the diagonal matrix of the sample variance matrice to replace $\bm{S}_n$
under the normality. These three tests are basically all targeted on
 a weighted $L_2$ norms between $\bm{\mu}_1$ and $\bm{\mu}_2$.
In a development in another direction,
% that using the max-norm rather than the $L_2$ norm,
 Cai, Liu and Xia (2014) proposed a test based on the max-norm of marginal $t$-statistics.  More importantly, they implemented a data transformation which is designed to increase the signal strength under sparsity as discovered early in Hall and Jin (2010) in their innovated higher criticism test for the one sample problem.

The $L_2$ norm based tests  are known to be effective in detecting dense signals in the sense that the differences between $\bm{\mu}_1$ and $\bm{\mu}_2$ are populated over
a large number of components. %, while the test of Liu and
However, the tests will encounter power loss under the sparse signal settings where only a small portion of
components of the two mean vectors are different. To improve the performance of these tests under the sparsity,  we propose a thresholding
test to remove the non-signal bearing dimensions. The idea of thresholding has been used in many applications, as demonstrated in Donoho and
Johnstone (1994) for selecting significant wavelet coefficient and Fan (1996) for testing the mean of random vectors  with IID normally
distributed components. See also Ji and Jin (2012) for variable selection in high dimensional regression model.
 We find that the thresholding can reduce the variance of the Chen and Qin (2010) (CQ) test statistic, and hence increases the power of the test under sparsity for non-Gaussian data.  We also confirm the effectiveness of the precision matrix transformation in increasing the signal strength of the CQ test. The transformation is facilitated by an estimator of the precision matrix via the Cholesky decomposition with the banding approach (Bickel and Levina, 2008a, 2008b).
It is shown that the test with the thresholding and the data transformation  has a lower detection boundary than that without the data transformation, and can be lower than the detection boundary of an Oracle test without data transformation.
% Hence, thresholding with the data transformation is recommended.
%, which is confirmed by our simulation study.
%and find the combination of the data transformation and thresholding produces the best power

The rest of the paper is organized as follows.  We  analyze the thresholding test and its relative power performance to the CQ test and the Oracle test in Section 2.
%the asymptotic distribution of the thresholding test is established and its asymptotic
%power is discussed.
A multi-level thresholding test is proposed in Section 3 for detecting faint signals. Section 4 considers a data transformation with an estimated precision matrix.
%Section 3 addresses the question of how to choose the level of threshold to achieve the optimality. In Sections 4, the similar results are obtained for the thresholding %test based on the transformed data when $\bm{\Omega}$ is both known and unknown.
Simulation results  are
 presented in Section 5.  Section 6 reports an empirical study  to select differentially
 expressed gene-sets for a human breast cancer data set. Section 7 concludes the paper with  discussions.  All technical details are relegated to
 the Appendix.

\setcounter{section}{2} \setcounter{equation}{0}
\section*{\large 2. Thresholding Test}

We first outline the CQ statistic before introducing the
thresholding approach.   The statistic (\ref{eq1-0})
can be written as $T_n=\sum_{k=1}^p T_{nk}$ where
\begin{eqnarray}
T_{nk}&=&\frac{1}{n_1(n_1-1)}\sum_{i\ne j}^{n_1}X_{1i}^{(k)}X_{1j}^{(k)}+\frac{1}{n_2(n_2-1)}\sum_{i\ne j}^{n_2}X_{2i}^{(k)}X_{2j}^{(k)}\nonumber\\
&-&\frac{2}{n_1n_2}\sum_{i}^{n_1}\sum_{j}^{n_2}X_{1i}^{(k)}X_{2j}^{(k)},
\label{eq1}
\end{eqnarray}
and $X_{ij}^{(k)}$ represents the $k$-th component of $\bm{X}_{ij}$.
It can be readily shown that $T_{nk}$ is unbiased to
$(\mu_{1k}-\mu_{2k})^2$, which may be viewed as the amount of signal
in the $k$-th dimension.

To facilitate simpler notations,
% the comparison easier between the CQ test and the thresholding test proposed later in this section,
 we modify the test statistic $T_n$ by standardizing each $T_{nk}$ by $\sigma_{1,kk}/n_1+\sigma_{2,kk}/n_2$, the variance of
 $\bar{X}_1^{(k)}-\bar{X}_2^{(k)}$, if both $\sigma_{1,kk}$ and $\sigma_{2,kk}$ are known.
 %\footnote{Should we write it as $\sigma_{nk}$ instead of $\sigma_{nk}^2$ since it is not really a variance? \bf{Dr. Chen: %$\frac{T_{nk}}{\sigma_{nk}^2}+1=(\frac{\bar{X}_1^{(k)}-\bar{X}_2^{(k)}}{\sigma_{nk}})^2+o_p(1)$. And $\sigma_{nk}^2$ is the variance of $\bar{X}_1^{(k)}-\bar{X}_1^{(k)}$. %Therefore, here we did not standardize $T_{nk}$ by its standard deviation but instead standardize $\bar{X}_1^{(k)}-\bar{X}_1^{(k)}$ by $\sigma_{nk}$, which is also what %you and Pingshou did in your paper. }}.
% We note that $Var(T_{n k}) = 2 \sigma_{nk}^2$ (?? check).  %    the half of its standard variance.
If $\sigma_{1,kk}$ and $\sigma_{2,kk}$ are unknown, we can use $
\hat{\sigma}_{1,kk}/n_1+ \hat{\sigma}_{2,kk}/n_2$ where
$\hat{\sigma}_{1,kk}$ and $\hat{\sigma}_{2,kk}$ are the usual sample
variance estimates at the $k$-th dimension. This will make the CQ
test invariant under the scale transformation; see Feng, Zou, Wang and
Zhu (2013) for a related investigation.
%Even though we assume $\sigma_{i,kk}$ are known, the same conclusion can be drawn by replacing %$\sigma_{nk}^2$ by $\hat{\sigma}_{nk}^2$ given above as demonstrated by Shao (1997), Jing, Shao and Zhou %(2008) and Wang and Hall (2009).
To expedite our discussion,  we assume  $\sigma_{i, k k}^2$ are
known and equal to one without loss of generality.  This leads to a
modified version of the CQ  statistic
\begin{eqnarray}
\tilde{T}_n=n\sum_{k=1}^pT_{nk}, \label{newCQ}
\end{eqnarray}
where $n=n_1 n_2/(n_1 + n_2)$.
Under the same setting, a modified version of the Bai and Saranadasa (BS) test statistic is
\begin{eqnarray}
\tilde{M}_n=n\sum_{k=1}^pM_{nk}-p,\label{eq:newBS}
\end{eqnarray}
where $M_{nk}=(\bar{X}_1^{(k)}-\bar{X}_2^{(k)})^2$.

%We first review its power performance when only a small portion of $\bm{\mu}_{1k}$ and $\bm{\mu}_{2k}$ are different.
Let $\delta_k=\mu_{1k}-\mu_{2k}$ and $S_{\beta}=\{k:  \delta_k \ne
0 \} \label{set_a}$
be the set of locations of the signals $\delta_k$ such that $|S_{\beta}|=p^{1-\beta}$ where $\beta \in (0,1)$ is the sparsity parameter. %We are particularly interested in the  sparse signal setting where $\beta \in (0.5,1)$.
Basically, the sparsity of the signal increases as $\beta$ is closer
to 1.        Under the sparsity, an overwhelming number of  $T_{nk}$
%$M_{nk}$ or
 carry no signals. However, including them increases the
variance of the test statistic, and  dilutes the signal to noise
ratio of the test; and thus hampers the power of the test.

Let us now analyze  the standardized CQ test under the sparsity.
%signal-to-noise ratio of the test procedure diminishes, which hampers its performance.
Define
%\footnote{Is it the covariance of the standardized $T_{nk}$ and $T_{nl}$ ? If, we should insert it in the next equatin.\bf{No. I filled in the blank in eqn(2.4) to %explain the definition of $\rho_{kl}$}}
\be
\rho_{kl}=
\mbox{Cov}\biggl\{\sqrt{n}(\bar{X}_1^{(k)}-\bar{X}_2^{(k)}),\sqrt{n}(\bar{X}_1^{(l)}-\bar{X}_2^{(l)})\biggr\}
= n(\sigma_{1,kl}/n_1+\sigma_{2,kl}/n_2). \label{correlation} \ee
%which becomes the regular correlation coefficient if $\bm{\Sigma}_1=\bm{\Sigma}_2$.\footnote{Why it has to be $\bm{\Sigma}_1=\bm{\Sigma}_2$ ?\bf{We do not assume $\bm{\Sigma}_1=\bm{\Sigma}_2$ when derive all the results. But here I just want to let readers know that $\rho_{kl}$ will be the regular correlation coefficient under a special case $\bm{\Sigma}_1=\bm{\Sigma}_2$  }. What if the two Simgas differ. Are not they are the cor coef as well ?}
Similar to the derivation in Chen and Qin (2010), the variance of
$\tilde{T}_n$ under $H_0$  is \be \sigma_{\tilde{T}_n,0}^2 =
2p+2\sum_{i \ne j}\rho_{ij}^2,\nonumber \ee and that under $H_1$ is
\be \sigma_{\tilde{T}_n,1}^2 = 2p+2\sum_{i \ne
j}\rho_{ij}^2+4n\sum_{k,l \in S_{\beta}}\delta_k\, \delta_l
\,\rho_{kl}.\label{ChenQin_alter} \ee
It can be seen that
$\sigma_{\tilde{T}_n,1}^2 \ge \sigma_{\tilde{T}_n,0}^2$ since the
last term of $\sigma_{\tilde{T}_n,1}^2$ is nonnegative due to
$R=(\rho_{ij})_{p \times p}$ being non-negative definite.

Under a general multivariate model and some conditions on the
covariance matrices, the asymptotic normality of $\tilde{T}_n$ can
be established (Chen and Qin, 2010):
\[
\frac{\tilde{T}_n-||\bm{\mu}_1-\bm{\mu}_2||^2}{\sigma_{\tilde{T}_n,1}} \xrightarrow{d} \mbox{N}(0,1), \, \mbox{as} \; p \to \infty \; \mbox{and}\; n\to \infty.
\]
This implies  the modified CQ  test that rejects $H_0$ if
$\tilde{T}_n/\hat{\sigma}_{\tilde{T}_n,0}>z_{\alpha}$ where
$z_{\alpha}$ is the upper $\alpha$ quantile of $\mbox{N}(0,1)$ and
$\hat{\sigma}_{\tilde{T}_n,0}$ is a consistent estimator of
$\sigma_{\tilde{T}_n,0}$.
% \footnote{Give the test procedure. {\bf Added in bold}}.

Let $\bar{\delta}^2=\sum_{k\in S_{\beta}}n\,\delta_k^2/p^{1-\beta}$
represent the average standardized signal. The  power of the test is
\begin{eqnarray}
\beta_{\tilde{T}_n}(||\bm{\mu}_1-\bm{\mu}_2||)
=\Phi\biggr(-\frac{\sigma_{\tilde{T}_n,0}}{\sigma_{\tilde{T}_n,1}}z_{\alpha}+\frac{ p^{1-\beta}\bar{\delta}^2}{\sigma_{\tilde{T}_n,1}}\biggl),\nonumber
\end{eqnarray}
where $\Phi(\cdot)$ is the  distribution function of $\mbox{N}(0,1)$. Since
$\sigma_{\tilde{T}_n,1}^2 \ge \sigma_{\tilde{T}_n,0}^2$, the first
term within $\Phi(\cdot)$ is bounded. Then,  the power of the test
is largely determined by the second term
\begin{eqnarray}
\mbox{SNR}_{\tilde{T}_n}=: \frac{ p^{1-\beta}\bar{\delta}^2}{\sqrt{2p+2\sum_{i \ne j}\rho_{ij}^2+4n\sum_{k,l \in S_{\beta}}\delta_k\,\delta_l\,\rho_{kl}}},\label{CQ_power}
\end{eqnarray}
which is called  the signal to noise ratio of the test since the numerator is the average signal strength and the denominator is the standard deviation of the test statistic under $H_1$.  An inspection reveals that  while the numerator of  $\mbox{SNR}_{\tilde{T}_n}$  is  contributed only by those signal bearing dimensions, the standard deviation in the denominator is contributed by all $T_{nk}$ including those with non-signals.

Specifically, if $\bm{\Sigma}_1=\bm{\Sigma}_2=\bm{I}_p$,
\[
\mbox{SNR}_{\tilde{T}_n}=\frac{ p^{1-\beta}\bar{\delta}^2}{\sqrt{2p+4p^{1-\beta}\bar{\delta}^2}}.
\]
Hence, if the sparsity $\beta> 1/2$ and the average signal
 $\bar{\delta}=o(p^{\beta/2-1/4})$,
$\mbox{SNR}_{\tilde{T}_n}=o(1)$. Then,  the test has little
power beyond the significant level. A reason for the power loss is
that the variance of $\tilde{T}_n$ is much inflated by including those
non-signal bearing $T_{nk}$.

To put the above analysis in prospective, we consider an Oracle test
which has the knowledge of the possible signal bearing set $S_{\beta}$ (with slight abuse of notation), which is much smaller than the entire set of dimensions.  The Oracle is only a semi-Oracle as he does not know the exact dimensions of the signals other than that they are within $S_{\beta}$.
%is not empty even under the null.  This Oracle knows that the signals are %within $S_{\beta}$;  but he does not know exactly which dimensions are %signal-bearing.

 The Oracle test statistic is
\begin{eqnarray}
O_n=n\sum_{k\in S_{\beta}} T_{nk},\label{oracle}.
\end{eqnarray}
%{\bf where the coordinates in the signal bearing set $S_{\beta}$ under $H_1$ %are also kept under $H_0$ in order to make the Oracle test statistic %meaningful under $H_0$.}
Similar to the derivation of (\ref{ChenQin_alter}), the variance of $O_n$ under  $H_0$ is
\[
\sigma_{O_n, 0}^2 = 2p^{1-\beta}+2\sum_{i \ne j\in S_{\beta}}\rho_{ij}^2,
\]
and that under $H_1$
%which implies that all the signal bearing dimensions are on $S_{\beta}$
 is
\be
\sigma_{O_n, 1}^2 = 2p^{1-\beta}+2\sum_{i \ne j\in S_{\beta}}\rho_{ij}^2+4n\sum_{k,l \in S_{\beta}}\delta_k\,\delta_l\,\rho_{kl}.\label{variance_Oracle}
\ee
Comparing $\sigma_{O_n,1}^2$ with $\sigma_{\tilde{T}_n,1}^2$ in (\ref{ChenQin_alter}),
we see that the first term of $\sigma_{O_n,1}^2$ is much smaller than that of $\sigma_{\tilde{T}_n,1}^2$.
% since the Oracle test statistics does not involve those $T_{nk}$ with $\delta_k=0$.
It may be shown that under the same conditions that establish the
asymptotic normality of $\tilde{T}_n$,
%the Oracle test has the asymptotic normality:
\[
\frac{O_n-||\bm{\mu}_1-\bm{\mu}_2||^2}{\sigma_{O_n, 1}} \xrightarrow{d} \mbox{N}(0,1), \, \mbox{as} \; p \to \infty \; \mbox{and}\; n\to \infty,
\]
which leads to the Oracle test that rejects $H_0$ if
$O_n/\hat{\sigma}_{O_n,0} > z_{\alpha}$ {where
$\hat{\sigma}_{O_n,0}$ is a ratio consistent estimator of
${\sigma}_{O_n,0} $.}

The asymptotic normality implies that the power of the Oracle test is
\begin{eqnarray}
\beta_{O_n}(||\bm{\mu}_1-\bm{\mu}_2||)
=\Phi\biggr(-\frac{\sigma_{O_n,0}}{\sigma_{O_n,1}}z_{\alpha}+\frac{
p^{1-\beta}\bar{\delta}^2}{\sigma_{O_n,1}}\biggl). \nonumber
\end{eqnarray}
%Similar to $\mbox{SNR}_{\tilde{T}_n}$ in (\ref{CQ_power}).
It is largely determined by
\begin{eqnarray}
\mbox{SNR}_{O_n}
=:\frac{ p^{1-\beta}\bar{\delta}^2}{\sqrt{2p^{1-\beta}+2\sum_{i\ne j \in S_{\beta}}\rho_{ij}^2+4n\sum_{k,l \in S_{\beta}}\delta_k\,\delta_l\,\rho_{kl}}},\label{Oracle_power}
\end{eqnarray}
which is much larger than $\mbox{SNR}_{\tilde{T}_n}$ since
$\sigma_{O_n,1}^2 \ll \sigma_{\tilde{T}_n,1}^2$. If
$\bm{\Sigma}_1=\bm{\Sigma}_2=\bm{I}_p$, \be
\mbox{SNR}_{O_n}=\frac{p^{1-\beta}\bar{\delta}^2}{\sqrt{2p^{1-\beta}+4p^{1-\beta}\bar{\delta}^2}}=\frac{p^{\frac{1-\beta}{2}}\bar{\delta}^2}{\sqrt{2+4\bar{\delta}^2}}
, \ee that tends to infinity for $\beta>1/2$  as long as
$\bar{\delta}$ is a large order of $p^{\beta/4-1/4}$, which is much
smaller than $p^{\beta/2-1/4}$ for the CQ test, indicating the test
is able to detect much fainter signal.

%The Oracle test has an ideal performance in powers due to the avoidance of the dimensions with $\delta_k=0$.
The reason that the Oracle test has better power  is that all the excluded dimensions are definitely non-signal bearing and those included have much smaller dimensions.
%dimensions are excluded.
In reality, the locations of those non-signal bearing dimensions are unknown.
However, thresholding can be carried out to exclude those non-signal bearing dimensions.
%{ We note that if $\delta_k=0$,   a derivation given in Lemma 1 shows that
%\begin{eqnarray}
%\mbox{P}\biggl(n\,T_{nk}+1 \ge 2\mbox{log}p\biggr)=\mbox{P}\biggl\{n(\bar{X}_1^{(k)}-\bar{X}_2^{(k)})^2 \ge 2\mbox{log}p\biggr\}\{1+o(1)\}. \label{prob}
%\end{eqnarray}
%For $\mbox{log}p=o(n^{1/3})$, applying the large deviation result
%(Petrov, 1995) leads to  that under $H_0$
%\begin{eqnarray}
%\mbox{P}\biggl(\max\limits_{1\le k \le p}n\,T_{nk}+1 \ge 2\mbox{log}p\biggr)&\le& p\mbox{P}\biggl\{n(\bar{X}_1^{(k)}-\bar{X}_2^{(k)})^2 \ge 2\mbox{log}p\biggr\}\{1+o(1)\}\nonumber\\
%&= &2 (2\pi)^{-1/2}(2\mbox{log}p)^{-1/2}\{1+o(1)\} \to 0,\label{step1}
%\end{eqnarray}
%which suggests  that thresholding at $2\mbox{log}p$ avoids all the non-signal bearing
%  $n\,T_{nk}+1$.  }
%Hence,  the threshodling  removes those $T_{nk}$ with $\delta_k=0$. }
%However,  thresholding at this level   also removes some $T_{nk}$ with weaker signals such that  $\delta_k^2 < 2 \mbox{log}p /n$.
 %\footnote{Shall we divide the variance $\sigma^2_{nk}$ {\bf yes, added.}}
%for $r<1$ as we will show later in this section.
Based on the large deviation results (Petrov, 1995),
%\footnote{I
%have simplify the discussion here by removing the previous (2.10)
%and (2.11). It won't affect the remaining content. (Jun)}
 we use a
thresholding level $\lambda_{n}(s) = 2s\mbox{log}p$ for $s\in
(0,1)$ to strike a balance between removing non-signal bearing
$T_{n k}$ while maintaining those with signals. The thresholding test statistic is
\begin{eqnarray}
L_1(s)=\sum_{k=1}^{p}n\,T_{nk}I\biggl\{n\,T_{nk}+1>\lambda_{n}(s)\biggr\},\label{thr1}
\end{eqnarray}
where $I(\cdot)$ is the indicator function.

We can also carry out the thresholding on BS test statistic
%\footnote{I want to remove the following segment to speed up. is it
%OK ?Dr. Chen: we cite equation (2.11) and (2.12) later in Section 4
%between equation (4.1) and (4.2).(Jun)}
 (\ref{eq:newBS}), which
leads to
\begin{eqnarray}
L_2(s)=\sum_{k=1}^{p}\biggl\{n(\bar{X}_1^{(k)}-\bar{X}_2^{(k)})^2-1 \biggr\}I\biggl\{n(\bar{X}_1^{(k)}-\bar{X}_2^{(k)})^2>\lambda_{n}(s)\biggr\}.\label{thr2}
\end{eqnarray}
As we will show later, both $L_1(s)$ and $L_2(s)$ have very similar properties. Therefore, we choose $L_n(s)$ to refer to either $L_1(s)$ or $L_2(s)$.

Before we show that the thresholding can reduce the variance
contributed from those non-signal bearing dimensions without harming
the signals, we introduce the notion of $\alpha$-mixing to quantify
the dependence among the components of the random vector
$\bm{X}=(X^{(1)}, \cdots, X^{(p)})^T$.

%Let $X^{(m)}$ represent $m$th component of $X$.
For any integers $a<b$, define $\mathcal{F}_{\bm{X},(a,b)}$ to be the $\sigma$-algebra generated by $\{{X}^{(m)}: m \in(a,b)\}$ and define the $\alpha$-mixing coefficient
%for a positive integer $k$ as
\[
\alpha_{\bm{X}}(k)= \sup \limits_{m\in \mathcal{N},A\in \mathcal{F}_{\bm{X},(1,m)}, B\in \mathcal{F}_{\bm{X},(m+k,\infty)}} |P(A\cap B)-P(A)P(B)|.
\]
%where $\mbox{Corr}(A, B)$ denotes the correlation coefficient between random variables $A$ and $B$ and
%$L^2\{\mathcal{F}_{(a,b)}\}$ denote the set of $\mathcal{F}_{(a,b)}$ random variables which have finite second moments.

The following conditions are assumed in our analysis.

\textbf{(C1)}: As $n\to \infty$, $p\to \infty$ and
$\mbox{log}p=o(n^{1/3})$.

\textbf{(C2)}: Let $\bm{X}_{ij}=\bm{\mu}_i+\bm{W}_{ij}$. There exists a positive
constant $H$ such that for $h \in [-H, H]^2$, $\mbox{E}\{e^{h^T
\cdot [(W_{ij}^{(k)})^2, (W_{ij}^{(l)})^2]}\} < \infty$ for $k \ne
l$.

\textbf{(C3)}: The sequence of random variables $\{X_{ij}^{(l)}\}_{l=1}^p$ is $\alpha$-mixing such that $\alpha_{\bm{X}}(k) \le C\alpha^k$ for some $\alpha \in (0,1)$ and a positive constant $C$, and $\rho_{kl}$ defined in (\ref{correlation}) are summable such that $\sum_{l=1}^p |\rho_{kl}| < \infty$ for any $k \in \{1,\cdots, p\}$.  %\footnote{Can we update relax the condition to $\sum_{k} k^{\alpha} \rho(k) < \infty$ %for some $\alpha >0$ ?: Prof. Chen: As we discussed, I have changed it to alpha- %mixing. \textbf{I also added a summable condition on the correlation of random %variables (Jun)}}

Condition (C1) specifies the growth rate of dimension $p$ relative to $n$ under which the large deviation results can be applied to derive the means and variances of the test statistics. Condition (C2) assumes that $(X_{ij}^{(k)}, X_{ij}^{(l)})$ has a bivariate  sub-Gaussian distribution, which {is more general than the Gaussian distribution}. %\footnote{Do you mean using the LD results ?I have changed sentence(Jun).}
Condition (C3) prescribes weak dependence among the column components of the random vector, which is commonly assumed in time series analysis.
%We take a time series view on the high dimensional random vectors.
% that $\rho_{kl}$ defined in (\ref{correlation}) satisfies $\sum_{l}^p |\rho_{kl}| < \infty$.
 %As we will show in Appendix, if we write $L_n(s)=\sum_{k=1}^p L_{n,k}(s)$, the leading order of $\mbox{Var}\{L_n(s)\}$ is contributed only by $\sum_{k=1}^p \mbox{Var}\{L_{n,k}(s)\}$.
%\footnote{Remove this sentence ? Say it later ? I have removed it (Jun)}

Derivations given in Appendix leading to (\ref{mean_alter}) and (\ref{variance_alter}) show that the mean of the thresholding test statistic $L_n(s)$ is
\begin{eqnarray}
\mu_{L_n(s)}&=&\biggl(\frac{2}{\sqrt{2\pi}}(2s\mbox{log}p)^{\frac{1}{2}} p^{1-s}+\sum_{k \in S_{\beta}}\{n\,\delta_k^2I(n\,\delta_k^2>2s\mbox{log}p)\nonumber\\
&& + (2s\mbox{log}p)\bar{\Phi}(\eta_k^-) I(n\,\delta_k^2<2s\mbox{log}p)\}\biggr)\{1+o(1)\},\label{mean_thr}
\end{eqnarray}
and the variance  is
\begin{eqnarray}
\sigma_{L_n(s)}^2
&=&\biggl(\frac{2}{\sqrt{2\pi}}\{(2s\mbox{log}p)^{\frac{3}{2}}+(2s\mbox{log}p)^{\frac{1}{2}}\}p^{1-s}+\sum_{k,l \in S_{\beta}}(4n\delta_k \delta_l\rho_{kl}
+2\rho_{kl}^2)\nonumber\\
&\times & I(n\delta_k^2>2s\mbox{log}p) I(n\,\delta_l^2>2s\mbox{log}p) +\sum_{k \in
S_{\beta}}(2s\mbox{log}p)^2\bar{\Phi}(\eta_k^-)\nonumber\\
&\times &I(n\,\delta_k^2<2s\mbox{log}p)\biggr)\{1+o(1)\}, \label{variance_thr}
\end{eqnarray}
where $\bar{\Phi}=1-\Phi$ and $\eta_k^-=(2s\mbox{log}p)^{1/2}-n^{1/2}\delta_k$.

\textbf{Theorem 1.} Assume Conditions \textbf{(C1)}-\textbf{(C3)}.  For any $s\in (0, 1)$,
\[
\sigma_{L_n(s)}^{-1} \bigl\{ L_n(s)-\mu_{L_n(s)}\bigr\}
 \xrightarrow{d} \mbox{N}(0,1).
\]

Let $\mu_{L_n(s),0}$ and $\sigma_{L_n(s),0}$ be the mean and
variance under $H_0$ which can be obtained by ignoring the
summation terms in (\ref{mean_thr}) and (\ref{variance_thr}).
 Then, Theorem 1 implies an asymptotic $\alpha$ level test that  rejects $H_0$ if
\begin{eqnarray}
L_n(s)> z_{\alpha} \hat{\sigma}_{L_n(s),0}+
\hat{\mu}_{L_n(s),0},\label{eq:thresholding-strong}
\end{eqnarray}
where $\hat{\mu}_{L_n(s),0}$ and $\hat{\sigma}_{L_n(s),0}$ are consistent estimators of ${\mu}_{L_n(s),0}$ and ${\sigma}_{L_n(s),0}$ satisfying
\begin{align}
\label{res}
\mu_{L_n(s),0}-\hat{\mu}_{L_n(s),0}= o\{{\sigma}_{L_n(s),0} \} \quad \hbox{and} \quad  \hat{\sigma}_{L_n(s),0} /{\sigma}_{L_n(s),0} \stackrel{p} \to 1.
\end{align}

If all the signals $\delta_k^2$ are strong such that $n\,\delta_k^2>2\mbox{log}p$, choosing $s=1^-$ such that $(1-s)\log(p)=o(1)$
leads to
%\footnote{Can we choose $s=1$ or $s=1 - L_p$ for a slow varying $L_p$ ? Do we stil have the CLT with this choice of $s$ ?: Prof. Chen: From the proof of the CLT we had, we can not use $s=1$. But we can use an $s$ that are very close to 1. For instance the $s$ used in Fan (1996) paper was $s=1-d\log(c\log(p))/\log(p)$. (Pingshou)  OK, then if we use Fan's level, can we still get the neat variance results ? \textbf{Choosing s as Fan's leads to $p^{1-s}$ is the smaller order of any function of log p. I checked and found the result does not change.(Jun)} {\bf The order of $p^{(1-s)/2}=c(\log(p))^{d/2}$ if we chose the same s as Fan (1996). Then in this case, we might need to choose $0<d<1/2$ such that the approximation error is ignorable.(Pingshou)}    This is confusing. Can we then still just use $s=1^{-}$ ?  PS: could you write in the main text on this ?: {Prof. Chen, %I have added a condition on $s$ such that we can still have the same result. A slightly %strong condition is given below when we consider estimations.}  }  leads to
 \begin{eqnarray}
\mu_{L_n(s)}&=&\biggl\{\frac{2}{\sqrt{2\pi}}(2\mbox{log}p)^{\frac{1}{2}}+\sum_{k \in S_{\beta}} n\,\delta_k^2\nonumber \biggr\}\{1+o(1)\},\label{mean_thr1}
\end{eqnarray}
and
\begin{eqnarray}
\sigma_{L_n(s)}^2
&=&\biggl(\frac{2}{\sqrt{2\pi}}\{(2\mbox{log}p)^{\frac{3}{2}}+(2\mbox{log}p)^{\frac{1}{2}}\}
+\sum_{k,l \in S_{\beta}}(4n\,\delta_k \,\delta_l\,\rho_{kl}+2\rho_{kl}^2)\biggr)\{1+o(1)\}.\nonumber
\end{eqnarray}
%where $L_p = a (\log p)^b$ is a slowly varying logarithm function
%of $p$.
%The first term of $\sigma_{L_n(s)}^2$ is only $2L_p$, much smaller than $2p$, the first variance term of the CQ statistic in (\ref{ChenQin_alter}),
%confirming that imposing the thresholding can significantly reduce the variance. Moreover,
Except for a slowly varying logarithm function of $p$, $\sigma_{L_n(s)}^2$ has the same leading order
variance of the Oracle statistic \be \sigma_{O_n,1}^2 = \sum_{k, l
\in S_{\beta}}\biggl(4n\,\delta_k\,\delta_l\,\rho_{kl}+2\rho_{kl}^2
\biggr), \nonumber \ee indicating the effectiveness of the
thresholding under the strong signal situation.
%which is free of the
%contribution from the non-signal bearing dimensions.
%For strong signals case, i.e. $n\delta_k^2 > 2 \mbox{log}p$, we can
%choose $s = 1^-$ such that $(1-s)\log(p)\to 0$.%\footnote{This
%contradicts to the statement early. {\bf Dr. Chen: The choice of $s$ is subtle. The second $s = 1^-$ such that $(1-s)\log(p)\to 0$ is stronger than the first $s=1^-$ such that $(1-s)\log(p)=O\{\log(\log(p ))\}$. The weaker restriction on $s$ leads that the first term on the right hand side of (2.13) becomes slowly varying function $L_p$ (not exact $\frac{2}{\sqrt{2\pi}}(2s\mbox{log}p)^{\frac{1}{2}} p^{1-s}$). To get exact $\frac{2}{\sqrt{2\pi}}(2s\mbox{log}p)^{\frac{1}{2}} p^{1-s}$ for implementing a test procedure, we need the second stronger restriction. But recommend we choose the second $s$ to avoid confusion. (Jun)}}
%where $1^{-}$ indicates a number arbitrarily close to 1 but less than 1.
{With the same choice of $s$ for strong signals case, $\mu_{L_n(s),0}$ and $\sigma^2_{L_n(s),0}$ can be respectively estimated by}
\[
\hat{\mu}_{L_n,0}=\frac{2}{\sqrt{2\pi}}(2\mbox{log}p)^{\frac{1}{2}}  \quad \mbox{and} \quad \hat{\sigma}_{L_n,0}^2
=\frac{2}{\sqrt{2\pi}}\{(2\mbox{log}p)^{\frac{3}{2}}+(2\mbox{log}p)^{\frac{1}{2}}\}.
\]
It can be shown that
% \[
%\frac{\mu_{L_1(s),0}-\hat{\mu}_{L_1,0}}{\hat{\sigma}_{L_1,0}}=O\{(\log p)^{-1/4}n^{-1/6}+(\log p)^{5/4}n^{-1/2}\},
%\]
%and
%\[
%\frac{\mu_{L_2(s),0}-\hat{\mu}_{L_2,0}}{\hat{\sigma}_{L_2,0}}=O\{(\log p)^{5/4}n^{-1/2}\},
%\]
%which both satisfy
(\ref{res}) is satisfied under (C1) {and thus can be employed in the
 formulation of a test procedure}.
%\footnote{Jun, can you provide the specific forms of the estimators for $\mu$ and $\sigma$ here ? I have added the estimators (Jun)} \footnote{Jun, could you supply the power of the test for $s=1^{-}$ when the signal is strong.I have added the power under strong %signals (Jun)}

{The asymptotic power of the thresholding test
(\ref{eq:thresholding-strong})  is}
\begin{eqnarray}
\beta_{L_n}(||\bm{\mu}_1-\bm{\mu}_2||)
=\Phi\biggl(-\frac{z_{\alpha}\sigma_{L_n(s),0}}{\sigma_{L_n(s),1}}+\frac{\mu_{L_n(s),1}-\mu_{L_n(s),0}}{\sigma_{L_n(s),1}}\biggr),\nonumber
\end{eqnarray}
{which, similar to the CQ and the Oracle tests, is largely
determined by} %its signal-to-noise ratio}
\bea \mbox{SNR}_{L_n} &=:&
\frac{\mu_{L_n(s),1}-\mu_{L_n(s),0}}{\sigma_{L_n(s),1}} \nn \\
%Under strong
%signals, choosing $s = 1^-$ leads to
%\begin{eqnarray}
%\mbox{SNR}_{L_n}
&= &
\frac{p^{1-\beta}\bar{\delta}^2}{\sqrt{2L_p+2p^{1-\beta}+2\sum_{k
\ne l \in S_{\beta}}\rho_{kl}^2+4n\,\sum_{k ,l \in
S_{\beta}}\delta_k\,\delta_l\,\rho_{kl}}},\label{test1_power1}
 \eea
which is much larger  than that of the CQ test in (\ref{CQ_power})
and differs from that of the Oracle test given in
(\ref{Oracle_power}) only by {a slowly varying multi-$\mbox{log} p$ function $L_p$}. This echoes
that established in Fan (1996) for Gaussian data with no dependence
among the column components of the data.
%{By choosing
%$s=1-d\log\{c\log(p)\}/\log(p)$ for some constants $c$ and $d$, Fan
%(1996) showed that a hard-threshold test can attain the power of an
%Oracle test within an order of a slowly varying factor
%
%$\sqrt{\log\{\log(p)\}}$. }

\setcounter{section}{3} \setcounter{equation}{0}
\section*{\large 3. Multi-Level Thresholding}
%\footnote{I think we may put this section first. I modified (Jun)} }

It is  shown in Section 2 that if all the  signals are strong such that $n \delta_k > 2 \mbox{log}p$,  a single thresholding with $s=1^{-}$
 improves significantly the power of the test and attains nearly the power of the Oracle test.
% is better than that of the CQ test.
However, if some signals are weak such that  $n\,\delta_k^2= 2r \mbox{log}p$ with $r < 1$ for some  $k \in S_{\beta}$,
the thresholding has to be administrated at smaller levels  $2s\mbox{log}p$ for $s \in (0,1)$.  In this case,
 the single-level thresholding does not work well. %since the signal strengths are  unknown and weaker.
One approach that provides a solution to such situation is the higher criticism test
(Donoho and Jin, 2004) which effectively combines many levels of
thresholding together to formulate a higher criticism (HC)
criterion.  Zhong, Chen and Xu (2013) proposed a more powerful  test
procedure than the HC test under sparsity and data dependence. Both
Donoho and Jin (2004)'s HC test and the test proposed in Zhong et
al. (2013) are for one sample, and both did not provide much details
on the power performance.

The multi-level thresholding statistic is
\begin{eqnarray}
M_{L_n}=\max \limits_{s\in (0,1-\eta)}\frac{L_n(s)- \hat{\mu}_{L_n(s),0}}{\hat{\sigma}_{L_n(s),0}}.\label{eqn:max1}
\end{eqnarray}
Maximizing over the thresholding statistics at multiple levels allows faint and unknown signals to be captured.
% by the test statistic. }
%One way to choose $s$ is to maximize $\sigma_{L_n(s),0}^{-1}\{L_n(s)-\mu_{L_n(s),0}\}$ in order to best differentiate the null and alternative hypothesis. The same idea has been applied {\bf for one sample tests} by Donoho and Jin (2004) for the Higher Criticism %test and
Since both $\hat{\mu}_{L_n(s),0}$ and $\hat{\sigma}_{L_n(s),0}$ are monotonically decreasing and $L_n(s)$ contains indicator functions, provided
 (\ref{res}) is satisfied, it can be shown that  the maximization in (\ref{eqn:max1})
 %$\hat{\sigma}_{L_n(s),0}^{-1}\{L_n(s)-\hat{\mu}_{L_n(s),0}\}$
 is attained over
$\mathcal{S}_n=\{s_k: s_k=n(\bar{X}_1^{(k)}-\bar{X}_1^{(k)})^2/(2\mbox{log}p), \mbox{for} \, k=1, \cdots, p\}\cap(0,1-\eta)$
so that
\begin{eqnarray}
M_{L_n}=\max \limits_{s\in \mathcal{S}_n}\frac{L_n(s)-\hat{\mu}_{L_n(s),0}}{\hat{\sigma}_{L_n(s),0}}.  \label{max1}
\end{eqnarray}
The following theorem shows that  $M_{L_n}$ is asymptotically Gumbel distributed.

\textbf{Theorem 2.} Assume Conditions \textbf{(C1)}-\textbf{(C3)}
and {condition (\ref{res}) is satisfied}. Then under $H_0$,
\[
\mbox{P}\biggl\{a(\mbox{log}p)M_{L_n}-b(\mbox{log}p,\eta) \le x  \biggr\} \to \mbox{exp}(-e^{-x}),
\]
where functions $a(y)=(2\mbox{log}y)^{\frac{1}{2}}$ and $b(y,\eta)=2\mbox{log}y+2^{-1}\mbox{log}\mbox{log}y-2^{-1}\mbox{log}\{\frac{4\pi}{(1-\eta)^2}\}$.

The theorem implies that a two-sample multi-level thresholding test of  asymptotic $\alpha$ level  rejects $H_0$ if
\be
M_{L_n} \ge G_{\alpha}=\{q_{\alpha}+b(\mbox{log}p,\eta)\}/a(\mbox{log}p), \label{eq:multi-test}
\ee
where $q_{\alpha}$ is the upper $\alpha$ quantile of the Gumbel distribution $\mbox{exp}(-e^{-x})$.

Define
\begin{eqnarray}
 \varrho(\beta) = \left\{
  \begin{array}{c  l}
    \beta-\frac{1}{2}, & \quad  \frac{1}{2}\le \beta \le \frac{3}{4};\\
    (1-\sqrt{1-\beta})^2, & \quad \frac{3}{4}< \beta <1.
\end{array} \right. \label{b1}
\end{eqnarray}
{Ingster (1997) shows that $r= \varrho(\beta)$ is the optimal
detection boundary for uncorrelated Gaussian data in the sense that
when $(r, \beta)$ lays above the phase diagram $r= \varrho(\beta)$,
there are tests whose probabilities of type I and type II errors
converge to zero simultaneously as $n \to \infty$, and if $(r,
\beta)$ is below the phase diagram, no such  test exists. Donoho and
Jin (2004) showed that the HC test attains $r= \varrho(\beta)$ as
the detection boundary when $\bm{X}_i$ are IID $\mbox{N}({\mu},
{I}_p)$ data. Zhong et al. (2013) showed that the $L_1$ and
$L_2$-versions of the HC tests also attain $r= \varrho(\beta)$ as
the detection boundary  for non-Gaussian data with column-wise
dependence,  and have more attractive power for $(r, \beta)$ further
above the detection boundary. }

\textbf{Theorem 3.} Assume Conditions \textbf{(C1)}-\textbf{(C3)} and $\hat{\mu}_{L_n(s),0}$ and $\hat{\sigma}_{L_n(s),0}$ satisfy (\ref{res}). If $r>\varrho(\beta)$, the sum of type I and II errors of the multi-level thresholding test converges to zero when $\alpha=\bar{\Phi}\{(\mbox{log}p)^{\epsilon}\}\to 0$ for an arbitrarily small $\epsilon>0$ as $n \to \infty$. If $r<\varrho(\beta)$, the sum of type I and II errors of the multi-level thresholding test converges to 1 as $\alpha \to 0$ and $n \to \infty$.

Theorem 3 implies  that the two-sample multi-level thresholding test also attains $r=\varrho(\beta)$ as the detection boundary in the current
two-sample test setting of nonparametric distributional assumption.  This means that the test can asymptotically distinguish $H_1$ from $H_0$
for any $(r, \beta)$  above the detection boundary. If the mean and variance estimators $\hat{\mu}_{L_n(s),0}$ and $\hat{\sigma}_{L_n(s),0}$
do not satisfy (\ref{res}), the detection boundary will be higher just like what will happen in Theorem 6 given in Section 4 when we consider
testing via data transformation with estimated precision matrix.

\setcounter{section}{4} \setcounter{equation}{0}
\section*{\large 4.  Test with Data Transformation}

%We have seen in the previous sections that the signal-to-noise ratio of the %test can be improved by the thresholding that reduces the variance of the test %statistic. Such variance reduction leads to a single level  thresholding test that %attains almost the same performance as the Oracle test in the case of strong %signals, and the multi-level thresholding test that attains the detection %boundary $r= \varrho(\beta)$ when the signals are weak.

We consider in this section another way for power improvement,  which involves enhancing the signal strength by data rotation, inspired by the works of Hall and Jin (2010) and Cai, Liu and Xia
(2014).  We will show in this section that the signal enhancement can be achieved by transforming the data via an estimate of the inverse of a mixture
 of $\bm{\Sigma}_1$ and $\bm{\Sigma}_2$.
%{\footnote{need to introduce the matrix class. I have copied over here. {\bf Since we do not assume exponential decay structure of original data, I deleted that sentence saying both have exponential decay structure (Jun).}}
Transforming data to achieve better power has been considered in
Hall and Jin (2010)  in their  innovated higher criticism test under
dependence and Cai, Liu and Xia (2014) in their max-norm based test.
The transformation used in Hall and Jin (2010) was via a banded
%\footnote{Did they consider banding? {\bf Yes. They band the
%Cholesky factor $U$ which satisfies $U\Sigma U^{\prime}$ to obtain
%$\bar{U}$. Then transform data by $\bar{U}^{\prime} U$. (Jun)}}
Cholesky factor, and that adopted in Cai, Liu and Xia (2014) was via
the CLIME estimator of the inverse of the covariance  matrix (the
precision matrix) proposed in Cai, Liu and Luo (2011).
%Both
%approaches essentially transform the data via the precision matrix.

Consider a bandable covariance matrix class
%\footnote{we need a discussion on the class, who used it, for what purposes, cite Bickel and Levina's papers for sure.  And better connect below. {\bf I added some discussions about this class (Jun)}}
\begin{eqnarray}
V(\epsilon_0, C, \alpha)&=&\biggl\{\bm{\Sigma}: 0<\epsilon_0\le \lambda_{\mbox{min}}(\bm{\Sigma}) \le \lambda_{\mbox{max}}(\bm{\Sigma}) \le \epsilon_0^{-1}, \alpha >0, \nonumber\\
&\qquad& |\sigma_{ij}|\le C(1+|i-j|)^{-(\alpha+1)} \quad \mbox{for all} \quad i, j: |i-j|\ge 1 \biggr\}. \nonumber
\end{eqnarray}
This class of matrices satisfies both the banding and thresholding conditions of Bickel and Levina (2008b). Hall and Jin (2010) also considered this class
when they proposed the innovated higher criticism test under dependence.
%Specially, it has been shown by Jaffard (1990), Sun (2005),
%Gr\"{o}chenig, and Leinert (2006) that if a matrix has polynomial off-diagonal decay, its inverse also has polynomial off-diagonal decay with the same
%rate.
%This property ensures that after the transformation, data are still weakly dependent.

\textbf{(C4)}: Both $\bm{\Sigma}_1$ and $\bm{\Sigma}_2$ belong to the matrix class  $V(\epsilon_0, C, \alpha)$.

%\textbf{(C5)}: The coordinates of those non-zero $\delta_l$ are randomly generated from $\{1,2,\cdots,p \}$.
%As we will show in Appendix, the variance of the transformed thresholding test statistic will be similar to that of the thresholding test statistic.
%\footnote{This may be moved later after we show that it can increase the signals. The %sentence is removed (Jun)}

{Although both \textbf{(C3)} and \textbf{(C4)} assume the weak dependence among the column components of the random vector $X_{ij}$, imposing \textbf{(C4)} ensures that the banding estimation of the covarianvce matrix which makes the transformed data are still weakly dependent. To appreciate this,  let $\bm{\Omega}=\{(1-\kappa)\bm{\Sigma}_1+\kappa\bm{\Sigma}_2\}^{-1}=
(\omega_{ij})_{p \times p}$. We first assume $\bm{\Omega}$ is known to
gain insight on the test.
 Rather than
transforming the data via $\bm{\Omega}$,  we transform it via
\[
\bm{\Omega}(\tau)=\biggl\{\omega_{ij}\mbox{I}(|i-j| \le \tau)\biggr\}_{p\times p},
\]
a banded version of $\bm{\Omega}$ for   an integer $\tau$  between $1$
and $p-1$.   There are two reasons to use $\bm{\Omega}(\tau)$.
% rather than $\bm{\Omega}$
 %for the transformation.
 One is that the signal
enhancement is facilitated mainly by elements of $\bm{\Omega}$ close to
the main diagonal.  Another is that the banding maintains the
$\alpha$-mixing structure of the transformed data provided $k-2\tau
\to \infty$.
{Since  both $\bm{\Sigma}_{1}$ and $\bm{\Sigma}_2$ have their off-diagonal
entries decaying to zero at polynomial rates,
 $\bm{\Omega}$
 has the same rate of decay as well (Jaffard, 1990; Sun, 2005; Gr\"{o}chenig, and Leinert,
 2006),
which ensures that the transformed data are still weakly
dependent.}

The two transformed samples are
%\footnote{What matrix classes assumed in HJ and CLX respectively? {\bf In HJ, the same polynomial decay structure was assumed for $\bm{\Sigma}$. However, in CLX, they assume weaker conditions on estimator $\hat{\bm{\Omega}}$: $||\hat{\bm{\Omega}}-\bm{\Omega}||_{L_1}=o_p\{1/log(p )\}$ and $\max \limits_{1\le i \le p}|\hat{\omega}_{ii}-\omega_{ii}|=o_p\{1/log(p )\}$ }.    We also need to cite a few more papers on the banding estimation of a precision matrix. {\bf I added some reference on page 18 when discussing estimation. (Jun)}}
$$\{\bm{Z}_{1j}(\tau)=\bm{\Omega}(\tau) \bm{X}_{1j}:1\le j \le n_1\} \quad \hbox{and} \quad \{\bm{Z}_{2j}(\tau)=\bm{\Omega}(\tau) \bm{X}_{2j}:1\le j \le n_2\}.$$
%of the original samples   $\{X_{1j}:1\le j \le n_1\}$ and $\{X_{2j}:1\le j \le n_2\}$.
Let
$\varpi_{kk}(\tau)=\mbox{Var}\{\sqrt{n}(\bar{Z}_{1}^{(k)}(\tau)-\bar{Z}_{2}^{(k)}(\tau))\}$ be
 the counterpart of $n(\sigma_{1, kk}/n_1 + \sigma_{2,
kk}/n_2)$ for the transformed data where $\bar{Z}_{i}^{(k)}(\tau)=n_i^{-1}\sum_{j=1}^{n_j}{Z}_{ij}^{(k)}(\tau)$ for $i=1,2$. Lemmas 5 and 7 in Appendix show that
there exists a constant $C>1$ such that
\be \varpi_{kk}(\tau)=\omega_{kk}+O(\tau^{-C}) \quad \hbox{and}
\quad \omega_{kk}
> 1. \label{eq:omegakk} \ee

We have two ways to construct the transformed thresholding test statistic by replacing $\bm{X}_{ij}$ with $\bm{Z}_{ij}(\tau)$ in either (\ref{thr1}) or (\ref{thr2}). Alhough both have similar properties, the latter which has the form
% expectation and thus is chosen to be the transformed thresholding test statistic:
\begin{eqnarray}
J_n(s,\tau)=\sum_{k=1}^{p}\biggl\{\frac{n(\bar{Z}_{1}^{(k)}(\tau)-\bar{Z}_{2}^{(k)}(\tau))^2}{\varpi_{kk}(\tau)}-1 \biggr\}I\biggl\{\frac{n(\bar{Z}_{1}^{(k)}(\tau)-\bar{Z}_{2}^{(k)}(\tau))^2}{\varpi_{kk}(\tau)}>\lambda_{n}(s)\biggr\}\label{eq2-new}
\end{eqnarray}
is easier to work with, which we will present in the following.

{Let  $\delta_{\bm{\Omega}(\tau)}=(\delta_{\bm{\Omega}(\tau),1},\cdots, \delta_{\bm{\Omega}(\tau),p})^T$ where
\be
\delta_{\bm{\Omega}(\tau),k}=\sum_{l}\bm{\Omega}_{kl}(\tau)\delta_l=\sum_{l\in S_{\beta}}\omega_{kl}\delta_l\mbox{I}(|k-l|\le \tau) \label{new_signal}
\ee
denotes the difference between the transformed means in the $k$-th dimension.
Similar to (\ref{mean_thr}) and (\ref{variance_thr}), the mean and
variance of the transformed statistic $J_n(s,\tau)$ are
\begin{eqnarray}
\mu_{J_n(s,\tau)}&=&\biggl(\frac{2}{\sqrt{2\pi}}(2s\mbox{log}p)^{\frac{1}{2}} p^{1-s}+\sum_{k \in S_{\bm{\Omega}(\tau),\beta}}\{n\,\frac{\delta_{\bm{\Omega}(\tau),k}^2}{\varpi_{kk}(\tau)}I(n\,\frac{\delta_{\bm{\Omega}(\tau),k}^2}{\varpi_{kk}(\tau)}>2s\mbox{log}p)\nonumber\\
&+& (2s\mbox{log}p)\bar{\Phi}(\eta_{\bm{\Omega}(\tau) k}^-) I(n\,\frac{\delta_{\bm{\Omega}(\tau),k}^2}{\varpi_{kk}(\tau)}<2s\mbox{log}p)\}\biggr)\{1+o(1)\}, \label{mean_thr2}
\end{eqnarray}
and
\begin{eqnarray}
\sigma_{J_n(s,\tau)}^2
&=&\biggl(\frac{2}{\sqrt{2\pi}}\{(2s\mbox{log}p)^{\frac{3}{2}}+(2s\mbox{log}p)^{\frac{1}{2}}\}p^{1-s}+\sum_{k,l \in S_{\bm{\Omega}(\tau),\beta}}(4n\frac{\delta_{\bm{\Omega}(\tau),k}}{\varpi_{kk}^{1/2}(\tau)} \frac{\delta_{\varpi(\tau),l}}{\varpi_{ll}^{1/2}(\tau)}\rho_{\bm{\Omega}, kl}\nonumber\\
&+&2\rho_{\bm{\Omega}, kl}^2)I(n\frac{\delta_{\bm{\Omega}(\tau),k}^2}{\varpi_{kk}(\tau)}>2s\mbox{log}p)I(n\,\frac{\delta_{\bm{\Omega}(\tau),l}^2}{\varpi_{ll}(\tau)}>2s\mbox{log}p)\nonumber\\
&+&\sum_{k \in
S_{\bm{\Omega}(\tau),\beta}}(2s\mbox{log}p)^2\bar{\Phi}(\eta_{\bm{\Omega}(\tau)
k}^-)I(n\,\frac{\delta_{\bm{\Omega}(\tau),k}^2}{\varpi_{kk}(\tau)}<2s
\mbox{log}p)\biggr)\{1+o(1)\}, \label{variance_thr2}
\end{eqnarray}
where {$S_{\bm{\Omega}(\tau),\beta}=\{k: \delta_{\bm{\Omega}(\tau),k}\ne 0 \}$ is the set of locations of the non-zero signals $\delta_{\bm{\Omega}(\tau),k}$}, $\eta_{\bm{\Omega}(\tau) k}^-=(2s\mbox{log}p)^{1/2}-n^{1/2}\delta_{\bm{\Omega}(\tau),k}/\varpi_{kk}(\tau)^{1/2}$ and
\[
\rho_{\bm{\Omega},kl}= \mbox{Cov}\biggl\{\frac{\sqrt{n}(\bar{Z}_1^{(k)}(\tau)-\bar{Z}_2^{(k)}(\tau))}{\sqrt{\varpi_{kk}(\tau)}},\frac{\sqrt{n}(\bar{Z}_1^{(l)}(\tau)-\bar{Z}_2^{(l)}(\tau))}{\sqrt{\varpi_{ll}(\tau)}}\biggr\}.
\]

In practice, the precision matrix $\bm{\Omega}$ is unknown and needs to
be estimated.
{We consider the Cholesky decomposition and the banding approach
similar to that in Bickel and Levina (2008a). Define
$\bm{Y}_{kl}=\bm{X}_{1k}-\sqrt{\frac{\kappa}{1-\kappa}} \bm{X}_{2l}$ for
$k=1,\cdots, n_1$ and $l=1,\cdots,n_2$, where $\kappa=\lim
\limits_{n\to \infty} n_1/(n_1+n_2)$. Then
$\mbox{Var}(Y_{kl})=\bm{\Sigma}_{w}\equiv\bm{\Sigma}_1+\frac{\kappa}{1-\kappa}\bm{\Sigma}_2$.
Thus, to estimate $\bm{\Omega}=(1-\kappa)^{-1}\bm{\Sigma}_{w}^{-1},$
we only need to estimate $\bm{\Sigma}_{w}^{-1}$.

%We first introduce the population version of Cholesky decomposition
%of $\bm{\Sigma}_{w}^{-1}$.
Let $\bm{Y}$ be an IID copy of $\bm{Y}_{kl}$ for any fixed $k$ and $l$ such
that  $\bm{Y}=(Y^{(1)},\cdots, Y^{(p)})^T$.
 For $j=1,\cdots,p$, define
$\hat{Y}^{(j)}={\bf a}_{j}^T{\bf W}^{(j)}$ where
${\bf a}_j=\{\mbox{Var}({\bf
W}^{(j)})\}^{-1}\mbox{Cov}(\hat{Y}^{(j)},{\bf W}^{(j)})$ and
%${\bf a}_j=(a_{j1},\cdots,a_{j,j-1})^T$ and
 ${\bf
W}^{(j)}=(Y^{(1)},\cdots,Y^{(j-1)})^T$.
% are vectors of length $j-1$.
Let
$\epsilon_{j}={Y}^{(j)}-\hat{Y}^{(j)}$ and
$d_j^2=\mbox{Var}(\epsilon_{j})$, and $\bm{A}$ be the lower
triangular matrix with the $j$-th row being $({\bf a}_j^T,{\bf
0}_{p-j+1})$ and $\bm{D}=\mbox{diag}(d_1^2,\cdots,d_p^2)$ where
$\mathbf{0}_{s}$ means a vector of $0$ with length $s$. Then,  the
population version of Cholesky decomposition is
$\bm{\Sigma}_{w}^{-1}=(I-\bm{A})^T\bm{D}^{-1}(I-\bm{A}).$

The banded estimators for $\bm{A}$
and $\bm{D}$ (Bickel and Levina, 2008a) can be used in the case of $p>\min\{n_1,n_2\}$.
Specifically, let $\bm{Y}_{n,kl}=\bm{X}_{1k}-\sqrt{\frac{n_1}{n_2}}
\bm{X}_{2l}:=(Y_{n,kl}^{(1)},\cdots, Y_{n,kl}^{(p)})^T$. Given a
$\tau$, regress $Y_{n,kl}^{(j)}$ on
$\mathbf{Y}_{n,kl,-\tau}^{(j)}=(Y_{n,kl}^{(j-\tau)},\cdots,Y_{n,kl}^{(j-1)})^T$
to obtain  the least square estimate of
$\mathbf{a}_{j,\tau}=(a_{j-\tau},\cdots,a_{j-1})^T$:
$$\hat{\mathbf{a}}_{j,\tau}=(\sum_{k=1}^{n_1}\sum_{l=1}^{n_2}\mathbf{Y}_{n,kl,-\tau}^{(j)}\mathbf{Y}_{n,kl,-\tau}^{(j)^T})^{-1}\sum_{k=1}^{n_1}\sum_{l=1}^{n_2}\mathbf{Y}_{n,kl,-\tau}^{(j)}\mathbf{Y}_{n,kl}^{(j)}.$$
Put
 $\hat{\mathbf{a}}_j^T=(\mathbf{0}_{\tau-1}^T,\hat{\mathbf{a}}_{j,\tau}^T,\mathbf{0}_{p-j+1}^T)$
be the $j$-th row of  a lower triangular matrix $\hat{A}_\tau$ and
$\hat{D}_{\tau}=\mbox{diag}(d_{1,\tau}^2,\cdots,d_{p,\tau}^2)$ where
$d_{j,\tau}^2=\frac{1}{n_1n_2}\sum_{k=1}^{n_1}\sum_{l=1}^{n_2}(Y_{n,kl}^{(j)}-\hat{\mathbf{a}}_{j,\tau}^T\mathbf{Y}_{n,kl,-\tau}^{(j)})^2.$
Thus,  the estimator of ${\bm{\Sigma}_{w}^{-1}}$ is
\begin{eqnarray}
\widehat{\bm{\Sigma}_{w}^{-1}}=(I-\hat{A}_\tau)^T\hat{D}_{\tau}^{-1}(I-\hat{A}_\tau),\label{chole_inv}
\end{eqnarray}
which results in
$\hat{\bm{\Omega}}_{\tau}=\{1-{n_1}/{(n_1+n_2})\}^{-1}\widehat{\bm{\Sigma}_{w}^{-1}}.$

The consistency of $\hat{\bm{\Omega}}_{\tau}$ to $\bm{\Omega}$ basically  follows
 the proof of Theorem 3 in Bickel and Levina (2008a) with a main
difference that  replaces the exponential tail inequality for a
sample mean in Lemma A.3 of their paper to an exponential inequality
of a two-sample U-statistics.  Moreover, if the banding parameter
$\tau \asymp (n^{-1}\mbox{log}p)^{-\frac{1}{2(\alpha+1)}}$
%\footnote{$n^{- {1\over 2 (\alpha+1)}}$?{\bf I made the expression clearer (Jun)} }
 and $n^{-1}\mbox{log}p=o(1)$,  it can be shown that
 %, similar to Bickel and Levina
 %(2008a), that
\[
|| \hat{\bm{\Omega}}_{\tau}-\bm{\Omega} ||= O_p\bigg\{
(\mbox{log}p/n)^{\frac{\alpha}{2(\alpha+1)}}\biggr\},
\]
where $\|\cdot\|$ is the spectral  norm.
%\footnote{Do you mean spectral norm? I think matrix $L_2$ norm is the spectral norm. They are exchangeable. I checked in Bickel and Levina, they called it matrix $L_2$ norm. But we can change. (Jun)}
%where $\hat{\Omega}_{\tau}$ can be obtained by the Cholesky decomposition and banding.

The transformed thresholding test statistic based on $\{\hat{\bm{Z}}_{1i}=\hat{\bm{\Omega}}_{\tau} \bm{X}_{1i}:1\le i \le n_1\}$ and $\{\hat{\bm{Z}}_{2i}=\hat{\bm{\Omega}}_{\tau} \bm{X}_{2i}:1\le i \le n_2\}$ is
\begin{eqnarray}
\hat{J}_n(s,\tau)=\sum_{k=1}^{p}\biggl\{\frac{n(\bar{\hat{Z}}_1^{(k)}-\bar{\hat{Z}}_2^{(k)})^2}{\hat{\omega}_{kk}}-1 \biggr\}I\biggl\{\frac{n(\bar{\hat{Z}}_1^{(k)}-\bar{\hat{Z}}_2^{(k)})^2}{\hat{\omega}_{kk}}>\lambda_{n}(s)\biggr\}.\label{eq2-estimate}
\end{eqnarray}
%which is obtained by replacing  $\Omega(\tau)$ %\footnote{Be more specific. It should be $\omega_{kk}(\tau)$ right by their counterparts of $\hat{\Omega} (\tau)$ {\bf Dr. Chen: we not only replace the denominator but also in transformed data $Z$.(Jun)}}
% with the estimated $\hat{\Omega}_{\tau}$ in (\ref{eq2-new}).

To consistently estimate $\bm{\Omega}$, we require that $\tau \asymp
(n^{-1}\mbox{log}p)^{-\frac{1}{2(\alpha+1)}}$. This requirement
leads to a modification on the range of the  thresholding level $s$
as shown in the next theorem.

\textbf{Theorem 4.} Assume Conditions
\textbf{(C1)}-\textbf{(C4)}.
%\footnote{I suggest move the part (a)
%into appendix to avoid the possible confusion on the thresholding
%parameters (Pingshou). Yes, two different choice of $\tau$ can cause
%confusion. So Dr. Chen, I just go ahead to move part (a) into
%appendix. It won't affect the main content. (Jun).   }
If $p=n^{1/\theta}$ for $0<\theta<1$ and $\tau \asymp (n^{-1}\mbox{log}p)^{-\frac{1}{2(\alpha+1)}}$, then for any $s\in (1-\theta, 1)$,
%where $\eta^{\star}$ is any arbitrary small number in $(0,1)$ such that $\eta^{\star}<\theta$ then
 %and $q$ is any number in $(1/(1+\alpha),1)$,
%\footnote{for any such $\eta^{\star}$ ?: {\bf If the Slustsky's theorem holds. Yes. the value of $\eta^\star$ should be the same as those in Theorem 4.}}
\[
\sigma_{J_n(s,\tau),0}^{-1} \biggl\{ \hat{J}_n(s,\tau)-\mu_{J_n(s,\tau),0}\biggr\}
 \xrightarrow{d} \mbox{N}(0,1).
\]
%\end{itemize}

The restriction on the thresholding level  $s$ in
Theorem 4 is to ensure the estimation error of
$\hat{\bm{\Omega}}_{\tau}$ is negligible. Similar restriction is
provisioned in
 Delaigle et al. (2011) and Zhong et al. (2013). Note that if  $\theta$ is arbitrarily close to 0, $p$ will  grow exponentially fast with $n$.
%As a result, the thresholding level $s$ should be higher.}

A single-level thresholding test based on the transformed data rejects $H_0$ if
\[
\hat{J}_n(s,\tau)> z_{\alpha}\hat{\sigma}_{J_n(s, \tau),0}+\hat{\mu}_{J_n(s, \tau),0},
\]
where $\hat{\mu}_{J_n(s, \tau),0}$ and $\hat{\sigma}^2_{J_n(s, \tau),0}$ are, respectively, consistent estimators of
\[
{\mu}_{J_n(s, \tau),0} =\biggl\{\frac{2}{\sqrt{2\pi}}(2s\mbox{log}p)^{\frac{1}{2}} p^{1-s}\biggr\}\{1+o(1)\},
\] and
\[
{\sigma}_{J_n(s, \tau),0}^2=\biggl\{\frac{2}{\sqrt{2\pi}}\{(2s\mbox{log}p)^{\frac{3}{2}}+(2s\mbox{log}p)^{\frac{1}{2}}\}p^{1-s}\biggr\}\{1+o(1)\},
\]
%\footnote{define both.. Where ? Please fill in above ???{\bf Done (Jun)}}
satisfying $
\mu_{J_n(s, \tau),0}-\hat{\mu}_{J_n(s, \tau),0}= o\{{\sigma}_{J_n(s,
\tau),0} \} \quad \hbox{and} \quad  \hat{\sigma}_{J_n(s, \tau),0}
/{\sigma}_{J_n(s, \tau),0} \stackrel{p} \to 1.$

From Theorem 4, the asymptotic power of the transformed thresholding test is %\footnote{The argument here is not clear enough. It is very much to Jun himself. {\bf Dr. Chen: You mean the argument about comparison in powers between thresholding test and the transformed thresholding test? (Jun)}}
\begin{eqnarray}
\beta_{\hat{J}_n(s,\tau)}(||\mu_1-\mu_2||)
=\Phi\biggl(-\frac{z_{\alpha}\sigma_{J_n(s, \tau),0}}{\sigma_{J_n(s, \tau),1}}+\frac{\mu_{J_n(s, \tau),1}-\mu_{J_n(s, \tau),0}}{\sigma_{J_n(s, \tau),1}}\biggr),\nonumber
\end{eqnarray}
which is determined by % its signal-to-noise ratio
\[
\mbox{SNR}_{\hat{J}_n(s,\tau)}=:\frac{\mu_{J_n(s,\tau),1}-\mu_{J_n(s,\tau),0}}{\sigma_{J_n(s,\tau),1}}.
\]
Therefore, to compare with the thresholding test without transformation,
 it is equivalent to compare $\mbox{SNR}_{\hat{J}_n(s,\tau)}$ to $\mbox{SNR}_{L_n}$.
  To this end, we assume the following regarding the distribution of the non-zero  $\delta_k$ in $S_{\beta}$.

  \bigskip

\textbf{(C5)}: The elements of $S_{\beta}$ are randomly distributed
among $\{1,2,\cdots,p \}$.

\bigskip

Under Conditions (C1)-(C5), Lemma 8 in the Appendix shows that with
probability approaching to 1, \be \mbox{SNR}_{\hat{J}_n(s,\tau)} \ge
\mbox{SNR}_{L_n}, \label{snr_comp} \ee which holds for both strong
and weak signals. Hence, the transformed thresholding test is
more powerful regardless of the underlying signal strength for
randomly allocated signals.}

Similar to  $M_{L_n}$ defined in (\ref{max1}) for weaker signals,  a multi-level thresholding statistic  for transformed data is
\begin{eqnarray}
M_{\hat{J}_n}=\max \limits_{s \in T_n} \frac{ \hat{J}_n(s, \tau)-\hat{\mu}_{J_n(s,\tau),0}}{\hat{\sigma}_{J_n(s,\tau),0}},\label{max2}
\end{eqnarray}
where $T_n=\{s_k:
s_k=n(\bar{\hat{Z}}_1^{(k)}-\bar{\hat{Z}}_2^{(k)})^2/(2\mbox{log}p\hat{\omega}_{kk})$
 for $k=1, \cdots, p\}\cap (1-\theta, 1-\eta^{\star})$ for
arbitrarily small $\eta^{\star}$. The asymptotic distribution of
$M_{\hat{J}_n}$ is given in the following Theorem.

\textbf{Theorem 5.} Assume Conditions \textbf{(C1)}-\textbf{(C4)}, $p=n^{1/\theta}$ for $0<\theta<1$ and $\tau \asymp (n^{-1}\mbox{log}p)^{-\frac{1}{2(\alpha+1)}}$. Then under $H_0$,
\[
\mbox{P}\biggl\{a(\mbox{log}p)M_{\hat{J}_n}-b(\mbox{log}p,\theta-\eta^{\star}) \le x  \biggr\} \to \mbox{exp}(-e^{-x}),
\]
where functions $a( \cdot )$ and $b(\cdot, \cdot)$ are defined in Theorem 2.

The theorem implies  an asymptotically $\alpha$ level test that
rejects $H_0$ if \be M_{\hat{J}_n} \ge
\{q_{\alpha}+b(\mbox{log}p,\theta-\eta^{\star})\}/a(\mbox{log}p).
\label{eq:multi-est-test} \ee

 It is  expected that the above test as well as the thresholding test without the data transformation will encounter size distortion. The size distortion is caused by the generally slow convergence to the extreme value distribution.
It may be also due to
%  Our investigation reveals that some of the size distortion comes from
the second order effects of the data dependence.  Our analyses have shown that the data dependence has no leading order effect on the asymptotic variance of the thresholding test statistics. However, a closer examination on the variance shows that the second order term is not that smaller than the leading order variance.
This can create a discrepancy when approximating the distribution of the multi-level thresholding statistics by the Gumbel distribution.
%Specifically,  let us
% recall that in (\ref{max1}), the test statistics $L_n(s)=\sum_{k=1}^p L_{n,k}
%(s)$ is standardized by $\hat{\mu}_{L_n(s), 0}$ and $\hat{\sigma}_{L_n(s), %0}$, %which are obtained by only keeping the leading order term.  Specifically  %$\hat{\sigma}_{L_n(s), 0}$ is obtained by keeping $\sum_{k=1}^p %\mbox{Var}\{L_{n,k}(s)\}$ and ignoring the part induced by the correlation.
%Although the covariance parts are asymptotically negligible, they can  %contribution significantly in finite samples. The same can be said to the test %statistics with data transformation given by (\ref{max2}).
To remedy the problem,
%the size of both Mult1 and Mult2 under finite dimensionality,
we proposed a parametric bootstrap approximation to the null distribution of the multi-level thresholding statistics 
with and without the data transformation.  We first estimate $\Sigma_i$ by $\hat{\Sigma}_i$ for $i=1, 2$ 
through the Cholesky decomposition which can be obtained by inverting the one-sample version of (\ref{chole_inv})
%\footnote{can we cite an equation number ? I think the Cholesky estimation of $\Sigma^{-1}$ can be easily adapted for 
%estimation of $\Sigma$. {\bf Dr. Chen: I cited an equation. You are right. If banding the inverse, 
%the resulted Cholesky decomposition is non-negative definite and thus invertible. Therefore, 
%$\hat{\Sigma}$ can be directly obtained by inverting $\hat{\Sigma}^{-1}$}} 
 based on the samples $\{\bm{X}_{1j}\}_{j=1}^{n_1}$ and $\{\bm{X}_{2j}\}_{j=1}^{n_2}$, respectively. 
 Bootstrap resamples are generated repeatedly from $\mbox{N}(0, \hat{\Sigma}_i)$ 
 which allows us to obtain the bootstrap copies of the statistic $M_{L_n}$ defined in (\ref{max1}), 
 namely  $M_{L_n}^{\ast, (1)}(s), \cdots, M_{L_n}^{\ast, (B)}(s)$, after $B$ repetitions.  %We use the mean and  variance of $\{L_2^{\ast, (b)}\}_{b=1}^B$  to replace $\hat{\mu}_{L_n(s), 0}$ and $\hat{\sigma}_{L_n(s), 0}$, respectively, in the standardization.
We use $\{M_{L_n}^{\ast, (b)}\}_{b=1}^B$ to obtain the empirical null distribution of the multi-level thresholding statistic.
The same parametric bootstrapping method can be also applied to the transformed multi-level thresholding statistic.

We have shown that   %\footnote{Jun, shall we move this part on transformation to the next section ? I have removed to the right section (Jun)}
 the transformed thresholding test has a better power performance than the thresholding test without the transformation.  We are to show that
%A carefully investigation as follows will demonstrate that the transformed
the transformed multi-level thresholding test has lower detection
boundary than the multi-level thresholding test without
transformation.
%A direct consequence is that  there exist some regions where the multi-level thresholding test is powerless but the transformed multi-level thresholding test is powerful.

To define the detection boundary of the transformed multi-level thresholding test, let
\[
\underline{\omega} = \underline{\lim}_{p\to \infty}
\biggl(\min\limits_{1\le k \le p}\omega_{kk}\biggr)  \quad
\mbox{and} \quad  \bar{\omega} = \overline{\lim}_{p\to \infty}
\biggl(\max\limits_{1\le k \le p}\omega_{kk}\biggr).
\]
Results in (4.1) imply that  $\underline{\omega}$ and $
\bar{\omega}$ $\ge 1$.
%Then the following theorem establishes the lower bound and upper bound of the detectable region for the transformed multi-level thresholding test {\bf $M_{J_n}$ which is defined in (\ref{max2}) but with known $\bm{\Omega}$.}
Define
\begin{eqnarray}
 \varrho_{\theta}(\beta) = \left\{
  \begin{array}{c  l}
    (\sqrt{1-\theta}-\sqrt{1-\beta-\frac{\theta}{2}})^2, & \quad  \frac{1}{2}\le \beta \le \frac{3-\theta}{4}; \\
    \beta-\frac{1}{2}, & \quad  \frac{3-\theta}{4}\le \beta \le \frac{3}{4};\\
    (1-\sqrt{1-\beta})^2, & \quad \frac{3}{4}< \beta <1.
\end{array} \right. \label{b2}
\end{eqnarray}

\bigskip

\textbf{Theorem 6.} Assume Conditions \textbf{(C1)}-\textbf{(C5)}.
\begin{itemize}
\item[(a)] When  $\bm{\Omega}$ is known, if $r<\bar{\omega}^{-1}\cdot \varrho(\beta)$, the sum of type I and II errors of the transformed multi-level thresholding test converges to 1 as $\alpha \to 0$ and $n \to \infty$; if $r>\underline{\omega}^{-1}\cdot \varrho(\beta)$,
the sum of type I and II errors of the transformed multi-level
thresholding test converges to zero when $\alpha=\bar{\Phi}\{(\mbox{log}p)^{\epsilon}\} \to 0$ for an arbitrarily small $\epsilon >0$ as $n \to \infty$.
\item[(b)] When $\bm{\Omega}$ is unknown and $p=n^{1/\theta}$ for
$0<\theta<1$, then if $r<\bar{\omega}^{-1}\cdot
\varrho_\theta(\beta)$, the sum of type I and II errors of the transformed multi-level thresholding test converges to 1 as $\alpha \to 0$ and $n \to \infty$; if $r>\underline{\omega}^{-1}\cdot
\varrho_\theta(\beta)$, the sum of type I and II errors of the transformed multi-level
thresholding test converges to zero when $\alpha=\bar{\Phi}\{(\mbox{log}p)^{\epsilon}\} \to 0$ for an arbitrarily small $\epsilon >0$ as $n \to \infty$.
 \end{itemize}

Hall and Jin (2010) has shown that utilizing the dependence can
lower the detection boundary $r= \varrho(\beta)$ for Gaussian data
with known covariance matrix. We demonstrate in Theorem 6 that the detection boundary can be lowered respectively for
the transformed multi-level thresholding test with $\bm{\Omega}$ being
known or unknown for sub-Gaussian data with estimated precision
matrix.  The theorem shows that there is a cost associated with using the estimated precision matrix in terms of a higher detetction boundary and
more restriction on the $p$ and $n$ relationship.

\setcounter{section}{5} \setcounter{equation}{0}
\section*{\large 5. Simulation Study}

In this section, the simulation was designed to confirm the
performance of the two multi-level thresholding tests defined in
(\ref{max1}) and (\ref{max2}) without and with transformation. We
also experimented the test of Chen and Qin (2010) given in
(\ref{eq1-0}), the Oracle test in (\ref{oracle}), and two tests
proposed by Cai, Liu and Xia (2014). The latter tests are based on
the max-norm statistics
\[
G(I)=\max \limits_{1\le k\le p} n(\bar{X}_1^{(k)}-\bar{X}_2^{(k)})^2 \quad \mbox{and} \quad G(\hat{\bm{\Omega}})=\max \limits_{1\le k\le p} \frac{n(\bar{\hat{Z}}_1^{(k)}-\bar{\hat{Z}}_2^{(k)})^2}{\hat{\omega}_{kk}},
\]
without and with transformation, where $\hat{\omega}_{kk}$ were
estimates of the diagonal elements of $\bm{\Omega}$. Cai, Liu and
Xia (2014)  showed that $G(I)$ and $G(\hat{\bm{\Omega}})$
converge to the type I extreme value distribution with cumulative
distribution function
$\mbox{exp}(-\frac{1}{\sqrt{\pi}}\mbox{exp}(-x/2))$, which was used
to formulate the test procedures based on  the two max-norm
statistics. Cai, Liu and Xia (2014) employed the CLIME estimator
based on a constrained $l_1$ minimization estimator of Cai, Liu and
Luo (2011) to estimate $\bm{\Omega}$.  Since we use the Cholesky
decomposition with banding to estimate $\bm{\Omega}$ in the
transformed thresholding test,  we used the estimated
$\hat{\omega}_{kk}$ from the approach in the formulation of the
max-norm statistics.

In the simulation experiments,  the two random samples
$\{\bm{X}_{1j}\}_{j=1}^{n_1}$ and $\{\bm{X}_{2j}\}_{j=1}^{n_2}$ were generated
according to the following multivariate model
%\footnote{Jun, check my writing below. {\bf Yes, Checked. (Jun)}}
$$\bm{X}_{ij}=\bm{\Sigma}^{1/2}_i \bm{Z}_{ij}+\bm{\mu}_i,$$
where the innovations $\bm{Z}_{ij}$ are IID $p$-dimensional random vectors with independent components such that $\mbox{E}(\bm{Z}_{ij})=0$ and $\mbox{Var}(\bm{Z}_{ij})=I_p$.  We considered two types of innovations: the Gaussian where $\bm{Z}_{ij} \sim N(0, I_p)$ and the Gamma where each component of $\bm{Z}_{ij}$ is standardized $\mbox{Gamma}(4,0.5)$ such that it has zero mean and unit variance.
%from the  multivariate normal and Gamma distributions
%\footnote{Can we try Gamma or other distributions ? I suspect CLX may not %do well there. Done (Jun)}
%with the mean vectors $\mu_1$ and $\mu_2$ respectively and common \footnote{These are ok for the gaussian, we need elaborate more for the gamma. Added (Jun)} %{\bf Specially, samples from multivariate Gamma distribution where $\Gamma \Gamma^{\prime}=\bm{\Sigma}$ and $p$ components of $Z_{ij}=(z_{ij1},\cdots,z_{imp})^{\prime}$ are independently and identically centralized $\mbox{Gamma}(4,0.5)$ distributed.}
For simplicity, we assigned   $\bm{\mu}_1= \bm{\mu}_2=0$ under $H_0$; and under $H_1$, $\bm{\mu}_1=0$  and $\bm{\mu}_2$
 had $[p^{1-\beta}]$ non-zero entries of equal value,   which were uniformly  allocated among $\{1,\cdots,p\}$. Here $[a]$ denotes the integer part of $a$.
The values of the  nonzero entries  were  $\sqrt{2r\mbox{log}p/n}$
for   a set
 of $r$-values ranging evenly from 0.1 to 0.4.  The
covariance matrices $\bm{\Sigma}_1=\bm{\Sigma}_2=:\bm{\Sigma}=(\sigma_{ij})$
where $\sigma_{ij}=\rho^{|i-j|}$ for $1\le i, j\le p$ and $\rho=0.6$.
%Clearly, the value of $\rho$ determines the level
%of dependence in $\bm{\Sigma}$, with a larger $\rho$ producing a denser $\bm{\Sigma}$ and vice versa.
%\footnote{Since we do not have the simulation for different values of $\rho$, this sentence may lead to questions from referees.}
The dimension $p$ was  $200$ and $600$, respectively and the sample sizes $n_1=30$ and $n_2=40$.
%\footnote{Can we try smaller $n_i$ ? Done (Jun). Two more plots with bigger n to produce Figs 3 and 4.
% {\bf PLEASE report the another set of sample size here.} I changed the original simulations to make all to have $n_1=30$ and $n_2=40$.(Jun)}

%The convergence of $\hat{\bm{\Omega}}_{\tau}$ based on the Cholesky
%decomposition depends on the choice of banding parameter $\tau$.
The banding width parameter  $\tau$  in the estimation of $\bm{\Omega}$
was chosen according to the data-driven procedure proposed by Bickel
and Levina (2008a), which is described as follows.  For a given data
set, we divided it into two subsamples by repeated ($N$ times)
random data split. For the $l$-th split,  $l \in \{1, \cdots,
N\}$, we let
$\hat{\bm{\Sigma}}_{\tau}^{(l)}=\{(I-\hat{A}_{\tau}^{(l)})^{\prime}\}^{-1}
\hat{D}_{\tau}^{(l)} (I-\hat{A}_{\tau}^{(l)})^{-1}$ be the Cholesky
decomposition of $\bm{\Sigma}$ obtained from the first subsample by
taking the same approach described in previous section for
$\hat{A}_{\tau}^{(l)}$ and $\hat{D}_{\tau}^{(l)}$. Also we let
$\bm{S}_n^{(l)}$ be the sample covariance matrix obtained from the second
subsample. Then the banding parameter $\tau$ is selected as \be
\hat{\tau}=\min \limits_{\tau} \frac{1}{N}\sum_{l=1}^{N}
||\hat{\bm{\Sigma}}_{\tau}^{(l)}-\bm{S}_n^{(l)} ||_{F}, \label{risk} \ee where
$||\cdot||_F$ denotes the Frobenius norm.

 Table 1 reports the empirical sizes of  the multi-thresholding tests with the data transformation (Mult2) and without the data transformation (Mult1), and Cai, Liu and Xia's max-norm tests with (CLX2)  and without (CLX1) the data transformation.
It also provides the empirical sizes for Mult1 and Mult2 with the bootstrap approximation of the critical values 
as described in Section 4.  
We observe that the empirical sizes of the two threshodling tests tended to be larger than the nominal 5\% level 
due to a slow convergence  to the extreme value distribution.  The proposed parametric bootstrap calibration 
can significantly improve the size.
%Mult1 displayed the significantly larger sizes than $0.05$. }

To make the power comparison fair,  we pre-adjusted the nominal
significant levels of all tests such that their empirical sizes were
all close to $0.05$. We obtain the average empirical power curves
(called power profiles) plotted with respect to $r$ and $\beta$
under each of the simulation settings outlined above based on 1000
simulations.  We observed only some very small change in the power
profiles when the underlying distribution was switched from the
Gaussian to the Gamma, which confirmed the nonparametric nature of
the tests considered. Due to the space limitation, we only display
in the following the power profiles based on the Gaussian data. 
%those for the Gamma innovations are given in the supplementary
%material.}

Figure 1 displays the empirical power profiles of
 the proposed multi-thresholding tests with data transformation (Mult2) and without data transformation (Mult1),
  and Cai, Liu and Xia's max-norm tests with (CLX2)  and without (CLX1) data transformation with respect to the signal strength $r$ at two given level of sparsity ($\beta=0.5$ and $0.6$) and $\rho=0.6$ for Gaussian data.
Figures 2-3 provide alternative views of the power profiles of these tests where the powers are displayed with respect to the sparsity $\beta$ at four levels of signal strength $r=0.1, 0.2, 0.3$ and $0.4$ for Gaussian data. These  figures also report the powers of  Chen and Qin (2010)'s test (CQ) and the Oracle test to provide some bench marks for the performance.

The basic trend of Figure 1 was that the powers of all the tests
were increasing as the signal strength $r$ was increased, and that
of Figures 2-3 is that the powers were decreasing as the sparsity
was increased. These are all expected. It is also expected to see in
each figure that the Oracle test had the best power among all the
tests since all the dimensions bearing noise were removed in
advance. A careful examination of the power profiles reveals that
the two tests that employed data transformation (Mult2 and CLX2)
were the top two performers among the non-Oracle tests, indicating
the effectiveness of the data transformation. The thresholding test
with data transformation (Mult2) had the best performance among all
the non-Oracle tests. This together with the observed  performance
of the thresholding test without transformation (Mult1) and the CLX2
shows that the combining the data transformation with the
thresholding leads to a quite powerful test performance. The CQ test
and the max-norm test without data transformation (CLX1) had the
least power among the tests, with the CLX1 being more powerful than
the CQ for the more sparse situation (large $\beta$)  and vice versa
for the faint signal case (smaller $r$). The CQ test
%\footnote{I
%think the CLX test was designed for sparse but strong signal}
was not designed for the sparse and faint signal settings of the
simulation, although it is a proper test under ultra high
dimensionality in the sense that the size of the test can be
attained and with reasonable power in non-sparse settings.
%Philisophically speaking,  the max-norm test and the CQ tests are similar as they are constructed based on different norms. One is the max-norm and the other the $L_2$ norm.
The above features became more pronounced when we increase the dimensionality to $p=600$ as shown in Figures 1 and 3.

\setcounter{section}{6} \setcounter{equation}{0}
\section*{\large 6. Empirical Study}

In this section, we demonstrate the performance of the multi-level
thresholding test defined in (\ref{max2}) on a human breast cancer
dataset, available at http://www.ncbi.nlm.\\nih.gov. The data have
been analyzed by Richardson et al. (2006) to provide insight into
the molecular pathogenesis of Sporadic basal-like cancers (BLC), a
distinct class of human breast cancers. The original {microarray gene expression} data consist of
7 normal specimens, 2 BRCA-associated breast cancer specimens, 18
sporadic BLC specimens and 20 non-BLC specimens. Since the most of
interests on this data set is to display the unique characteristics
of BLC relative to non-BLC specimens, we formed two samples. One
consists of $n_1=18$ BLC cases and another consists of $n_2=20$
non-BLC specimens for analysis which form two samples respectively.

Biologically speaking,  each gene does not function individually in
isolation. Rather,  genes  tend to work collectively to perform
their biological functions. Gene-sets are technically defined in
Gene Ontology (GO) system  that provides structured vocabularies
which produce names of gene-sets (also called GO terms), see
Ashburner et al. (2000) for more details.

There were  9918 GO terms, which were obtained from the original
data set after we excluded some GO terms with missing information. To accommodate high dimensionality, we further removed those GO
terms with the number of genes less than 20 and the number of
remaining GO terms varied by chromosomes. In order to take
advantage of the inter-gene correlation, we first selected genes
from one of 23 chromosomes and then ordered them by their locations
on the chromosome. By doing this,  genes with adjacent locations are
more strongly correlated than genes far away from each other. This
would also facilitate the bandable assumption for the covariance
matrices. A major motivation in our analysis is to identify sets of
genes which are significantly different
 between the BLC and the non-BLC specimens.

As discussed in Richardson et al. (2006), BLC specimens display $X$
chromosome abnormalities in the sense that most of the BLC cases
lack markers of a normal inactive X chromosome, which are rare in
non-BLC specimens. Moreover, single nucleotide polymorphism
%\footnote{what is it ? Full name added. (Jun)}
 array analysis
demonstrated loss of heterozygosity (loss of a normal and functional
allele at a heterozygous locus) in chromosome 14 and 17 was quite
frequent in BLC specimens, a phenomenon largely missing among
non-BLC specimens.
% Abnormalities of chromosome 14 and 17 are also special characteristic of BLC.} \footnote{Could we rewrite to
%make it more easily understood? Added some description. (Jun)}
Therefore, our main interest  was on chromosomes X, 14 and 17.

We applied the multi-level thresholding test based on the data
transformation on each of gene-sets in chromosomes $X$, $14$ and
$17$ by first transforming the data with estimated $\bm{\Omega}$
 through the Cholesky decomposition discussed in Section 4. We also
applied the CQ test to serve as contrasts.
%the ordered data on a chosen chromosome and for a given
%GO term were transformed by
%$\bm{\Omega}=(\frac{n_2}{n_1+n_2}\bm{\Sigma}_1+\frac{n_1}{n_1+n_2}\bm{\Sigma}_2)^{-1}$
%for, which was estimated
By controlling the false discovery rate (Benjamini and Hochberg, 1995) at $0.05$, the CQ test declared 81 GO terms significant on
chromosome $X$, 80 out of which were also declared significant by
the multi-level thresholding test. However, the multi-thresholding
test found 4 more significant GO terms not found significant by the
CQ test.  Similarly, on chromosome 14, CQ test declared 76 GO terms
significant which were all included by the 86 GO terms declared
significant by the multi-level thresholding test. On chromosome 17,
5 out of 166 GO terms declared significant by the CQ test were not
declared significant  by the multi-level thresholding test. On the
other hand, 14 out of 175 GO terms declared significant by the
multi-level thresholding test were not declared significant by the
CQ test.

Table 2 lists the top
ten most significant GO terms declared by  the multi-level
thresholding test %at 5\% false discovery rate
on the three chromosomes, respectively.  The table also marks those gene-sets
which were not tested significant by the CQ test. There were three
gene-sets in the top ten which were not declared significant by the
CQ test in chromosomes $X$ and $14$, and two gene-sets in
chromosomes 17.  These empirical results  support our theoretically
findings that the multi-level thresholding test with data
transformation is more powerful than the CQ test by conducting both thresholding and utilizing data dependence.

\setcounter{section}{7} \setcounter{equation}{0}
\section*{\large 7. Discussion}

Our analysis in this paper shows that the thresholding combined with the data transformation via the estimated precision matrix leads to a very powerful
test procedure.  The analysis also shows that thresholding alone is not sufficient in lifting the power when there is sufficient amount of dependence
 in the covariance, and the data transformation is quite crucial.  The latter confirms the benefit of the transformation discovered by
 Hall and Jin (2010) for the higher criticism test and Cai, Liu and Xia (2014) for the max-norm based test.
The proposed test of thresholding with data transformation can be
viewed as  a significant improvement of the test of Chen and Qin (2010) for sparse and faint signals. The CQ
test is similar to the max-norm test without data transformation, except that
it is  based on the $L_2$ norm. Generally speaking, the max-norm
test works better for more sparse and stronger signals whereas  the CQ test
is for denser but fainter signals.  These aspects were confirmed by our simulations.
 A reason for the proposed test (with both thresholding and data transformation)
having better power than the test of Cai, Liu and Xia (2014) with
data transformation is due to the thresholding conducted on the
$L_2$ formulation of the test statistics since the proposed test has
both thresholding and data transformation whereas CLX test has only
 the data transformation. The max-norm formulation does not accommodate
  the need to threshold. This reveals an adavntage of the $L_2$ formulation.
 %\footnote{I think the maximum norm is a special case of thresholding by taking the threshold to the second largest marginal test statistic}

The results that the proposed test with the 
estimated covariance can produce lower detection boundary
than that of the standard higher criticism test using asymptotic p-values (Delaigle et al., 2011) is another advantage
of the proposal.  We want to point out that the study carried out  in this paper is not a direct extension from that in Zhong, Chen and Xu (2013).
Zhong et al. (2013) considered an alternative $L_2$-formulation  to the higher criticism (HC) test of
 Donoho and Jin (2004) for one-sample hypotheses. They showed that,  although the  $L_2$ formulation 
 attains the same detection boundary as the HC test, the $L_2$ formulation is more advantageous to the HC 
 when the sparsity and signal strength combination $(\beta, r)$  is  above the detection boundary.  
 However,  Zhong et al. (2013) did not study the specific benefits of the thresholding in improving the power of 
 the high dimensional multivariate test and the relative performance to the Oracle test; nor did they considered the data transformation via the precision matrix.

%\clearpage

\setcounter{equation}{0}
\def\theequation{A.\arabic{equation}}
\def\thesection{A}

\section*{\large Appendix: Technical Details.}

Throughout the Appendix, we assume $n_1\to \infty$, $n_2 \to \infty$ and let $n=\frac{n_1n_2}{n_1+n_2}$.

\noindent{\bf A.1. Lemmas}

\textbf{Lemma 1.} We denote $\delta_k={\mu}_{1k}-{\mu}_{2k}$. As $x=o(n^{\frac{1}{3}})$, $T_{nk}$ in (\ref{eq1}) satisfies
\begin{eqnarray}
\mbox{P}(n\,T_{nk}+1>x)&=&\{1+o(1)\}I(\sqrt{n}|\delta_k|>\sqrt{x})\nonumber\\
&+&\biggl[\bar{\Phi}(\sqrt{x}-\sqrt{n}|\delta_k|) +\bar{\Phi}(\sqrt{x}+\sqrt{n}|\delta_k|)\biggr]\{1+O(n^{-1/6})\nonumber\\
&+&O(\frac{x^{3/2}}{n^{1/2}})\}I(\sqrt{n}|\delta_k|<\sqrt{x}).\nonumber
\end{eqnarray}

Proof. We first denote
\begin{eqnarray}
T_{nk,1}&=&n(\bar{X_1}^{(k)}-\bar{X_2}^{(k)})^2,\nonumber\\
T_{nk,2}&=&1-n\biggl\{\sum_{i=1}^{n_1}(X_{1i}^{(k)})^2/n_1^2+\sum_{i=1}^{n_2}(X_{2i}^{(k)})^2/n_2^2\biggr\},\nonumber\\
T_{nk,3}&=&n\biggl\{(\bar{X_1}^{(k)})^2/(n_1-1)+(\bar{X_2}^{(k)})^2/(n_2-1)\biggr\},\nonumber\\
T_{nk,4}&=&-n\biggl\{\sum_{i=1}^{n_1}(X_{1i}^{(k)})^2/(n_1^3-n_1^2)+\sum_{i=1}^{n_2}(X_{2i}^{(k)})^2/(n_2^3-n_2^2)\biggr\}.\nonumber
\end{eqnarray}

Then, the modified CQ test statistic satisfies
\begin{eqnarray}
n\,T_{nk}+1=T_{nk,1}+T_{nk,2}+T_{nk,3}+T_{nk,4},\nonumber
\end{eqnarray}
and with the modified version of BS statistic, we know
\[
nM_{nk}=T_{nk,1}.
\]

Application of Slutsky argument yields
\begin{eqnarray}
\mbox{P}(T_{nk,1}&>&x+n^{-\frac{1}{6}})-\mbox{P}(|T_{nk,2}+T_{nk,3}+T_{nk,4}|>n^{-\frac{1}{6}})\nonumber\\
&\le& \mbox{P}(n\,T_{nk}+1>x)
\le \mbox{P}(T_{nk,1}>x-n^{-\frac{1}{6}})+\mbox{P}(|T_{nk,2}+T_{nk,3}+T_{nk,4}|>n^{-\frac{1}{6}}).\nonumber
\end{eqnarray}

If $\sqrt{n}|\delta_k|<\sqrt{x}$, the result below follows Theorem 5.23 of Petrov (1995),
\begin{eqnarray}
&\quad&\mbox{P}(T_{nk,1}>x\pm n^{-\frac{1}{6}})\nonumber\\
&=&\mbox{P}\biggl\{\sqrt{n}(\bar{X_1}^{(k)}
-\bar{X_2}^{(k)})-\sqrt{n}|\delta_k|>\sqrt{x\pm n^{-\frac{1}{6}}}
-\sqrt{n}|\delta_k|\biggr\}\nonumber\\
&+&\mbox{P}\biggl\{\sqrt{n}(\bar{X_1}^{(k)}
-\bar{X_2}^{(k)})-\sqrt{n}|\delta_k|<-\sqrt{x\pm n^{-\frac{1}{6}}}
-\sqrt{n}|\delta_k|\biggr\}\nonumber\\
&=&\biggl[\bar{\Phi}(\sqrt{x\pm n^{-\frac{1}{6}}}-\sqrt{n}|\delta_k|) +\bar{\Phi}(\sqrt{x\pm n^{-\frac{1}{6}}}+\sqrt{n}|\delta_k|)\biggr]\{1+O(\frac{x^{3/2}}{n^{1/2}})\}\nonumber\\
&=&\biggl[\bar{\Phi}(\sqrt{x}-\sqrt{n}|\delta_k|) +\bar{\Phi}(\sqrt{x}+\sqrt{n}|\delta_k|)\biggr]\{1+O(n^{-1/6})+O(\frac{x^{3/2}}{n^{1/2}})\}\nonumber
\end{eqnarray}

On the other hand, if $\sqrt{n}|\delta_k|>\sqrt{x}$,
\begin{eqnarray}
\mbox{P}(T_{nk,1}>x\pm n^{-\frac{1}{6}})&=&1-\mbox{P}\biggl\{\sqrt{n}(\bar{X_1}^{(k)}
-\bar{X_2}^{(k)})-\sqrt{n}|\delta_k|<\sqrt{x\pm n^{-\frac{1}{6}}}
-\sqrt{n}|\delta_k|\biggr\}\nonumber\\
&+&\mbox{P}\biggl\{\sqrt{n}(\bar{X_1}^{(k)}
-\bar{X_2}^{(k)})-\sqrt{n}|\delta_k|<-\sqrt{x\pm n^{-\frac{1}{6}}}
-\sqrt{n}|\delta_k|\biggr\}\nonumber\\
&=&1+o(1).\nonumber
\end{eqnarray}

To show $\mbox{P}(|T_{nk,2}+T_{nk,3}+T_{nk,4}|>n^{-\frac{1}{6}})=\biggl[\mbox{P}(T_{nk,1}>x\pm n^{-\frac{1}{6}})\biggr]o(1)$, we only need to show that each of $T_{nk,2}$, $T_{nk,3}$ and $T_{nk,4}$ satisfies $\mbox{P}(|T_{nk,l}|>n^{-\frac{1}{6}})=\biggl[\mbox{P}(T_{nk,1}>x\pm n^{-\frac{1}{6}})\biggr]o(1)$ for $l=2,3,4$. Since $T_{nk,2}+T_{nk,3}+T_{nk,4}$ is invariant under location transformation, we also assume $\mu_1=\mu_2=0$. For $T_{nk,2}$, we notice that
\begin{eqnarray}
\mbox{P}(|T_{nk,2}|>n^{-\frac{1}{6}})&\le& \mbox{P}(|\frac{1}{\sqrt{n_1}}\sum_{i=1}^{n_1}[(X_{1i}^{(k)})^2-1]|>a n^{\frac{1}{3}})\nonumber\\
&+&\mbox{P}(|\frac{1}{\sqrt{n_2}}\sum_{i=1}^{n_2}[(X_{2i}^{(k)})^2-1]|>b n^{\frac{1}{3}}),\nonumber
\end{eqnarray}
for some constant $a$ and $b$. Then, under Sub-Gaussian assumption, the first probability on the right side of above inequality can be bounded, i.e.,
\begin{eqnarray}
 \mbox{P}(|\frac{1}{\sqrt{n_1}}\sum_{i=1}^{n_1}[(X_{1i}^{(k)})^2-1]|>a n^{\frac{1}{3}})\le c e^{-\frac{n^{1/3}}{M}},\nonumber
\end{eqnarray}
for some constant $c$ and $M$. The similar result also holds for the second probability. By comparing this upper bound with the leading order of function $\bar{\Phi}(\sqrt{x})$, we see that
\begin{eqnarray}
\mbox{P}(|T_{nk,2}|>n^{-\frac{1}{6}})=\biggl[\mbox{P}(T_{nk,1}>x\pm n^{-\frac{1}{6}})\biggr]o(1).\nonumber
\end{eqnarray}
The similar result holds for $T_{nk,3}$ and $T_{nk,4}$. This completes the proof of Lemma 1.

\bigskip
\textbf{Lemma 2.} Assume Conditions (C1)-(C2). The mean of the thresholding test statistic $L_n(s)$ is
\begin{eqnarray}
\mu_{L_n(s)}&=&\sum_{i\in S^c_{\beta}} \mbox{E}\{L_{n,i}(s)\}+ \sum_{k \in S_{\beta}}\mbox{E}\{L_{n,k}(s)\}\nonumber\\
&=&\biggl(\frac{2}{\sqrt{2\pi}}\sqrt{2s\mbox{log}p}p^{1-s}+\sum_{k \in S_{\beta}}\biggl[n\,\delta_k^2I\biggl\{n\,\delta_k^2>\lambda_{n}(s)\biggr\}\nonumber\\
&+&(2s\mbox{log}p)\bar{\Phi}(\eta_k^-) I\biggl\{n\,\delta_k^2<\lambda_{n}(s)\biggr\}\biggr]\biggr)\{1+o(1)\}.\label{mean_alter}
\end{eqnarray}

Proof. To obtain the mean of the thresholding test statistic $L_n(s)$, we only give detailed derivation for $L_1(s)$ since the derivation for $L_2(s)$ is similar. We first apply Fubini's theorem which turns the expectation of $L_n(s)$ into the tail probability, namely
\begin{eqnarray}
\mbox{E}\{L_n(s)\}&=&\sum_{k=1}^p\mbox{E}\{L_{n,k}(s)\}\nonumber\\
&=&\sum_{k=1}^{p}\biggl[\lambda_{n}(s)\mbox{P}\biggl\{n\,T_{nk}+1>\lambda_{n}(s)\biggr\}+\int_{\lambda_{n}(s)}^{\infty} \mbox{P}(n\,T_{nk}+1>z)dz\biggr].\nonumber
\end{eqnarray}

Since the underlying distribution of $X_{ij}$ is not specified, the calculation of probability above needs to be approximated by using large deviation results from Petrov (1995).  We first consider the case where $n\,\delta_k^2< \lambda_{n}(s)$.  Then the following result can be derived from Lemma 1:
\begin{eqnarray}
\mbox{E}(L_{n,k})&=&\biggl\{\lambda_{n}[\bar{\Phi}(\sqrt{\lambda_{n}}+\sqrt{n}\delta_k)+\bar{\Phi}(\sqrt{\lambda_{n}}-\sqrt{n}\delta_k)]\nonumber\\
&+&\int_{\lambda_{n}}^{n^{\frac{1}{3}-\epsilon}} [\bar{\Phi}(\sqrt{z}+\sqrt{n}\delta_k)+\bar{\Phi}(\sqrt{z}-\sqrt{n}\delta_k)]dz\nonumber\\
&+&\int_{n^{\frac{1}{3}-\epsilon}}^{\infty} \mbox{P}(n\,T_{nk}+1>z)dz\biggr\}\biggl\{1+O(n^{-1/6})+O(\frac{{\lambda_{n}}^{3/2}}{n^{1/2}}) \biggr\},\label{apx1}
\end{eqnarray}
for an arbitrary small constant $\epsilon$. First we have the following inequality
\begin{eqnarray}
\mbox{P}(n\,T_{nk}+1>z)
\le \mbox{P}(T_{nk,1}>\frac{z}{2})+\mbox{P}(|T_{nk,2}+T_{nk,3}+T_{nk,4}|>\frac{z}{2}).\nonumber
\end{eqnarray}

By sub-Gaussian assumption, for a given constant $M$ and $c$,
\begin{eqnarray}
\mbox{P}(\sqrt{n}|\bar{X}_1^{(k)}-\bar{X}_2^{(k)}-|\delta_k||>x)\le ce^{-x^2/M}.\nonumber
\end{eqnarray}
Then,
\begin{eqnarray}
\int_{n^{\frac{1}{3}-\epsilon}}^{\infty}\mbox{P}(T_{nk,1}>\frac{z}{2})dz&\le&2\int_{n^{\frac{1}{3}-\epsilon}}^{\infty}\mbox{P}(\sqrt{n}|\bar{X}_1^{(k)}-\bar{X}_2^{(k)}-|\delta_k||> \sqrt{\frac{z}{2}}-\sqrt{n}|\delta_k|)dz\nonumber\\
&\le&8c\int_{n^{\frac{1}{6}-\frac{\epsilon}{2}}-\sqrt{n}|\delta_k|}^{\infty}ye^{-\frac{y^2}{M}}dy+8c\sqrt{n}|\delta_k|\int_{n^{\frac{1}{6}-\frac{\epsilon}{2}}-\sqrt{n}|\delta_k|}^{\infty}e^{-\frac{y^2}{M}}dy\nonumber\\
&\le&8cMe^{-\frac{(n^{\frac{1}{6}-\frac{\epsilon}{2}}-\sqrt{n}|\delta_k|)^2}{M}}\{1+o(1)\},\nonumber
\end{eqnarray}
which is smaller order of the first term on the right hand side of (\ref{apx1}). Similarly, we can show that $\mbox{P}(|T_{nk,2}+T_{nk,3}+T_{nk,4}|>\frac{z}{2})=o\biggl\{\lambda_{n}[\bar{\Phi}(\sqrt{\lambda_{n}}+\sqrt{n}|\delta_k|)+\bar{\Phi}(\sqrt{\lambda_{n}}-\sqrt{n}|\delta_k|)]\biggr\}$. Therefore, the last integration in (\ref{apx1}) satisfies
\begin{eqnarray}
\int_{n^{\frac{1}{3}-\epsilon}}^{\infty} \mbox{P}(n\,T_{nk}+1>z)dz=\biggl\{\lambda_{n}[\bar{\Phi}(\sqrt{\lambda_{n}}+\sqrt{n}\delta_k)+\bar{\Phi}(\sqrt{\lambda_{n}}-\sqrt{n}\delta_k)]\biggr\}o(1),\nonumber
\end{eqnarray}
which is smaller order of $\frac{\lambda_{n}^{3/2}}{n^{1/2}}$ since it decays exponentially. Therefore, we have
\begin{eqnarray}
\mbox{E}(L_{n,k})&=&\biggl\{\lambda_{n}[\bar{\Phi}(\sqrt{\lambda_{n}}+\sqrt{n}|\delta_k|)+\bar{\Phi}(\sqrt{\lambda_{n}}-\sqrt{n}|\delta_k|)]\nonumber\\
&+&\int_{\lambda_{n}}^{\infty} [\bar{\Phi}(\sqrt{z}+\sqrt{n}|\delta_k|)+\bar{\Phi}(\sqrt{z}-\sqrt{n}|\delta_k|)]dz\biggr\}\biggl\{1+O(n^{-1/6})+O(\frac{\lambda_{n}^{3/2}}{n^{1/2}}) \biggr\},\nonumber
\end{eqnarray}
which leads to the following result by the partial integration
\begin{eqnarray}
\mbox{E}(L_{n,k})&=&\biggl\{(\sqrt{\lambda_{n}}+\sqrt{n}|\delta_k|)\phi(\sqrt{\lambda_{n}}-\sqrt{n}|\delta_k|)\nonumber\\
&+&(\sqrt{\lambda_{n}}-\sqrt{n}|\delta_k|)\phi(\sqrt{\lambda_{n}}+\sqrt{n}|\delta_k|)
+n\,\delta_{k}^2[\bar{\Phi}(\sqrt{\lambda_{n}}-\sqrt{n}|\delta_k|)\nonumber\\
&+&\bar{\Phi}(\sqrt{\lambda_{n}}+\sqrt{n}|\delta_k|)] \biggr\}\biggl\{1+O(n^{-1/6})+O(\frac{\lambda_{n}^{3/2}}{n^{1/2}})\biggr\}.\nonumber
\end{eqnarray}

We can simplify the result using the relationship $\bar{\Phi}(y)=\phi(y)/y$ for a sufficient large $y$. As a result, we have
\begin{eqnarray}
\mbox{E}(L_{n,k})
&=&\lambda_{n}\biggl[\bar{\Phi}(\sqrt{\lambda_{n}}-\sqrt{n}|\delta_k|)+\bar{\Phi}(\sqrt{\lambda_{n}}+\sqrt{n}|\delta_k|) \biggr]\biggl\{1+O(n^{-1/6})+O(\frac{\lambda_{n}^{3/2}}{n^{1/2}})\biggr\}.\nonumber
\end{eqnarray}

Next, we consider $n\,\delta_k^2>\lambda_{n}$. For this case, a direct application of Lemma 1 leads to
\begin{eqnarray}
\mbox{E}(L_{n,k})=\lambda_{n}\{1+o(1)\}+\int_{\lambda_{n}}^{n\,\delta_k^2}\mbox{P}(n\,T_{nk}+1>z)dz+\int_{n\,\delta_k^2}^{\infty}\mbox{P}(n\,T_{nk}+1>z)dz, \nonumber
\end{eqnarray}
where the first integration gives $(n\,\delta_k^2-\lambda_{n})\{1+o(1)\}$. Similarly, for given constant $c$ and $M$, we can show that
\[
\int_{n\,\delta_k^2}^{\infty}\mbox{P}(n\,T_{nk}+1>z)dz \le 8cM,
\]
which is smaller order of $n\,\delta_k^2$. Hence,
\begin{eqnarray}
\mbox{E}(L_{n,k})=n\,\delta_k^2\{1+o(1)\}I(n\,\delta_k^2>\lambda_{n}).\nonumber
\end{eqnarray}

In summary, the expectation of $L_{n,k}$ is
\begin{eqnarray}
\mbox{E}\{L_{n,k}(s)\}&=&n\,\delta_k^2\{1+o(1)\}I\biggl\{n\,\delta_k^2>\lambda_{n}(s)\biggr\}+\lambda_{n}(s)\biggl\{\bar{\Phi}(\eta_k^+)+\bar{\Phi}(\eta_k^-)\biggr\}\nonumber\\
&\quad&\{1+O(n^{-\frac{1}{6}})+O(\frac{(\mbox{log}p)^{3/2}}{n^{1/2}})\}I\biggr\{n\,\delta_k^2<\lambda_{n}(s)\biggr\},\nonumber
\end{eqnarray}
where $\bar{\Phi}=1-\Phi$, $\eta^{\pm}_k=\sqrt{2s\mbox{log}p}\pm |\sqrt{n}\delta_k|$.
Specially, under $H_0$,
\begin{eqnarray}
\mbox{E}\{L_{n,k}(s)\}=2\sqrt{2s\mbox{log}p}\phi(\sqrt{2s\mbox{log}p})\{1+O(n^{-\frac{1}{6}})+O(\frac{(\mbox{log}p)^{3/2}}{n^{1/2}})\},\nonumber
\end{eqnarray}
which leads to the expectation of $L_n$ under $H_0$:
\begin{eqnarray}
\mu_{L_n(s),0}=\frac{2}{\sqrt{2\pi}}\sqrt{2s\mbox{log}p}p^{1-s}\{1+O(n^{-\frac{1}{6}})+O(\frac{(\mbox{log}p)^{3/2}}{n^{1/2}})\}.\label{mean_null}
\end{eqnarray}
And the expectation of $L_n$ under $H_1$ is
\begin{eqnarray}
\mu_{L_n(s),1}&=&\sum_{i\in S^c_{\beta}} \mbox{E}\{L_{n,i}(s)\}+ \sum_{k \in S_{\beta}}\mbox{E}\{L_{n,k}(s)\}\nonumber\\
&=&\biggl(\frac{2}{\sqrt{2\pi}}\sqrt{2s\mbox{log}p}p^{1-s}+\sum_{k \in S_{\beta}}\biggl[n\,\delta_k^2I\biggl\{n\,\delta_k^2>\lambda_{n}(s)\biggr\}\nonumber\\
&+&(2s\mbox{log}p)\bar{\Phi}(\eta_k^-) I\biggl\{n\,\delta_k^2<\lambda_{n}(s)\biggr\}\biggr]\biggr)\{1+o(1)\}.\label{mean_alter}
\end{eqnarray}

\bigskip
\textbf{Lemma 3.} Assume Conditions (C1)-(C3). The variance of the thresholding test statistic $L_n(s)$ is
\begin{eqnarray}
\sigma_{L_n(s),1}^2
&=&\biggl(\frac{2}{\sqrt{2\pi}}[(2s\mbox{log}p)^{\frac{3}{2}}+(2s\mbox{log}p)^{\frac{1}{2}}]p^{1-s}+4p\bar{\Phi}(\sqrt{2s\mbox{log}p})\nonumber\\
&+&\sum_{k,l \in S_{\beta}}(4n\,\delta_k\,\delta_l\,\rho_{kl}+2\rho_{kl}^2)I\biggl\{n\,\delta_k^2>\lambda_{n}(s)\biggr\}I\biggl\{n\,\delta_l^2>\lambda_{n}(s)\biggr\}\nonumber\\
&+&\sum_{k \in S_{\beta}}(2s\mbox{log}p)^2\bar{\Phi}(\eta_k^-)I\biggl\{n\,\delta_k^2<\lambda_{n}(s) \biggr\}\biggr)\{1+o(1)\}.\label{variance_alter}
\end{eqnarray}

Proof. Recall that $L_n(s)=\sum_{k=1}^p L_{n,k}(s)$. Then,
\[
\mbox{Var}\{L_n(s) \}=\sum_{k=1}^p\mbox{Var}\{L_{n,k}(s) \}+\sum_{k \ne l}\mbox{Cov}\{L_{n,k}(s), L_{n,l}(s)\}.
\]
We first find  the variance of $L_{n,k}$ given by
\[
\mbox{Var}(L_{n,k})=\mbox{E}(L_{n,k}^2)-\mbox{E}^2(L_{n,k}),
\]
where by Fubini's theorem,
\[
\mbox{E}(L_{n,k}^2)=\lambda^2_n(s)\mbox{P}(n\,T_{nk}+1>\lambda_n(s))+2\int_{\lambda_n(s)}^{\infty}z\mbox{P}(n\,T_{nk}+1>z)dz.
\]

By the same techniques as we derive the mean of $L_{n,k}$, the variance of $L_{n,k}$ is
\begin{eqnarray}
\mbox{Var}(L_{n,k})&=&(4n\,\delta_k^2+2)\{1+o(1)\}I(n\,\delta_k^2>\lambda_{n})\nonumber\\
&+&v(\eta_k^+,\eta_k^-)\{1+o(1)\}I(n\,\delta_k^2<\lambda_{n}),\label{apx4}
\end{eqnarray}
where $v(\eta_k^+,\eta_k^-)$ is given by
\begin{eqnarray}
&\quad&v(\eta_k^+,\eta_k^-)\nonumber\\
&=&
[(\eta^-_k)^3+4\sqrt{n}|\delta_k|(\eta^-_k)^2+3\eta^-_k+6n\,\delta_k^2\eta^-_k
+8\sqrt{n}|\delta_k|
+4(\sqrt{n}|\delta_k|)^3-2\eta^+_k]\phi(\eta^-_k)\nonumber\\
&+&[(\eta^+_k)^3-4\sqrt{n}|\delta_k|(\eta^+_k)^2+3\eta^+_k+6n\,\delta_k^2\eta^+_k
-8\sqrt{n}|\delta_k|
-4(\sqrt{n}|\delta_k|)^3-2\eta^-_k]\phi(\eta^+_k)\nonumber\\
&+&(n^2\delta_k^4+4n\,\delta_k^2+2)[\bar{\Phi}(\eta^-_k)
+\bar{\Phi}(\eta^+_k)]-[\eta^+_k\phi(\eta^-_k)+\eta^-_k\phi(\eta^+_k)
+n\,\delta_{k}^2\bar{\Phi}(\eta^-_k)\nonumber\\
&+&n\,\delta_{k}^2\bar{\Phi}(\eta^+_k)]^2.\nonumber
\end{eqnarray}
The covariance between $L_{n,k}$ and $L_{n,l}$ depends on the values of $n\,\delta_k^2$ and $n\,\delta_l^2$. To show this dependence explicitly, we denote $\mbox{Cov}(L_{n,k}, L_{n,l})$ by $\gamma(\sqrt{n}\delta_k,\sqrt{n}\delta_l,\rho_{kl})$. Then,
\begin{eqnarray}
&\quad&\gamma(\sqrt{n}\delta_k,\sqrt{n}\delta_l,\rho_{kl})\nonumber\\
&=&(4\rho_{kl}n\delta_k\,\delta_l+2\rho_{kl}^2)\{1+o(1)\}I(n\,\delta_k^2>\lambda_{n})\nonumber\\
&+&\Upsilon_1(\eta_k^-,\sqrt{n}|\delta_k|,\sqrt{n}|\delta_l|;\rho_{kl})Q(\eta_k^-,\eta_l^-;\rho_{kl})+\Upsilon_1(\eta_l^-,\sqrt{n}|\delta_l|,\sqrt{n}|\delta_k|;\rho_{kl})\nonumber\\
&\times&Q(\eta_l^-,\eta_k^-;\rho_{kl})
+\Upsilon_2(\eta_k^-,\eta_l^-,\sqrt{n}|\delta_k|,\sqrt{n}|\delta_l|;\rho_{kl})q(\eta_k^-,\eta_l^-;\rho_{kl})\nonumber\\
&+&\Upsilon_3(\sqrt{n}|\delta_k|,\sqrt{n}|\delta_l|;\rho_{kl})U(\eta_k^-,\eta_l^-;\rho_{kl})
+\Upsilon_1(\eta_k^+,-\sqrt{n}|\delta_k|,\sqrt{n}|\delta_l|;-\rho_{kl})\nonumber\\&\times&Q(\eta_k^+,\eta_l^-;-\rho_{kl})
+\Upsilon_1(\eta_l^-,\sqrt{n}|\delta_l|,-\sqrt{n}|\delta_k|;-\rho_{kl})Q(\eta_l^-,\eta_k^+;-\rho_{kl})\nonumber\\
&+&\Upsilon_2(\eta_k^+,\eta_l^-,-\sqrt{n}|\delta_k|,\sqrt{n}|\delta_l|;-\rho_{kl})q(\eta_k^+,\eta_l^-;-\rho_{kl})+\Upsilon_3(-\sqrt{n}|\delta_k|,\sqrt{n}|\delta_l|;-\rho_{kl})\nonumber\\
&\times&U(\eta_k^+,\eta_l^-;-\rho_{kl})
+\Upsilon_1(\eta_k^-,\sqrt{n}|\delta_k|,-\sqrt{n}|\delta_l|;-\rho_{kl})Q(\eta_k^-,\eta_l^+;-\rho_{kl})\nonumber\\&+&\Upsilon_1(\eta_l^+,-\sqrt{n}|\delta_l|,\sqrt{n}|\delta_k|;-\rho_{kl})Q(\eta_l^+,\eta_k^-;-\rho_{kl})\nonumber\\
&+&\Upsilon_2(\eta_k^-,\eta_l^+,\sqrt{n}|\delta_k|,-\sqrt{n}|\delta_l|;-\rho_{kl})q(\eta_k^-,\eta_l^+;-\rho_{kl})+\Upsilon_3(\sqrt{n}|\delta_k|,-\sqrt{n}|\delta_l|;-\rho_{kl})\nonumber\\
&\times&U(\eta_k^-,\eta_l^+;-\rho_{kl})
+\Upsilon_1(\eta_k^+,-\sqrt{n}|\delta_k|,-\sqrt{n}|\delta_l|;\rho_{kl})Q(\eta_k^+,\eta_l^+;\rho_{kl})\nonumber\\&+&\Upsilon_1(\eta_l^+,-\sqrt{n}|\delta_l|,-\sqrt{n}|\delta_k|;\rho_{kl})Q(\eta_l^+,\eta_k^+;\rho_{kl})\nonumber\\
&+&\Upsilon_2(\eta_k^+,\eta_l^+,-\sqrt{n}|\delta_k|,-\sqrt{n}|\delta_l|;\rho_{kl})q(\eta_k^+,\eta_l^+;\rho_{kl})+\Upsilon_3(-\sqrt{n}|\delta_k|,-\sqrt{n}|\delta_l|;\rho_{kl})\nonumber\\&\times&U(\eta_k^+,\eta_l^+;\rho_{kl})
-\biggl\{\eta_k^+\phi(\eta_k^-)+\eta_k^-\phi(\eta_k^+)+n\,\delta_k^2[\bar{\Phi}(\eta_k^-)
+\bar{\Phi}(\eta_k^+)] \biggr\}\nonumber\\
&\times&\biggl\{\eta_l^+\phi(\eta_l^-)+\eta_l^-\phi(\eta_l^+)+n\,\delta_l^2[\bar{\Phi}(\eta_l^-)
+\bar{\Phi}(\eta_l^+)] \biggr\}I(n\,\delta_k^2<\lambda_{n}),\nonumber
\end{eqnarray}
where
\begin{eqnarray}
\Upsilon_1(a,b_1,b_2;\rho)&=&a^3\rho^2+2a^2(b_1\rho^2+b_2\rho)+a(b_1^2\rho^2+4b_1b_2\rho+b_2^2+\rho^2)\nonumber\\
&+&2b_1^2b_2\rho+2b_1b_2^2+2b_1\rho^2+2b_2\rho,\nonumber
\end{eqnarray}
\begin{eqnarray}
\Upsilon_2(a_1,a_2,b_1,b_2;\rho)&=&\sqrt{1-\rho^2}(a_1^2\rho+a_2^2\rho+a_1a_2+2a_1b_1\rho+2a_2b_2\rho\nonumber\\
&+&2a_1b_2+2a_2b_1+b_1^2\rho+b_2^2\rho+4b_1b_2+\rho),\nonumber
\end{eqnarray}
\begin{eqnarray}
\Upsilon_3(b_1,b_2;\rho)&=&b_1^2b_2^2+2\rho^2+4\rho b_1b_2,\nonumber
\end{eqnarray}
and functions of $U$, $Q$ and $q$ are defined as follows.
\begin{eqnarray}
U(a,b;\rho)=\biggl\{2\pi (1-\rho^2)^{\frac{1}{2}}\biggr\}\int_{a}^{\infty}\int_{b}^{\infty}
\mbox{exp}\biggl\{-\frac{x^2+y^2-2\rho xy}{2(1-\rho^2)} \biggr\}dx dy,\nonumber
\end{eqnarray}
and
\begin{eqnarray}
Q(a,b;\rho)=\phi(a)\bar{\Phi}\biggl(\frac{b-\rho a}{\sqrt{1-\rho^2}}\biggr), \quad
q(a,b;\rho)=\phi(a)\phi\biggl(\frac{b-\rho a}{\sqrt{1-\rho^2}}\biggr).\nonumber
\end{eqnarray}

As $n\to \infty$, let $\lambda_{n}(s)=2s\mbox{log}p$ and $L_p$  be a slowly varying function, the following results are established:

(1). If $\delta_k=0$ and $\delta_l=0$,
\begin{eqnarray}
\gamma(0,0,\rho_{kl})
&=& \biggl\{\rho_{kl}L_p p^{-\frac{2s}{1+\rho_{kl}}}+4\rho_{kl}^2\biggl[U(\sqrt{2s\mbox{log}p},\sqrt{2s\mbox{log}p};\rho_{kl})\nonumber\\
&+&U(\sqrt{2s\mbox{log}p},\sqrt{2s\mbox{log}p};-\rho_{kl})\biggr]\biggr\}\{1+o(1)\};\nonumber
\end{eqnarray}

(2). If $n\,\delta_k^2=n\,\delta_l^2=2r\mbox{log}p$ with $0<r<s$,
\begin{eqnarray}
\gamma(n\,\delta_k^2,n\,\delta_l^2,\rho_{kl})= \rho_{kl}L_p p^{-\frac{2}{1+\rho_{kl}}(\sqrt{s}-\sqrt{r})^2}\{1+o(1)\};\nonumber
\end{eqnarray}

(3). If $n\,\delta_k^2=n\,\delta_l^2>\lambda_{n}$,
\begin{eqnarray}
\gamma(n\,\delta_k^2,n\,\delta_l^2,\rho_{kl})=( 2\rho_{kl}^2+4\rho_{kl}n\,\delta_k\,\delta_l)\{1+o(1)\} ;\nonumber
\end{eqnarray}

(4). If $n\,\delta_k^2=2r\mbox{log}p$ with $0<r<s$ and $n\,\delta_l^2=0$,
\begin{eqnarray}
\gamma(n\,\delta_k^2,0,\rho_{kl})= \rho_{kl}L_p p^{-s}\{1+o(1)\};\nonumber
\end{eqnarray}

(5). If $n\,\delta_k^2>\lambda_{n}$ and $n\,\delta_l^2=0$,
\begin{eqnarray}
\gamma(n\,\delta_k^2,0,\rho_{kl})
&=& \biggl\{n\,\rho_{kl}L_p\,\delta_k^2\, p^{-s}+2\rho_{kl}^2\biggl[U(\eta_k^-,\sqrt{2s\mbox{log}p};\rho_{kl})\nonumber\\&+&U(\eta_k^-,\sqrt{2s\mbox{log}p};-\rho_{kl})\biggr]\biggr\}\{1+o(1)\}.\nonumber
\end{eqnarray}

Combining everything together, we can evaluate the variance of $L_n(s)$. To this end,
we first write $L_n(s)$ as the sum of two parts: one has all indices with $\delta_i=0$ and the other includes all indices with $\delta_k\ne 0$:
\[
L_n(s)=\sum_{i\in S_{\beta}^c} L_{n,i}+ \sum_{k \in S_{\beta}} L_{n,k}.
\]
Then,
\begin{eqnarray}
\mbox{Var}\{L_n(s)\}&=&\sum_{k=1}^p \mbox{Var}(L_{n,k})+ \sum_{i, j \in S_{\beta}^c}\mbox{Cov}(L_{n,i},L_{n,j})+ \sum_{k, l \in S_{\beta}}\mbox{Cov}(L_{n,k},L_{n,l})\nonumber\\
&+& \sum_{i \in S_{\beta}^c, k \in S_{\beta}}\mbox{Cov}(L_{n,i},L_{n,k})\nonumber\\
&:=&I_{(1)}+I_{(2)}+I_{(3)}+I_{(4)},\nonumber
\end{eqnarray}
where according to (\ref{apx4}), $I_{(1)}$ can be written as
\begin{eqnarray}
I_{(1)}
&=&\biggl\{\frac{2}{\sqrt{2\pi}}[(2s\mbox{log}p)^{\frac{3}{2}}+(2s\mbox{log}p)^{\frac{1}{2}}]p^{1-s}+4p\bar{\Phi}(\sqrt{2s\mbox{log}p})\nonumber\\
&+&\sum_{k \in S_{\beta}}(4n\,\delta_k^2+2)I(n\,\delta_k^2>\lambda_{n})
+\sum_{k \in S_{\beta}}(2s\mbox{log}p)^2\bar{\Phi}(\eta_k^-)I(n\,\delta_k^2<\lambda_{n})\biggr\}\{1+o(1)\}.\nonumber
\end{eqnarray}

As for $I_{(2)}$, the result in Lemma 3 gives
\begin{eqnarray}
|I_{(2)}| \le \sum_{i, j \in S_{\beta}^c}|\gamma(0,0,\rho_{ij})|
\le |L_p|\sum_{i, j \in S_{\beta}^c}|\rho_{kl}| p^{-\frac{2s}{1+|\rho_{kl}|}}\{1+o(1)\}.\nonumber
\end{eqnarray}
Since $\sum_{l} |\rho_{kl}| < \infty$ for any $k$ given in condition (C3), there is a finite number of $\rho_{kl} > \epsilon$ for arbitrary small $\epsilon\in (0,1)$. Therefore, we see that
\[
|I_{(2)}| \le |L_p| p^{1-\frac{2s}{1+\epsilon}}\{1+o(1)\},
\]
which is the smaller order of $I_{(1)}$.

The evaluation of $I_{(3)}$ depends on if the signal is greater than the threshold level $\lambda_n(s)$. If $n\,\delta_k^2=n\,\delta_l^2=2r\mbox{log}p$ with $0<r<s$, then
\[
|I_{(3)}| \le |L_p|\sum_{k,l \in S_{\beta}}|\rho_{kl}|p^{-\frac{2}{1+|\rho_{kl}|}(\sqrt{s}-\sqrt{r})^2}\le |L_p| p^{1-\beta-\frac{2}{1+\epsilon}(\sqrt{s}-\sqrt{r})^2},
\]
which again, is the smaller order of $I_{(1)}$. On the other hand,
if $n\,\delta_k^2=n\,\delta_l^2
> \lambda_n(s)$, $I_{(3)}$ is the same order as $I_{(1)}$ based on the similar derivation. Finally, let us consider $I_{(4)}$, which, according to Lemma 3, is
\[
|I_{(4)}| \le |L_p| p^{1-\beta-s},
\]
if $n\,\delta_k^2 < \lambda_n(s)$. And
\[
|I_{(4)}| \le n\,|L_p| \delta_k^2 p^{1-\beta-s},
\]
if $n\,\delta_k^2 >\lambda_n(s)$. Therefore, for both cases, we see that $I_{(4)}$ is the smaller order of $I_{(1)}$. In summary, the following result holds for the variance of $L_n(s)$:
\begin{eqnarray}
\mbox{Var}\{L_n(s)\}&=&\biggl\{\frac{2}{\sqrt{2\pi}}[(2s\mbox{log}p)^{\frac{3}{2}}+(2s\mbox{log}p)^{\frac{1}{2}}]p^{1-s}+4p\bar{\Phi}(\sqrt{2s\mbox{log}p})\nonumber\\
&+&\sum_{k,l \in S_{\beta}}(4n\,\delta_k\,\delta_l\,\rho_{kl}+2\rho_{kl}^2)I(n\,\delta_k^2>\lambda_{n})I(n\,\delta_l^2>\lambda_{n})\nonumber\\
&+&\sum_{k \in S_{\beta}}(2s\mbox{log}p)^2\bar{\Phi}(\eta_k^-)I(n\,\delta_k^2<\lambda_{n})\biggr\}\{1+o(1)\},\label{var_Ln}
\end{eqnarray}
which under $H_0$ is
\begin{eqnarray}
\sigma_{L_n(s),0}^2=\biggl\{\frac{2}{\sqrt{2\pi}}[(2s\mbox{log}p)^{\frac{3}{2}}+(2s\mbox{log}p)^{\frac{1}{2}}]p^{1-s}+4p\bar{\Phi}(\sqrt{2s\mbox{log}p})\biggr\}\{1+o(1)\}
,\label{variance_null}
\end{eqnarray}
and under $H_1$ is
\begin{eqnarray}
\sigma_{L_n(s),1}^2
&=&\biggl(\frac{2}{\sqrt{2\pi}}[(2s\mbox{log}p)^{\frac{3}{2}}+(2s\mbox{log}p)^{\frac{1}{2}}]p^{1-s}+4p\bar{\Phi}(\sqrt{2s\mbox{log}p})\nonumber\\
&+&\sum_{k,l \in S_{\beta}}(4n\,\delta_k\,\delta_l\,\rho_{kl}+2\rho_{kl}^2)I\biggl\{n\,\delta_k^2>\lambda_{n}(s)\biggr\}I\biggl\{n\,\delta_l^2>\lambda_{n}(s)\biggr\}\nonumber\\
&+&\sum_{k \in S_{\beta}}(2s\mbox{log}p)^2\bar{\Phi}(\eta_k^-)I\biggl\{n\,\delta_k^2<\lambda_{n}(s) \biggr\}\biggr)\{1+o(1)\}.\label{variance_alter}
\end{eqnarray}

\bigskip
\textbf{Lemma 4.} Suppose $\{\bm{Z}_i\}_{i=1}^p$ is a sequence of $\alpha$-mixing random variables with zero mean and satisfying
\[
M_{2l+\delta}=\mbox{sup}_i \biggl[\mbox{E}(\bm{Z}_i)^{2l+\delta} \biggr]^{1/(2l+\delta)}< \infty,
\]
for $\delta>0$ and $l \ge 1$. Let $\alpha(i)$ be $\alpha$-mixing coefficient. % satisfies $\sum_{i=1}^{\infty} i^{l-1}\alpha(i)^{\delta/(2l+\delta)}< \infty$,
Then,
\[
\mbox{E}(\sum_{i=1}^p\bm{Z}_i)^{2l}\le Cp^l \biggl[M_{2l}^{2l}+M_{2l+\delta}^{2l}\sum_{i=1}^{\infty}i^{l-1}\alpha(i)^{\delta/(2l+\delta)} \biggr],
\]
where $C$ is a finite constant positive constant depending only on $l$.

Proof. The detailed proof can be found in Kim (1994).

\bigskip
\textbf{Lemma 5.} Under condition (C4), the following relationship holds:
\[
\varpi_{kk}(\tau)=\omega_{kk}+O(\tau^{-C}), \quad \mbox{for} \quad C>1,
\]
where $\omega_{kk}=\{(1-\kappa)\bm{\Sigma}_1+\kappa\bm{\Sigma}_2\}^{-1}_{kk}$ and $\varpi_{kk}(\tau)=\mbox{Var}\{\sqrt{n}(\bar{\bm{Z}}_{1}^{(k)}(\tau)-\bar{\bm{Z}}_{2}^{(k)}(\tau))\}$.

Proof. At first, we have
\begin{eqnarray}
&\quad&\mbox{Var}\{\sqrt{n}(\bm{\bar{Z}}_{1}(\tau)-\bm{\bar{Z}}_{2}(\tau))\}\nonumber\\&=&\bm{\Omega}(\tau)\biggl(\frac{n_2}{n_1+n_2}\bm{\Sigma_1}+\frac{n_1}{n_1+n_2}\bm{\Sigma_2} \biggr) \bm{\Omega}(\tau)\nonumber\\
&=&\{\bm{\Omega}+\bm{\Omega}(\tau)-\bm{\Omega} \}\biggl(\frac{n_2}{n_1+n_2}\bm{\Sigma}_1+\frac{n_1}{n_1+n_2}\bm{\Sigma}_2 \biggr)\{\bm{\Omega}+\bm{\Omega}(\tau)-\bm{\Omega} \},\nonumber
\end{eqnarray}
which implies that
\[
\varpi_{kk}(\tau)=\omega_{kk}+\biggl(\{\bm{\Omega}(\tau)-\bm{\Omega} \}\biggl(\frac{n_2}{n_1+n_2}\bm{\Sigma}_1+\frac{n_1}{n_1+n_2}\bm{\Sigma}_2 \biggr)\{\bm{\Omega}(\tau)-\bm{\Omega} \} \biggr)_{kk},
\]
where the cross term $2\{(\Omega(\tau)-\Omega)\}_{kk}$ is gone since $\Omega(\tau)$ has the same diagonal elements as $\Omega$.
If we let $A=\{\bm{\Omega}(\tau)-\bm{\Omega} \}$ and $B=\biggl(\frac{n_2}{n_1+n_2}\bm{\Sigma}_1+\frac{n_1}{n_1+n_2}\bm{\Sigma}_2 \biggr)$, then the above equation can be written as
\[
\varpi_{kk}(\tau)=\omega_{kk}+\sum_{l,m}A_{kl}A_{km}B_{lm}.
\]
Therefore,
\begin{eqnarray}
|\varpi_{kk}(\tau)| &\le& |\omega_{kk}| + \sum_m |A_{km}| \sum_l |A_{kl} B_{lm}|\nonumber\\
&\le&|\omega_{kk}| + \sum_m |A_{km}| \max \limits_{l} |A_{kl}| \sum_l |B_{lm}|\nonumber\\
&\le&|\omega_{kk}| + C_1^2 \max \limits_{l} |A_{kl}|\nonumber\\
&\le&|\omega_{kk}| + C_1^2  \tau^{-C}, \nonumber
\end{eqnarray}
where from line 3 to 4, we use the fact that both $A$ and $B$ have polynomial off-diagonal decay structure, and from line 4 to 5, we use the fact that
\[
A=\{\bm{\Omega}(\tau)-\bm{\Omega} \}=-\biggl\{\omega_{ij}\mbox{I}(|i-j|>\tau) \biggr\}_{p \times p}
\]
and
\[
\max \limits_{l} |A_{kl}|=|\omega_{kl}|\mbox{I}(|k-l|=\tau+1),
\]
since $A$ has polynomial decay structure described by the matrix class $V(\epsilon_0, C, \alpha)$.
This completes the proof of Lemma 5.

\bigskip
\textbf{Lemma 6.} If $\beta > 1/2$ and the banding parameter $\tau= L_p$ for a slowly varying function $L_p$, then under condition (C5), with probability approaching 1,
\[
\delta_{\bm{\Omega}(\tau), k}\approx \omega_{kk}\delta_k \quad \mbox{for}\;\; k\in S_{\beta},
\]

Proof. Let $l_1< l_2< \cdots < l_m$ be indices randomly selected from $(1, \cdots, p)$. Then for any $1 \le k \le p$, the following result is proved in Lemma A.8. of Hall and Jin (2010):
\[
\mbox{P} \{\min \limits_{1\le i \le m-1}\{|l_{i+1}-l_i | \le k \} \le \frac{km(m+1)}{p},
\]
which implies that if $l_1< l_2< \cdots < l_m \in S_{\beta}$ and $\tau=L_p$, the probability that the minimum of inter-distance between two signals is greater than $\tau$ is $L_p p^{1-2\beta}$ which converges to 0 for $\beta > 1/2$. This combines with the fact that
\[
\delta_{\bm{\Omega}(\tau),k}=\sum_{l}\Omega_{kl}(\tau)\delta_l=\sum_{l\in S_{\beta}}\omega_{kl}\delta_l\mbox{I}(|k-l|\le \tau),
\]
leads to the result in Lemma 6.

\bigskip
\textbf{Lemma 7.} For any positive definite matrix $A_{p, p}=(a_{ij})_{p \times p}$ and its inverse $B_{p, p}=(b_{ij})_{p \times p}$, the following inequality holds
\[
a_{ii}\cdot b_{ii} \ge 1 \quad i=1, \cdots, p.
\]

Proof. We first show that $a_{pp}\cdot b_{pp} \ge 1$. To this end, we write
\[
A_{p,p} =
 \begin{pmatrix}
  A_{p-1, p-1} & a_{p-1, 1}  \\
  a^{\prime}_{p-1 ,1} & a_{pp}
   \end{pmatrix}.
\]

Then using the result from matrix inversion in block form, we have
\be
b_{pp}=(a_{pp}-a^{\prime}_{p-1 ,1}A_{p-1,p-1}^{\prime}a_{p-1,1})^{-1}, \label{le5}
\ee
which implies that $a_{pp}\cdot b_{pp} \ge 1$ since $a_{pp}-a^{\prime}_{p-1 ,1}A_{p-1,p-1}^{\prime}a_{p-1,1} >0$.

For any $i$, we can switch $a_{ii}$ from its original position to the position $(p,p)$ using the permutation matrix $P_{p,p}$. Accordingly, $b_{ii}$ is moved from its original location to $(p,p)$ by the same matrix $P_{p,p}$. By the fact that the permutation matrix is also the orthogonal matrix, we have
\[
P_{p,p} A_{p,p} P_{p,p} P_{p,p} B_{p,p} P_{p,p}=I_{p,p}.
\]
Therefore, from (\ref{le5}), we have $a_{ii}\cdot b_{ii} \ge 1$ for any $i$. This completes the proof of Lemma 7.

\bigskip
\textbf{Lemma 8.} Under the conditions assumed in Theorem 4 and condition (C5), with probability approaching to 1,
\[
\beta_{\hat{J}_n(s,\tau)} \ge \beta_{L_n(s)}.
\]

Proof. Since the power of thresholding test or transformed thresholding test is determined by its signal-to-noise ratio, it is sufficient to compare $\mbox{SNR}_{\hat{J}_n(s,\tau)}$ to $\mbox{SNR}_{L_n(s)}$. Recall that
\[
\mbox{SNR}_{\hat{J}_n(s,\tau)}=\frac{\mu_{J_n(s,\tau),1}-\mu_{J_n(s,\tau),0}}{\sigma_{J_n(s,\tau),1}}.
\]

If we let $A_k=n\,\frac{\delta_{\Omega(\tau),k}^2}{\varpi_{kk}(\tau)}I(n\,\frac{\delta_{\Omega(\tau),k}^2}{\varpi_{kk}(\tau)}>2s\mbox{log}p)+ (2s\mbox{log}p)\bar{\Phi}(\eta_{\Omega(\tau) k}^-) I(n\,\frac{\delta_{\Omega(\tau),k}^2}{\varpi_{kk}(\tau)}<2s\mbox{log}p)$, then
 \[
\mu_{J_n(s,\tau),1}-\mu_{J_n(s,\tau),0}=\sum_{k \in S_{\beta}}A_k+ \sum_{k \in S_{\Omega(\tau), \beta} \cap S_{\beta}^c } A_k.
\]
Similarly, if we let
\[
B_{kl}=(4n\frac{\delta_{\Omega(\tau),k}}{\varpi_{kk}^{1/2}(\tau)} \frac{\delta_{\varpi(\tau),l}}{\varpi_{ll}^{1/2}(\tau)}\rho_{\Omega, kl}+2\rho_{\Omega, kl}^2)I(n\frac{\delta_{\Omega(\tau),k}^2}{\varpi_{kk}(\tau)}>2s\mbox{log}p)I(n\,\frac{\delta_{\Omega(\tau),l}^2}{\varpi_{ll}(\tau)}>2s\mbox{log}p),
\]
and
\[
C_k=(2s\mbox{log}p)^2\bar{\Phi}(\eta_{\Omega(\tau) k}^-)I(n\,\frac{\delta_{\Omega(\tau),k}^2}{\varpi_{kk}(\tau)}<2s\mbox{log}p),
\]

then
\begin{eqnarray}
\sigma_{J_n(s,\tau),1}^2&=&\frac{2}{\sqrt{2\pi}}\{(2s\mbox{log}p)^{\frac{3}{2}}+(2s\mbox{log}p)^{\frac{1}{2}}\}p^{1-s}\nonumber\\
&+&\sum_{k,l \in S_{\beta}}B_{kl}+\sum_{k\in S_{\beta}}C_k+\sum_{k,l \in S_{\Omega(\tau), \beta} \cap S_{\beta}^c}B_{kl}+\sum_{k\in S_{\Omega(\tau), \beta} \cap S_{\beta}^c}C_k.\nonumber
\end{eqnarray}

First, due to the fact that $\Omega$ has polynomial off-diagonal decay and $\varpi_{kk}(\tau)=\omega_{kk}+O(\tau^{-C})$ as shown in Lemma 5,  we know that $\rho_{\Omega, kl}$  is summable, i.e., $\sum_{l=1}^p \rho_{\Omega, kl} < \infty$ for any $k \in \{1, \cdots, p\}$. It follows that as $p \to \infty$,
\[
\frac{\sum_{k \in S_{\Omega, \beta} \cap S_{\beta}^c } A_k}{\sum_{k,l \in S_{\Omega(\tau), \beta} \cap S_{\beta}^c}B_{kl}+\sum_{k \in S_{\Omega(\tau), \beta} \cap S_{\beta}^c}C_k} \to 1 \quad \mbox{and}\quad \frac{\sum_{k \in S_{\beta} } A_k}{\sum_{k,l \in S_{\beta}}B_{kl}+\sum_{k \in S_{\beta}}C_k} \to 1.
\]
As a result,
\be
\mbox{SNR}_{\hat{J}_n(s,\tau)} \ge \frac{\sum_{k \in S_{\beta}}A_k}{\sqrt{\frac{2}{\sqrt{2\pi}}\{(2s\mbox{log}p)^{\frac{3}{2}}+(2s\mbox{log}p)^{\frac{1}{2}}\}p^{1-s}+\sum_{k,l \in S_{\beta}}B_{kl}+\sum_{k\in S_{\beta}}C_k}},\label{power_comp}
\ee
where $A_k$, $B_{kl}$ and $C_k$ depend on $\frac{\delta_{\Omega(\tau),k}^2}{\varpi_{kk}(\tau)}$.
Using Lemmas 5, 6 and 7, we know that for $k \in S_{\beta}$, with probability approaching to 1,
\[
\frac{\delta_{\Omega(\tau),k}^2}{\varpi_{kk}(\tau)} \approx \omega_{kk}\delta_k^2 \ge \delta_k^2,
\]
which shows that the right hand side of (\ref{power_comp}) is no less than $\mbox{SNR}_{L_n}$. Hence, we have
\[
\mbox{SNR}_{\hat{J}_n(s,\tau)} \ge \mbox{SNR}_{L_n}.
\]
This completes the proof of Lemma 8.

\bigskip

\noindent{\bf A.2. Proofs of Theorems 1, 2 and 3}

\textbf{Theorem 1.} Assume Conditions \textbf{(C1)}-\textbf{(C3)}.  For any $s\in (0, 1)$,
\[
\sigma_{L_n(s)}^{-1} \biggl\{ L_n(s)-\mu_{L_n(s)}\biggr\}
 \xrightarrow{d} \mbox{N}(0,1),
\]
where $\mu_{L_n(s)}$ and $\sigma_{L_n(s)}$ are given in (2.15) and (2.16) in the main paper, respectively.

Proof. We only give the proof for the asymptotic normality of $L_n(s)$ under $H_0$ since the proof for the asymptotic normality under $H_1$ will be similar. We use Bernstein's block method to show the central limit theorem. We first partition $\{\sigma_{L_n(s),0}^{-1}(L_{n,k}-\mu_{L_{n,k},0})\}_{k=1}^p$ into $r$ blocks, each block having $b$ variables such that $rb \le p \le (r+1)b$. Then the remaining $(p-rb)$ terms are grouped into $S_3$:
\begin{eqnarray}
S_3=\sigma_{L_n(s),0}^{-1}\sum_{k=rb+1}^p(L_{n,k}-\mu_{L_{n,k},0}).\label{apx8}
\end{eqnarray}

We further divide each of $r$ blocks into two sub-blocks with larger sub-block $\Lambda_{j,1}$ having the first $b_1$ variables and smaller sub-block $\Lambda_{j,2}$ having the second $b_2$ variables:
\[
\Lambda_{j,1}=\sum_{k=1}^{b_1}(L_{n, (j-1)b+k}-\mu_{L_{n,(j-1)b+k}}),
\]
and
\[
\Lambda_{j,2}=\sum_{k=1}^{b_2}(L_{n, (j-1)b+b_1+k}-\mu_{L_{n,(j-1)b+b_1+k}}).
\]
Also we require that as $p \to \infty$, $rb_1/p \to 1$ and $r b_2 /p \to 0$.

As a result,
\[
\sigma_{L_n(s),0}^{-1}[L_n(s)-\mu_{L_n(s),0}]=S_1+S_2+S_3,
\]
where $S_1=\sigma_{L_n(s),0}^{-1}\sum_{j=1}^r \Lambda_{j,1}$, $S_2=\sigma_{L_n(s),0}^{-1}\sum_{j=1}^r \Lambda_{j,2}$ and $S_3$ are given in (\ref{apx8}).

We want to show that both $S_2=o_p(1)$ and $S_3=o_p(1)$. To this end, we first notice that $\mbox{E}(S_2)=0$ and $\mbox{E}(S_3)=0$.
Also,
\begin{eqnarray}
\mbox{Var}(S_2)&=&\sigma_{L_n(s),0}^{-2}\mbox{Var}\biggl\{ \sum_{j=1}^r \Lambda_{j,2}\biggr\}\nonumber\\
&\le&\sigma_{L_n(s),0}^{-2}\biggl\{ rb_2\frac{\sigma_{L_n(s),0}^{2}}{p}+2r \sum_{k_1=1}^{b_2}\sum_{k_2=1}^{b_2}|\gamma(0,0,\rho_{k_1 k_2})|\nonumber\\
&+&2\sum_{j_1=1}^{r}\sum_{j_2=1}^{r}\sum_{k_1=1}^{b_2}\sum_{k_2=1}^{b_2}|\gamma(0,0,\rho_{(j_1b+b_1+k_1)(j_2b+b_1+k_2)})|\biggr\}\nonumber\\
&\le&\sigma_{L_n(s),0}^{-2}\biggl\{5rb_2\frac{\sigma_{L_n(s),0}^{2}}{p}\biggr\}\nonumber\\
&=&O(\frac{rb_2}{p}),\label{thm1_1}
\end{eqnarray}
which goes to 0 as $p$ diverges to infinity. Hence, we have $S_2=o_p(1)$. Similarly, we can show that $S_3=o_p(1)$. Therefore,  we have the following result:
\[
\sigma_{L_n(s),0}^{-1}[L_n(s)-\mu_{L_n(s),0}]=S_1+ o_p(1).
\]
As long as we can show $S_1 \to \mbox{N}(0,1)$, the central limit theorem is proved by Slutsky theorem.

By Bradley's lemma, there exist independent random variables $W_j$ such that $W_j$ and $\Lambda_{j,1}$ are identically distributed. Further, we let
\[
\Delta_n(s)=\sigma_{L_n(s),0}^{-1}\sum_{j=1}^r \Lambda_{j,1}-\sigma_{L_n(s),0}^{-1}\sum_{j=1}^rW_j.
\]
Then for any $\epsilon>0$,
\begin{eqnarray}
\mbox{P}(|\Delta_n(s)|>\epsilon)&\le& r\mbox{P}(|\Lambda_{j,1}-W_j |>\epsilon\sigma_{L_n(s),0}/r)\nonumber\\
&\le&18r(\epsilon\sigma_{L_n(s),0}/r)^{-2/5}\{\mbox{E}(\Lambda_{j,1}^2)\}^{1/5}\alpha_X^{4/5}(b_2)\nonumber\\
&\le&C\epsilon^{-2/5}r^{6/5}\alpha_X^{4/5}(b_2).\label{thm1_2}
\end{eqnarray}
Using the fact that $rb_1\approx p$, we let $r=p^a$ for $a \in (0,1)$,$b_1=p^{1-a}$ and $b_2=p^{c}$ for $c \in (0, 1-a)$. Then, since $\alpha_X(k) < c \alpha^k$ for $\alpha \in (0,1)$ under condition (C3), we have
\[
r^{6/5}\alpha_X^{4/5}(b_2)=p^{6a/5}\alpha^{4/5p^c}\to 0,
\]
as $p \to \infty$. Therefore, we know $\Delta_n(s)=o_p(1)$ and
\[
\sigma_{L_n(s),0}^{-1}S_1=\sigma_{L_n(s),0}^{-1}\sum_{j=1}^r W_j+o_p(1).
\]
Therefore, we only need to show that for independent random variables $\{W_j\}_{j=1}^p$,
\[
\sigma_{L_n(s),0}^{-1}\sum_{j=1}^r W_j \xrightarrow{d} \mbox{N}(0,1).
\]
which can be proved by checking the Lyapounov condition, i.e.,
\begin{eqnarray}
\lim_{r\to \infty}\sum_{j=1}^r \mbox{E}(\sigma_{L_n(s),0}^{-1}W_j )^4=0.\label{Lip}
\end{eqnarray}

To check the Lyapounov condition, we can apply Lemma 4 by letting $Z_i= L_{n,i}-\mu_{L_{n,i,0}}$ and $l=2$. Then,
\begin{eqnarray}
\mbox{E}(\sigma_{L_n(s),0}^{-1}W_j )^4&=&\sigma_{L_n(s),0}^{-4}\mbox{E}\biggl[\sum_{k=1}^{b_1}(L_{n,k}-\mu_{L_{n,k,0}}) \biggr]^4\nonumber\\
&\le&Cb_1^2\sigma_{L_n(s),0}^{-4}\biggl[M_{4}^{4}+M_{4+\delta}^{4}\sum_{i=1}^{\infty}i\alpha_X(i)^{\delta/(4+\delta)} \biggr].\nonumber
\end{eqnarray}
By the fact that $M_4^4=O\{(\mbox{log}p)^{\frac{7}{2}}p^{-s}\}$,$M_{4+\delta}^{4}=O\{(\mbox{log}p)^{\frac{4\delta+14}{4+\delta}}p^{-\frac{4s}{4+\delta}}\}$ and
\be
\sum_{i=1}^{\infty}i\alpha_X(i)^{\delta/(4+\delta)}\le \sum_{i=1}^{\infty}i
(c\alpha^i)^{\delta/(4+\delta)}< \infty, \label{seq}
\ee
we have
\[
\mbox{E}(\sigma_{L_n(s),0}^{-1}W_j )^4 = O\{L_p r^{-2}p^{\frac{(4+2\delta)s}{4+\delta}}\}.
\]

Therefore, to have (\ref{Lip}) satisfied, we need $r^{-1}p^{\frac{(4+2\delta)s}{4+\delta}} \to 0$. If we choose $r=p^a$ with $a=1-\eta$ for an arbitrarily small $\eta \in (0,1)$, then this is equivalent to
\[
s<1-\eta-(1-\eta)\frac{\delta}{4+2\delta}.
\]
Since (\ref{seq}) holds for any small constant $\delta$, this means that $s<1-\eta$. Then, the Liyapounov condition is satisfied which completes the proof of Theorem 1.

\textbf{Theorem 2.} Assume Conditions \textbf{(C1)}-\textbf{(C3)}. Then under $H_0$,
\[
\mbox{P}\biggl\{a(\mbox{log}p)M_{L_n}-b(\mbox{log}p,\eta) \le x  \biggr\} \to \mbox{exp}(-e^{-x}),
\]
where functions $a(y)=(2\mbox{log}y)^{\frac{1}{2}}$ and $b(y,\eta)=2\mbox{log}y+2^{-1}\mbox{log}\mbox{log}y-2^{-1}\mbox{log}\{\frac{4\pi}{(1-\eta)^2}\}$.

Proof. The proof of Theorem 2 is similar to that of Theorem 2 in Zhong, Chen and Xu (2013). Therefore, we omit it.

\textbf{Theorem 3.} Assume Conditions \textbf{(C1)}-\textbf{(C3)} and $\hat{\mu}_{L_n(s),0}$ and $\hat{\sigma}_{L_n(s),0}$ satisfy (3.3). If $r>\varrho(\beta)$, the sum of type I and II errors of the multi-level thresholding test converges to zero when $\alpha=\bar{\Phi}\{(\mbox{log}p)^{\epsilon}\}\to 0$ for an arbitrarily small $\epsilon>0$ as $n \to \infty$. If $r<\varrho(\beta)$, the sum of type I and II errors of the multi-level thresholding test converges to 1 as $\alpha \to 0$ and $n \to \infty$.

Proof. Recall that the signal-to-noise ratio of the thresholding test is
\begin{eqnarray}
\mbox{SNR}_{L_n}=\frac{\mu_{L_n(s),1}-\mu_{L_n(s),0}}{\sigma_{L_n(s),1}}.\nonumber
\end{eqnarray}
To find the detection boundary above which the thresholding test is powerful, we let $n\delta_k^2=2r \mbox{log}p$ and consider the following cases among $r$, $s$ and $\beta$.
%where the numerator is determined by the difference between (\ref{mean_alter}) and (\ref{eqn5}) and the denominator is determined by (\ref{eqn10}).

\textbf{case 1}: $s\le r$ and $s\le \beta$. Then,
\[
\mu_{L_n(s),1}-\mu_{L_n(s),0}=L_pp^{1-\beta},
\]
and $\sigma_{L_n(s),1}=L_p p^{(1-s)/2}$, which gives us
\be
\mbox{SNR}_{L_n}=L_pp^{\frac{1+s-2\beta}{2}}. \label{p_case1}
\ee
To have $\mbox{SNR}_{L_n} \to \infty$, we need $s>2\beta-1$. Therefore, the detectable region for this case is $r \ge 2\beta-1$ and the best power rate $L_pp^{\{1+\min(r, \beta)-2\beta\}/2}$ can be obtained by choosing $s=\min(r, \beta)$.

\textbf{case 2}: $s\le r$ and $s > \beta$. For this case, we have
\[
\mu_{L_n(s),1}-\mu_{L_n(s),0}=L_pp^{1-\beta},
\]
and $\sigma_{L_n(s),1}=L_p p^{(1-\beta)/2}$. Therefore,
\be
\mbox{SNR}_{L_n}=L_pp^{\frac{1-\beta}{2}}, \label{p_case2}
\ee
which implies that the detectable region is $r> \beta$ and the best power rate is $L_pp^{(1-\beta)/2}$ for any $\beta<s\le r$.

\textbf{case 3}: $s>r$ and $s\le (\sqrt{s}-\sqrt{r})^2+\beta$. It is equivalent to have $\sqrt{r}<\sqrt{s}\le (r+\beta)/(2\sqrt{r})$. As a result,
\[
\mu_{L_n(s),1}-\mu_{L_n(s),0}=L_pp^{1-(\sqrt{s}-\sqrt{r})^2-\beta},
\]
and $\sigma_{L_n(s),1}=L_p p^{(1-s)/2}$. Therefore,
\be
\mbox{SNR}_{L_n}=L_pp^{\frac{1}{2}-\beta+r-\frac{(\sqrt{s}-2\sqrt{r})^2}{2}}. \label{p_case3}
\ee
To have $\mbox{SNR}_{L_n} \to \infty$, we need
\[
2\sqrt{r}-\sqrt{1-2\beta+2r}<\sqrt{s}<2\sqrt{r}+\sqrt{1-2\beta+2r}.
\]
Therefore, the detectable region is
\[
\beta-\frac{1}{2}<r<\beta \;\; \mbox{and} \;\; r\ge (1-\sqrt{1-\beta})^2 \;\; \mbox{and}\;\; \mbox{either}\;\; r\le \frac{\beta}{3} \;\; \mbox{or}\;\; r>\frac{\beta}{3}.
\]

\textbf{case 4}: $s>r$ and $s > (\sqrt{s}-\sqrt{r})^2+\beta$. It is equivalent to $\sqrt{s}>\max\{(r+\beta)/(2\sqrt{r}, \sqrt{r})\}$. For this case,
\[
\mu_{L_n(s),1}-\mu_{L_n(s),0}=L_pp^{1-(\sqrt{s}-\sqrt{r})^2-\beta},
\]
and $\sigma_{L_n(s),1}=L_p p^{1-(\sqrt{s}-\sqrt{r})^2-\beta/2}$. Therefore,
\be
\mbox{SNR}_{L_n}=L_pp^{\frac{1-(\sqrt{s}-\sqrt{r})^2-\beta}{2}}. \label{p_case4}
\ee
To have $\mbox{SNR}_{L_n} \to \infty$, we need
\[
\sqrt{r}-\sqrt{1-\beta}<\sqrt{s}<\sqrt{r}+\sqrt{1-\beta},
\]
which gives the detectable region
\[
r> (1-\sqrt{1-\beta})^2.
\]

Combining all four cases, we obtain the detectable region:
\begin{eqnarray}
 \varrho(\beta) = \left\{
  \begin{array}{c  l}
    \beta-\frac{1}{2}, & \quad  \frac{1}{2}\le \beta \le \frac{3}{4};\\
    (1-\sqrt{1-\beta})^2, & \quad \frac{3}{4}< \beta <1,
\end{array} \right. \nonumber
\end{eqnarray}
such that when $r> \varrho(\beta)$, the power of the thresholding test tends to 1.

Let's first prove the first statement in Theorem 3 where $r>\varrho(\beta)$.
Since the maximal thresholding test is of asymptotic $\alpha$ level, it suffices to show that its power tends to 1 above the detection boundary as $p\to \infty$ and $\alpha \to 0$. To this end, we first notice that with $\hat{\mu}_{L_n(s)}$ and $\hat{\sigma}_{L_n(s),0}$ satisfying (3.3) and for any $s$,
\begin{eqnarray}
\mbox{P}(M_{L_n}\ge G_{\alpha}|H_1) &\ge& \mbox{P}\biggl\{\biggl(\frac{L_n(s)-\mu_{L_n(s),0}}{\sigma_{L_n(s),0}}+\frac{\mu_{L_n(s),0}-\hat{\mu}_{L_n(s),0}}{\sigma_{L_n(s),0}}\biggr)\frac{\sigma_{L_n(s),0}}{\hat{\sigma}_{L_n(s),0}}\ge G_{\alpha}|H_1\bigg\}\nonumber\\
&=&\Phi\bigg(-\frac{\sigma_{L_n(s),0}}{\sigma_{L_n(s),1}}G_{\alpha}+\frac{\mu_{L_n(s),1}-\mu_{L_n(s),0}}{\sigma_{L_n(s),1}}\biggr)\{1+o(1)\}.\nonumber
\end{eqnarray}
Then, we can choose $\alpha_n=\bar{\Phi}\{(\mbox{log}p)^{\epsilon}\} \to 0$ as $p \to \infty$ for any small number $\epsilon>0$ such that $G_{\alpha}=O\{(\mbox{log}\mbox{log}p)^{1/2}\}$. If $r> \varrho(\beta)$, we can find a $s$ satisfying one of cases given by (\ref{p_case1}), (\ref{p_case2}), (\ref{p_case3}) and (\ref{p_case4}) such that the second term in $\Phi(\cdot)$ dominates and tends to infinity, which leads to $\Phi(\cdot)\to 1$ as long as $s$ is chosen adaptive to $(r, \beta)$ above $r= \varrho(\beta)$. 

Then we show that the second statement is true in Theorem 3. Note that if $r<\varrho(\beta)$, we have $r<\beta$ and $(r+\beta)^2/(4r)>1$.  The multi-level thresholding test statistic $M_{L_n}$ can be written as
\begin{eqnarray}
M_{L_n}&=&\max \limits_{s \in S_n}\biggl\{ \biggl(\frac{L_n(s)-\mu_{L_n(s),1}}{\sigma_{L_n(s),1}}\frac{\sigma_{L_n(s),1}}{\sigma_{L_n(s),0}}+\frac{\mu_{L_n(s),1}-\mu_{L_n(s),0}}{\sigma_{L_n(s),1}}\frac{\sigma_{L_n(s),1}}{\sigma_{L_n(s),0}} \biggr)\frac{\sigma_{L_n(s),0}}{\hat{\sigma}_{L_n(s),0}}\nonumber\\
&+&\frac{\mu_{L_n(s),0}-\hat{\mu}_{L_n(s),0}}{\hat{\sigma}_{L_n(s),0}}\biggr\}.\nonumber
\end{eqnarray}

Using the fact that $r<\beta$ and $(r+\beta)^2/(4r)>1$, we can show that $\sigma_{L_n(s),1}/\sigma_{L_n(s),0}=1+o(1)$. Based on the argument in cases 1-4 above, we know $(\mu_{L_n(s),1}-\mu_{L_n(s),0})/\sigma_{L_n(s),1}\\ \to 0$.  
All, together with the assumption (3.3) in the main paper, we know that
\[
M_{L_n}=\max \limits_{s \in S_n}\tilde{L}_n(s)\{1+o_p(1)\},
\]
where $\tilde{L}_n(s)=(L_n(s)-\mu_{L_n(s),1})/\sigma_{L_n(s),1}$.
It can be shown that (see Zhong, Chen and Xu (2013))
\[
\mbox{P}\{a(\mbox{log}p)\max \limits_{s \in S_n}\tilde{L}_n(s)-b(\mbox{log}p,c) \le x\} \to exp(-e^{-x}),
\]
where $c=\mbox{max}(\eta-r+2r\sqrt{1-\eta}-\beta, \eta)I(r<1-\eta)+\mbox{max}(1-\beta,\eta)I(r>1-\eta)$. Therefore,
\begin{eqnarray}
\mbox{P}(M_{L_n}\ge G_{\alpha}|H_1)&=&\mbox{P}(M_{L_n}\ge \{q_{\alpha}+b(\mbox{log}p,c)/a(\mbox{log}p)\}|H_1)\{1+o(1)\}\nonumber\\
&=&\alpha\{1+o(1)\}\to 0,\nonumber
\end{eqnarray}
which implies that the type II error tends to 1 as $\alpha \to 0$. This completes the second statement in Theorem 3.

\noindent{\bf A.3. Proof of Theorem 4}

We first derive the mean and variance of the transformed thresholding test statistics. Note that by the relationship $\bm{Z}_{ij}(\tau)=\bm{\Omega}(\tau) \bm{X}_{ij}$ and $\sum_{l}|\omega_{kl}|<\infty$, for a given constant $C$,
$Z_{ij}^{(k)}(\tau)=\sum_{l}\omega_{kl} X_{ij}^{(l)}\mbox{I}(|k-l|<\tau)$. Since $X_{ij}^{(l)}$ is sub-Gaussian for any $l=1, \cdots, p$, $Z_{ij}^{(l)}(\tau)$ is sub-Gaussian by using H\"{o}lder inequality and mathematical induction. Hence, the large derivation results can be applied to derive the mean and variance of the transformed thresholding test statistic defined in (\ref{eq2-new}) by following the similar steps when we derive the mean and variance of the thresholding step. Therefore,  to obtain the mean and variance of the transformed thresholding test, we can simply replace $\delta_k$ by $\delta_{\bm{\Omega}(\tau),k}$ and $S_{\beta}$ by $S_{\bm{\Omega}(\tau),\beta}$ in (\ref{mean_thr}) and (\ref{variance_thr}), respectively, where after the transformation, nonzero signals $\delta_k$ becomes $\delta_{\bm{\Omega}(\tau),k}$ and the set $S_{\beta}$ including these nonzero signals becomes $S_{\bm{\Omega}(\tau),\beta}$.

We first establish the asymptotic normality of transformed thresholding test defined in (\ref{eq2-new}) where the banding parameter $t$ is chosen to be a slowly varying function. To this end, we first show that both $\{Z_{1i}^{(k)}(t)\}_{k=1}^p$ and $\{Z_{2i}^{(k)}(t)\}_{k=1}^p$ are $\alpha$-mixing sequences. By condition (C3), $\{X_{1j}^{(k)}\}_{k=1}^p$ and $\{X_{2j}^{(k)}\}_{k=1}^p$ are $\alpha$-mixing sequences. Then any event $A\in \sigma(\mathcal{F}_{\bm{X},(1, a)}^{(1)},\mathcal{F}_{\bm{X},(1, a)}^{(2)})$ and $B\in \sigma(\mathcal{F}_{\bm{X},(a+k, \infty)}^{(1)}, \mathcal{F}_{\bm{X},(a+k, \infty)}^{(2)})$,
\[
|P(A\cap B)-P(A)P(B)|\to 0\quad\mbox{as}\quad k\to\infty.
\]
By the relationship between $\bm{Z}_{1i}(t)$ and $\bm{X}_{1i}$, for any $t$,
\[
Z_{1i}^{(a)}(t) \in\sigma(\mathcal{F}_{\bm{X}, (a-t,a+t)}^{(1)}), \quad \mbox{and} \quad Z_{1i}^{(a+k)}(t)\in\sigma(\mathcal{F}_{\bm{X},(a+k-t,a+k+t)}^{(1)}).
\]
Then as long as $k-2t \to\infty$, $|P(A' \cap B')-P(A')P(B')|\to 0$ for any $A' \in \sigma(\mathcal{F}_{\bm{Z},(1, a)}^{(1)},\mathcal{F}_{\bm{Z},(1, a)}^{(2)})$ and $B' \in \sigma(\mathcal{F}_{\bm{Z},(a+k, \infty)}^{(1)},\mathcal{F}_{\bm{Z},(a+k, \infty)}^{(2)})$. It follows that
\[
\alpha_{\bm{Z}_{1}(t)}(k)=\alpha_{\bm{X}_1}(k-2t) \quad \mbox{if} \quad k> 2t.
\]
Therefore, $\alpha_{\bm{Z}_{1}(t)}(k)\to 0$ as $k-2t\to\infty$ where $\alpha_{\bm{Z}_{1}(t)}$ is the $\alpha$-mixing coefficient for the sequence $\{Z_{1j}^{(k)}(t)\}_{k=1}^p$. Similarly, it can be shown that $\alpha_{Z_{2}(t)}(k)\to 0$ as $k-2t\to\infty$. Thus, both $\{Z_{1i}^{(k)}(t)\}_{k=1}^p$ and $\{Z_{2i}^{(k)}(t)\}_{k=1}^p$ are $\alpha$-mixing sequences. Then the asymptotic normality of $J_n(s,t)$ can be established by applying the Bernstein's blocking method as we have done in the proof of Theorem 1. To further establish the normality of $\hat{J}_n(s,\tau)$, we note that our $\hat{J}_n$ can be written as
\begin{eqnarray}
\hat{J}_n&=& J_n+\sum_{k=1}^p(\frac{\hat{S}_{nk}}{\hat{\omega}_{kk}}-\frac{S_{nk}}{\varpi_{kk}})I(\frac{S_{nk}}{\varpi_{kk}}>\lambda_{n})+\sum_{k=1}^p(\frac{S_{nk}}{\varpi_{kk}}+1)[I(\frac{\hat{S}_{nk}}{\hat{\omega}_{kk}}>\lambda_{n})\nonumber\\
&-&I(\frac{S_{nk}}{\varpi_{kk}}>\lambda_{n})]+\sum_{k=1}^p(\frac{\hat{S}_{nk}}{\hat{\omega}_{kk}}-\frac{S_{nk}}{\varpi_{kk}})[I(\frac{\hat{S}_{nk}}{\hat{\omega}_{kk}}>\lambda_{n})-I(\frac{S_{nk}}{\varpi_{kk}}>\lambda_{n})]\nonumber\\
&=&J_n+ \mbox{I} +\mbox{II} +\mbox{III},\nonumber
\end{eqnarray}
where $\hat{S}_{nk}=n(\bar{\hat{Z}}_1^{(k)}-\bar{\hat{Z}}_2^{(k)})^2$ and $S_{nk}=n(\bar{Z}_1^{(k)}(t)-\bar{Z}_2^{(k)}(t))^2$.
To show the asymptotic normality of $\hat{J_n}$ under $H_0$, we only need to show that $\mbox{I}/\sigma_{J_n,0}=o_p(1)$ and $\mbox{II}/\sigma_{J_n,0}=o_p(1)$ since $\mbox{III}$ is smaller order of $\mbox{I}$ or $\mbox{II}$.

We first consider $\mbox{I}$, which can be bounded by
\begin{eqnarray}
\mbox{I}\le \mbox{max}_{1\le k \le p}|\frac{\hat{S}_{nk}}{\hat{\omega}_{kk}}-\frac{S_{nk}}{\varpi_{kk}}|\sum_{k=1}^p \mbox{I}(\frac{S_{nk}}{\varpi_{kk}}> \lambda_{n}).\nonumber
\end{eqnarray}

Using $\mbox{E}\{\sum_{k=1}^p\mbox{I}(\frac{S_{nk}}{\varpi_{kk}}> \lambda_{n})\}=\sum_{k=1}^p\mbox{P}(\frac{S_{nk}}{\varpi_{kk}}> \lambda_{n})$, and from Lemma 1,
\[
\sum_{k=1}^p\mbox{P}(\frac{S_{nk}}{\varpi_{kk}}> \lambda_{n})= O(\frac{p^{1-s}}{\sqrt{2s\mbox{log}p}}),
\]
we have $\sum_{k=1}^p\mbox{I}(\frac{S_{nk}}{\varpi_{kk}}> \lambda_{n})=O_p(\frac{p^{1-s}}{\sqrt{2s\mbox{log}p}})$. Recall that $\hat{S}_{nk}=n\{\sum_l \hat{\omega}_{kl}(\bar{X}_1^{(l)}-\bar{X}_2^{(l)}) \}^2$ and $S_{nk}=n\{\sum_l \omega_{kl}(t)(\bar{X}_1^{(l)}-\bar{X}_2^{(l)}) \}^2$. Then,
\begin{eqnarray}
\max \limits_{k} |\frac{\hat{S}_{nk}}{\hat{\omega}_{kk}}-\frac{S_{nk}}{\varpi_{kk}}| &\le &\max \limits_{l}n(\bar{X}_1^{(l)}-\bar{X}_2^{(l)})^2 \max \limits_{k}\frac{(\sum_l \omega_{kl})^2}{\omega_{kk}^2}|\hat{\omega}_{kk}-\omega_{kk}|\{1+o(1)\}\nonumber\\
&+&\max \limits_{l}n(\bar{X}_1^{(l)}-\bar{X}_2^{(l)})^2 \max \limits_{k}\frac{\sum_l|\omega_{kl}+\sqrt{\frac{\omega_{kk}}{\varpi_{kk}}}\omega_{kl}(t) |}{\omega_{kk}}\biggl\{\sum_l |\hat{\omega}_{kl}-\omega_{kl}|\nonumber\\
&+&\sum_l|\omega_{kl}-\sqrt{\frac{\omega_{kk}}{\varpi_{kk}}}\omega_{kl}(t)|\biggr\}\nonumber\\
&\le& M \max \limits_{l}n(\bar{X}_1^{(l)}-\bar{X}_2^{(l)})^2 \max \limits_{k}\biggl\{\sum_{l=1}^p|\hat{\omega}_{kl}-\omega_{kl} |+t^{-a}+O(t^{-C}) \biggr\},\nonumber
\end{eqnarray}
where $M>0$, $a >0$ and we use the fact that $\varpi_{kk}=\omega_{kk}+O(t^{-C})$ from Lemma 5. From the fact that $\max \limits_{l}n(\bar{X}_1^{(l)}-\bar{X}_2^{(l)})^2= O_p(\mbox{log}p)$ and $\mbox{max}_{k}\sum_{l=1}^p|\hat{\omega}_{kl}-\omega_{kl} |=O_p[(\frac{{log} p}{n})^{q/2}] $ for any $q$ such that $1/(\alpha+1)<q<1$ (See Bickel and Levina, 2008b), we know
\be
\mbox{max}_{k}|\frac{\hat{S}_{nk}}{\hat{\omega}_{kk}}-\frac{S_{nk}}{\varpi_{kk}}|\sum_{k=1}^p\mbox{I}(\frac{S_{nk}}{\varpi_{kk}}> \lambda_{n})=O_p\{L_pp^{1-s}n^{-q/2}+\mbox{log}p(t^{-a}+t^{-C})\},\nonumber
\ee
where $L_p$ and $t$ are slowly varying functions. We can choose $t$ such that $\mbox{log}p(t^{-a}+t^{-C})=o(1)$. Therefore, we have
$\mbox{I}=O_p(L_pp^{1-s}n^{-q/2})$.
By assumption that $p=n^{1/\theta}$ and $s>1-q\theta$, then $\mbox{I}/\sigma_{J_n,0}=o_p(1)$.

For the second term II, we have
\begin{eqnarray}
\mbox{II} &\le&  \max \limits_{k} |\frac{S_{nk}}{\varpi_{kk}} +1| \sum_{k=1}^p |I(\frac{\hat{S}_{nk}}{\hat{\omega}_{kk}}>\lambda_{n})-I(\frac{S_{nk}}{\varpi_{kk}}>\lambda_{n}) |\nonumber\\
&\le&  \max \limits_{k} |\frac{S_{nk}}{\varpi_{kk}}+1 |  \max \limits_{k} I\biggl\{\frac{\hat{S}_{nk}}{\hat{\omega}_{kk}}>\lambda_{n}\biggr\} \sum_{k=1}^pI\biggl\{|\frac{\hat{S}_{nk}}{\hat{\omega}_{kk}}-\frac{S_{nk}}{\varpi_{kk}}| >|\frac{S_{nk}}{\varpi_{kk}}-\lambda_n|  \biggr\}\nonumber \\
&+&  \max \limits_{k} |\frac{S_{nk}}{\varpi_{kk}}+1 | \mbox{max}_k I\biggl\{\frac{S_{nk}}{\varpi_{kk}}>\lambda_{n}\biggr\} \sum_{k=1}^pI\biggl\{|\frac{S_{nk}}{\varpi_{kk}}-\frac{\hat{S}_{nk}}{\hat{\omega}_{kk}}|>|\frac{S_{nk}}{\varpi_{kk}}-\lambda_n|  \biggr\}\nonumber\\
&:=&\mbox{II}_{1}+\mbox{II}_{2}.\nonumber
\end{eqnarray}
Because the proofs for $\mbox{II}_{1}$ and $\mbox{II}_{2}$ are similar, we only show $\mbox{II}_{2}$ in the following.

First, we note that
\begin{eqnarray}
 \max \limits_{k} |\frac{S_{nk}}{\varpi_{kk}}+1 | &\le&1+ \max \limits_{k} \frac{(\sum_l \omega_{kl}(t))^2}{\varpi_{kk}}  \max \limits_{l}n(\bar{X}_1^{(l)}-\bar{X}_2^{(l)})^2=O_p(\mbox{log}p).\nonumber
\end{eqnarray}
And,
\begin{eqnarray}
&\quad&\sum_{k=1}^pI\biggl\{|\frac{\hat{S}_{nk}}{\hat{\omega}_{kk}}-\frac{S_{nk}}{\varpi_{kk}} |>|\frac{S_{nk}}{\varpi_{kk}}-\lambda_n(s) |  \biggr\}\nonumber\\
&\le& \sum_{k=1}^p I(|\frac{\hat{S}_{nk}}{\hat{\omega}_{kk}}-\frac{S_{nk}}{\varpi_{kk}} |>h)+ \sum_{k=1}^p I (|\frac{S_{nk}}{\varpi_{kk}}-\lambda_n(s) | <h).\label{apx12}
\end{eqnarray}
The second indicator function on the right above can be evaluated by the following:
\begin{eqnarray}
\mbox{E}\biggl\{ \sum_{k=1}^p I (|\frac{S_{nk}}{\varpi_{kk}}-\lambda_n(s) | <h) \biggr\} &=&\sum_{k=1}^p\mbox{P}(|\frac{S_{nk}}{\varpi_{kk}}-\lambda_n(s) | <h)\nonumber\\
&=& \sum_{k=1}^p\biggl\{\bar{\Phi}(\sqrt{\lambda_n(s)-h})-\bar{\Phi}(\sqrt{\lambda_n(s)+h})\biggr\}\nonumber\\
&=&\frac{h}{\sqrt{2s\mbox{log}p}}p^{1-s}.\nonumber
\end{eqnarray}
Therefore, in (\ref{apx12}), $\sum_{k=1}^p I (|\frac{S_{nk}}{\varpi_{kk}}-\lambda_n(s) | <h)=O_p(\frac{h}{\sqrt{2s\mbox{log}p}}p^{1-s})$. To evaluate $\sum_{k=1}^p I(|\frac{\hat{S}_{nk}}{\hat{\omega}_{kk}}-\frac{S_{nk}}{\varpi_{kk}} |>h) $ in (\ref{apx12}), we use the same approach. First, notice that
\begin{eqnarray}
|\frac{\hat{S}_{nk}}{\hat{\omega}_{kk}}-\frac{S_{nk}}{\varpi_{kk}} | \le M \max \limits_{l}n(\bar{X}_1^{(l)}-\bar{X}_2^{(l)})^2 \biggl\{\sum_{l=1}^p|\hat{\omega}_{kl}-\omega_{kl} |\biggr\}+o(1).\nonumber
\end{eqnarray}
Then,
\begin{eqnarray}
&\quad&\mbox{E}(\sum_{k=1}^p I(|\frac{\hat{S}_{nk}}{\hat{\omega}_{kk}}-\frac{S_{nk}}{\varpi_{kk}} |>h))\nonumber\\
&\le&\sum_{k=1}^p\mbox{P}\biggl\{ M \max \limits_{l}n(\bar{X}_1^{(l)}-\bar{X}_2^{(l)})^2\sum_{l=1}^p|\hat{\omega}_{kl}-\omega_{kl} |>h\biggr\}\nonumber\\
&\le&\sum_{k=1}^p\mbox{P}(\sum_{l=1}^p|\hat{\omega}_{kl}-\omega_{kl} |>\frac{h}{MnT^2})+\sum_{k=1}^p\mbox{P}(\max \limits_{l}|\bar{X}_1^{(l)}-\bar{X}_2^{(l)}|>T),\nonumber
\end{eqnarray}
where, if we choose $T=C\sqrt{\mbox{log}p/n}$,
$\sum_{k=1}^p\mbox{P}(\max \limits_{l}|\bar{X}_1^{(l)}-\bar{X}_2^{(l)}|>T)\le p^{2-C} \to 0$,
for sufficient large $C$. If $h=C^*\mbox{log}p (\frac{\mbox{log}p}{n})^{q/2}$, there exists a $a>0$ such that
\begin{eqnarray}
\sum_{k=1}^p\mbox{P}(\sum_{l=1}^p|\hat{\omega}_{kl}-\omega_{kl} |>\frac{h}{MnT^2})&=&\sum_{k=1}^p\mbox{P}(\sum_{l=1}^p|\hat{\omega}_{kl}-\omega_{kl} |>M^{\prime}(\frac{\mbox{log}p}{n})^{q/2})
\le p^{1-a}.\nonumber
\end{eqnarray}
Therefore, by choosing $C^*$ large enough such that $a>q\theta/2$, $\sum_{k=1}^p I(|\frac{\hat{S}_{nk}}{\hat{\omega}_{kk}}-\frac{S_{nk}}{\varpi_{kk}} |>h)=O_p(p^{1-a})=o_p(pn^{-q/2})$ for $p=n^{1/\theta}$. Under the choice $h=C^*\mbox{log}p (\frac{\mbox{log}p}{n})^{q/2}$, $\sum_{k=1}^p I (|\frac{S_{nk}}{\varpi_{kk}}-\lambda_n(s) | <h)=O_p(L_p n^{-\frac{q}{2}}p^{1-s})$. In addition, we know that $\mbox{max}_k I\{\frac{S_{nk}}{\varpi_{kk}}>\lambda_{n}(s)\}=O_p(p^{-s})$. Therefore, we know that
$\mbox{II}_2=o_p(L_pp^{1-s}n^{-q/2})$.
Similarly, one can show that $\mbox{II}_1=o_p(L_pp^{1-s}n^{-q/2})$. In summary, $\mbox{II}/\sigma_{J_n,0}=o_p(\mbox{I}/\sigma_{J_n,0})=o_p(1)$.
This completes the proof of Theorem 4.
%In summary, we see that $\mbox{I}/\sigma_{J_n,0}= O_p(L_pp^{(1-s)/2}n^{-q/2})$ and $\mbox{II}/\sigma_{J_n,0}= O_p(L_pp^{(1-s)/2}n^{-q/2})$. To have $\mbox{I}/\sigma_{J_n,0}=o_p(1)$ and $\mbox{II}/\sigma_{J_n,0}=o_p(1)$, we require that $L_pp^{(1-s)/2}n^{-q/2}\to 0$.

\medskip

\noindent{\bf A.4. Proof of Theorem 5}

The proof of Theorem 5 is similar to that of Theorem 2. We omit it.

\bigskip
\noindent{\bf A.5. Proof of Theorem 6}

We first consider $\bm{\Omega}$ is known. We know that the power of the transformed thresholding test is determined by
\[
\mbox{SNR}_{J_n(s,\tau)}=\frac{\mu_{J_n(s,\tau),1}-\mu_{J_n(s,\tau),0}}{\sigma_{J_n(s,\tau),1}}.
\]
%where
%\begin{eqnarray}
%\mu_{J_n(s,\tau),1}-\mu_{J_n(s,\tau),0}&=&\sum_{k \in S_{\bm{\Omega}(\tau),\beta}}\biggl\{n\,\frac{\delta_{\bm{\Omega}(\tau),k}^2}{\varpi_{kk}(\tau)}I(n\,\frac{\delta_{\bm{\Omega}(\tau),k}^2}{\varpi_{kk}(\tau)}>2s\mbox{log}p)\nonumber\\
%&+& (2s\mbox{log}p)\bar{\Phi}(\eta_{\bm{\Omega}(\tau) k}^-) I(n\,\frac{\delta_{\bm{\Omega}(\tau),k}^2}{\varpi_{kk}(\tau)}<2s\mbox{log}p)\biggr\}.\nonumber
%\end{eqnarray}
%And $\sigma_{J_n(s,\tau),1}^2$ is given by (\ref{var_trans_alter}).

Recall that for $k \in S_{\beta}$,
$\underline{\omega}\delta_k^2\le \frac{\delta_{\bm{\Omega}(\tau),k}^2}{\varpi_{kk}(\tau)} \le \bar{\omega}\delta_k^2$. Then, we have the following inequality
\be
\frac{\mu_{J_n(s,\tau),1}-\mu_{J_n(s,\tau),0}}{\sigma_{J_n(s,\tau),1}} \ge \frac{M_1}{V_1},\label{ine1}
\ee
where $M_1=\sum \limits_{k \in S_{\beta}}\biggl\{ n\underline{\omega}\delta_{k}^2I(n \underline{\omega}\delta_{k}^2>2s\mbox{log}p)
+(2s\mbox{log}p)\bar{\Phi}(\eta_{k}^-) I(n \underline{\omega}\delta_{k}^2<2s\mbox{log}p)\biggr\}$ and
\begin{eqnarray}
V_1^2&=&\frac{2}{\sqrt{2\pi}}\{(2s\mbox{log}p)^{\frac{3}{2}}+(2s\mbox{log}p)^{\frac{1}{2}}\}p^{1-s}\nonumber\\
&+&\sum_{k,l \in S_{\beta}}(4n\underline{\omega}^2 \delta_{k}\,\delta_{l}\,\rho_{\bm{\Omega},kl}+2\rho_{\bm{\Omega},kl}^2)I(n \underline{\omega}\delta_{k}^2>2s\mbox{log}p)I(n \underline{\omega}\delta_{l}^2>2s\mbox{log}p)\nonumber\\
&+&\sum_{k\in S_{\beta}}(2s\mbox{log}p)^2\bar{\Phi}(\eta_{k}^-)I(n\underline{\omega}\delta_{k}^2<2s\mbox{log}p).\nonumber
\end{eqnarray}

Note that $M_1/V_1$ is the signal-to-noise ratio of the thresholding test without the transformation. But the signal $n\underline{\omega}\delta_k^2=2\underline{\omega}r\mbox{log}p$. From the proof of Theorem 3, we know that $M_1/V_1\to \infty$ as long as $s$ is properly chosen and $\underline{\omega}r> \varrho(\beta)$. Therefore,
\[
\frac{\mu_{J_n(s,\tau),1}-\mu_{J_n(s,\tau),0}}{\sigma_{J_n(s,\tau),1}}\to \infty,
\]
as long as $\underline{\omega}r> \varrho(\beta)$. This establishes the upper bound of the detectable region.

%To show the power of the transformed multi-level thresholding test $M_{J_n}$ tends to 1, we notice that
To show the second statement in part (a) of Theorem 6, we notice that the maximal transformed thresholding test is of asymptotic $\alpha$ level. Therefore, it is sufficient to show that its power tends to 1 above the detection boundary as $n \to \infty$ and $\alpha \to 0$. To this end, we notice that
\begin{eqnarray}
\mbox{P}(M_{J_n}\ge G_{\alpha}|H_1) &\ge& \mbox{P}\biggl(\frac{J_n(s,\tau)-\mu_{J_n(s,\tau),0}}{\sigma_{J_n(s, \tau),0}}\ge G_{\alpha}|H_1\biggr)\nonumber\\
&=&\Phi\bigg(-\frac{\sigma_{J_n(s, \tau),0}}{\sigma_{J_n(s, \tau),1}}G_{\alpha}+\frac{\mu_{J_n(s, \tau),1}-\mu_{J_n(s, \tau),0}}{\sigma_{J_n(s, \tau),1}}\biggr)\nonumber\\
&\ge& \Phi\bigg(-\frac{\sigma_{J_n(s, \tau),0}}{\sigma_{J_n(s, \tau),1}}G_{\alpha}+\frac{M_1}{V_1}\biggr).\nonumber
\end{eqnarray}
Then, we can choose $\alpha_n=\bar{\Phi}\{(\mbox{log}p)^{\epsilon}\} \to 0$ as $p \to \infty$ for any small number $\epsilon>0$ such that $G_{\alpha}=O\{(\mbox{log}\mbox{log}p)^{1/2}\}$. If $\underline{\omega}r> \varrho(\beta)$, we can find a $s$ satisfying one of cases given in the proof of Theorem 3 such that the second term in $\Phi(\cdot)$ dominates and tends to infinity, which leads to $\Phi(\cdot)\to 1$. %This shows that if $\underline{\omega}r> \varrho(\beta)$, the power of $M_{J_n}$ tends to 1.

Then we consider the first statement in part (a) of Theorem 6. Note that
\[
\frac{\mu_{J_n(s,\tau),1}-\mu_{J_n(s,\tau),0}}{\sigma_{J_n(s,\tau),1}} \le \frac{M_2}{V_2},
\]
where $M_2=\sum \limits_{k \in S_{\beta}}\biggl\{ n\bar{\omega}\delta_{k}^2I(n \bar{\omega}\delta_{k}^2>2s\mbox{log}p)
+(2s\mbox{log}p)\bar{\Phi}(\eta_{k}^-) I(n \bar{\omega}\delta_{k}^2<2s\mbox{log}p)\biggr\}$ and
\begin{eqnarray}
V_2^2&=&\frac{2}{\sqrt{2\pi}}\{(2s\mbox{log}p)^{\frac{3}{2}}+(2s\mbox{log}p)^{\frac{1}{2}}\}p^{1-s}\nonumber\\
&+&\sum_{k,l \in S_{\beta}}(4n\bar{\omega}^2 \delta_{k}\,\delta_{l}\,\rho_{\bm{\Omega},kl}+2\rho_{\bm{\Omega},kl}^2)I(n \bar{\omega}\delta_{k}^2>2s\mbox{log}p)I(n \bar{\omega}\delta_{l}^2>2s\mbox{log}p)\nonumber\\
&+&\sum_{k\in S_{\beta}}(2s\mbox{log}p)^2\bar{\Phi}(\eta_{k}^-)I(n\bar{\omega}\delta_{k}^2<2s\mbox{log}p).\nonumber
\end{eqnarray}

We also note that $M_2/V_2$ is the signal-to-noise ratio of the thresholding test with $n\bar{\omega}\delta_k^2=2\bar{\omega}r\mbox{log}p$, which converges to 0 for any $s$ if $\bar{\omega}r< \varrho(\beta)$, i.e.,
\[
\frac{\mu_{J_n(s,\tau),1}-\mu_{J_n(s,\tau),0}}{\sigma_{J_n(s,\tau),1}}\to 0.
\]
Similar to the proof for the second statement of Theorem 3, we can show that
\[
M_{J_n}=\max \limits_{s \in T_n}\tilde{J}_n(s)\{1+o_p(1)\},
\]
where $\tilde{J}_n(s)=(J_n(s)-\mu_{J_n(s,\tau),1})/\sigma_{J_n(s, \tau),1}$.
Since
\[
\mbox{P}\{a(\mbox{log}p)\max \limits_{s \in T_n}\tilde{J}_n(s)-b(\mbox{log}p,c) \le x\} \to exp(-e^{-x}),
\]
where $c=\mbox{max}(\eta-r+2r\sqrt{1-\eta}-\beta, \eta)I(r<1-\eta)+\mbox{max}(1-\beta,\eta)I(r>1-\eta)$. Then, similar to the proof in Theorem 3, we have
\begin{eqnarray}
\mbox{P}(M_{J_n}\ge G_{\alpha}|H_1)=\alpha\{1+o(1)\}\to 0,\nonumber
\end{eqnarray}
which implies that the type II error tends to 1 as $\alpha \to 0$.

%To see the transformed multi-level thresholding test is powerless when $\bar{\omega}r< \varrho(\beta)$, we notice that there exists an $s^{\star}$ such that
%\begin{eqnarray}
%\mbox{P}(M_{J_n}\ge G_{\alpha}|H_1) &=& \mbox{P}\biggl(\frac{J_n(s^{\star},\tau)-\mu_{J_n(s^{\star},\tau),0}}{\sigma_{J_n(s^{\star}, \tau),0}}\ge G_{\alpha}|H_1\biggr)\nonumber\\
%&=&\Phi\bigg(-\frac{\sigma_{J_n(s^{\star}, \tau),0}}{\sigma_{J_n(s^{\star}, \tau),1}}G_{\alpha}+\frac{\mu_{J_n(s^{\star}, \tau),1}-\mu_{J_n(s^{\star}, \tau),0}}{\sigma_{J_n(s^{\star}, \tau),1}}\biggr)\nonumber\\
%&\le& \Phi\bigg(-\frac{\sigma_{J_n(s^{\star}, \tau),0}}{\sigma_{J_n(s^{\star}, \tau),1}}G_{\alpha}+\frac{M_2(s^{\star})}{V_2(s^{\star})}\biggr).\nonumber
%\end{eqnarray}
%From the above discussion, we have seen that no matter what value of $s$ is, $M_2/V_2$ is powerless as long as $\bar{\omega}r< \varrho(\beta)$. This implies that  the transformed multi-level thresholding test is powerless when $\bar{\omega}r< \varrho(\beta)$.

Next we consider $\bm{\Omega}$ is unknown. Let $G_{\alpha}^{\star}=\{q_{\alpha}+b(\mbox{log}p, \eta^{\star}-\theta)\}/a(\mbox{log}p)$. If we choose $\alpha_n=\bar{\Phi}\{(\mbox{log}p)^{\epsilon}\} \to 0$ as $p \to \infty$ for any small number $\epsilon>0$, $G_{\alpha}^{\star}=O\{(\mbox{log}\mbox{log}p)^{1/2}\}$. We only show that if $r >\underline{\omega}^{-1}\varrho_{\theta}(\beta)$, the sum of type I and II of $M_{\hat{J}_n}$ converges to 0, since the proof that the sum of type I and II of $M_{\hat{J}_n}$ tends to 1 if $r <\bar{\omega}^{-1}\varrho_{\theta}(\beta)$ is similar to the proof for $M_{J_n}$. We notice that
\begin{eqnarray}
\mbox{P}(M_{\hat{J}_n}\ge G_{\alpha}^{\star}|H_1) &\ge& \mbox{P}\biggl(\frac{\hat{J}_n(s,\tau)-\hat{\mu}_{J_n(s,\tau),0}}{\hat{\sigma}_{J_n(s, \tau),0}}\ge G_{\alpha}^{\star}|H_1\biggr)\nonumber\\
&=&\mbox{P}\biggl\{\biggl(\frac{J_n(s,\tau)-\mu_{J_n(s,\tau),0}}{\sigma_{J_n(s, \tau),0}}+\frac{\mu_{J_n(s,\tau),0}-\hat{\mu}_{J_n(s,\tau),0}}{\sigma_{J_n(s, \tau),0}}\nonumber\\
&+&o_p(1)\biggr)\frac{\sigma_{J_n(s, \tau),0}}{\hat{\sigma}_{J_n(s, \tau),0}}\ge G_{\alpha}^{\star}|H_1\biggr\},\label{det_1}
\end{eqnarray}
where we have used the fact that if $p=n^{1/\theta}$ for $0<\theta<1$, $(\hat{J}_n(s,\tau)-\mu_{J_n(s,\tau),0})/\sigma_{J_n(s, \tau),0}\\=(J_n(s,\tau)-\mu_{J_n(s,\tau),0})/\sigma_{J_n(s, \tau),0}+o_p(1)$ given in the proof of Theorem 5. Moreover, as shown in Zhong, Chen and Xu (2013), with $p=n^{1/\theta}$ for $0<\theta<1$,
\[
\frac{\mu_{J_n(s,\tau),0}-\hat{\mu}_{J_n(s,\tau),0}}{\sigma_{J_n(s, \tau),0}} \to 0, \quad \mbox{and} \quad \frac{\sigma_{J_n(s, \tau),0}}{\hat{\sigma}_{J_n(s, \tau),0}} \to 1.
\]
Then the probability in (\ref{det_1}) is determined by
\begin{eqnarray}
\frac{J_n(s,\tau)-\mu_{J_n(s,\tau),0}}{G_{\alpha}^{\star} \sigma_{J_n(s, \tau),0}}=\biggl(\frac{J_n(s,\tau)-\mu_{J_n(s,\tau),1}}{G_{\alpha}^{\star} \sigma_{J_n(s, \tau),1}}+\frac{\mu_{J_n(s,\tau),1}-\mu_{J_n(s,\tau),0}}{G_{\alpha}^{\star} \sigma_{J_n(s, \tau),1}} \biggr)\frac{\sigma_{J_n(s, \tau),1}}{\sigma_{J_n(s, \tau),0}}, \nonumber
\end{eqnarray}
where $(J_n(s,\tau)-\mu_{J_n(s,\tau),1})/(G_{\alpha}^{\star} \sigma_{J_n(s, \tau),1})=o_p(1)$, and $\sigma_{J_n(s, \tau),1}>\sigma_{J_n(s, \tau),0}$. Therefore, as long as we can show $(\mu_{J_n(s,\tau),1}-\mu_{J_n(s,\tau),0})/(G_{\alpha}^{\star} \sigma_{J_n(s, \tau),1}) \to \infty$, (\ref{det_1}) tends 1. From inequality (\ref{ine1}), we only need to show that with properly chosen $s$, $M_1/(G_{\alpha}^{\star} V_1) \to \infty$. As we have shown in Theorem 5, we need to choose the level of the threshold $s \in (1-\theta, 1)$ if $\bm{\Omega}$ is unknown such that the asymptotic normality of the transformed thresholding test with $\hat{\bm{\Omega}}$ can be established. The modification on the detection boundary can be derived by adding the additional restriction $s> 1-\theta$ on the four cases in the proof of Theorem 3. Similar to the result in Delaigle, Hall and Jin (2011), and Zhong, Chen and Xu (2013), the modified detection boundary is given by (\ref{b2}). As a result, we know that $M_1/(G_{\alpha}^{\star} V_1) \to \infty$ if $\underline{\omega}r >\varrho_{\theta}(\beta)$. This shows that if $r >\underline{\omega}^{-1}\varrho_{\theta}(\beta)$, the power of $M_{\hat{J}_n}$ tends to 1.

\bigskip

\section*{Reference}
\parindent 0.2in
%\textsc{Anderson, T.W.} (2003). \textit{An introduction to multivariate statistical analysis}. Third edition. Wiley- Interscience.
\textsc{Ashburner, M., Ball, C., Blake, J., Botstein, D., Butler, H., Cherry, J., Davis, A., Dolinski, K., Dwight, S., Eppig, J., Harris, M., Hill, D., Issel-Tarver, L., Kasarskis, A., Lewis, S., Matese, J., Richardson, J., Ringwald, M., Rubin, G. and Sherlock, G.} (2000). Gene ontology: tool for the unification of biology. \textit{Nature Genetics}, \textbf{25}, 25-29.

\parindent -0.2in
\leftskip 0.2in

\textsc{Bai, Z. and Saranadasa, H.} (1996). Effect of high dimension: by an example of a two sample problem. \textit{Statistic Sinica}, \textbf{6}, 311-329.

\textsc{Benjamini, Y. and Hochberg, Y.} (1995). Controlling the false discovery rate: A practical and powerful approach to multiple testing. \textit{Journal of the Royal Statistical Society: Series B}, \textbf{57} 289-300.

\textsc{Bickel, P. and Levina, E.} (2008a). Regularized estimation of large covariance matrices. \textit{The Annals of Statistics}, \textbf{36}, 199-227.

\textsc{Bickel, P. and Levina, E.} (2008b). Covariance regularization by thresholding. \textit{The Annals of Statistics}, \textbf{36}, 2577-2604.

\textsc{Cai, T., Liu, W. and Luo, X.} (2011). A constrained $l_1$ minimization approach to sparse precision matrix estimation. \textit{Journal of the American Statistical Association}, \textbf{106}, 594-607.

\textsc{Cai, T., Liu, W. and Xia, Y.} (2014). Two-sample test of high dimensional means under dependence. \textit{Journal of the Royal Statistical Society: Series B}, \textbf{76}, 349-372.

%\textsc{Cai, T., Zhang, C. and Zhou, H.} (2012). Optimal rates of convergence for covariance matrix estimation. \textit{The Annals of Statistics}, \textbf{38}, 2118-2144.

\textsc{Chen, S. X. and Qin, Y. } (2010). A two sample test for high dimensional data with applications to gene-set testing. \textit{The Annals of Statistics}, \textbf{38}, 808-835.

\textsc{Delaigle, A., Hall, P. and Jin, J.} (2011). Robustness and accuracy of methods for high dimensional data analysis based on Students t-statistic. \textit{Journal of the Royal Statistical Society: Series B}, \textbf{73}, 283-301.

\textsc{Donoho, D. and Jin, J.} (2004). Higher criticism for detecting sparse heterogeneous mixtures. \textit{The Annals of Statistics}, \textbf{32}, 962-994.

\textsc{Donoho, D. and Johnstone, I.} (1994). Ideal spatial adaptation by wavelet shrinkage. \textit{Biometrika}, \textbf{81}, 425-455.

%\textsc{EL Karoui, N.} (2008). Operator norm consistent estimation of large dimensional sparse covariance matrices. \textit{The Annals of Statistics}, \textbf{36}, 2717-2756.

\textsc{Fan, J.} (1996). Test of significance based on wavelet thresholding and Neyman's truncation. \textit{Journal of the American Statistical Association}, \textbf{91}, 674-688.

\textsc{Feng, L., Zou, C., Wang, Z. and Zhu, L.} (2013). Two-sample Behrens-Fisher problem for high-dimensional data. Manuscript.

\textsc{Gr\"{o}chenig, K. and Leinert, M.} (2006). Symmetry and inverse-closedness of matrix algebra and functional calculus for infinite matrices. \textit{Transactions of the American Mathematical Society}, \textbf{358}, 2695-2711.

%\textsc{Hall, P. and Jin, J.} (2008). Properties of higher criticism under strong dependence. \textit{The Annals of Statistics}, \textbf{36}, 381-402.

\textsc{Hall, P. and Jin, J.} (2010). Innovated higher criticism for detecting sparse signals in correlated noise. \textit{The Annals of Statistics}, \textbf{38}, 1686-1732.

%\textsc{Huang, J., Liu, N., Pourahmadi, M., and Liu, L.} (2006). Covariance matrix selection and estimation via penalized normal likelihood. \textit{Biometrika}, \textbf{93}, 85-98.

\textsc{Ingster, Y. I.} (1997). Some problems of hypothesis testing leading to in�nitely divisible distributions. \textit{Mathematical Methods of Statistics}, \textbf{6}, 47-69.

\textsc{Jaffard, S.} (1990). Propri\'{e}t\'{e}s des matrices ``bien localis\'{e}es" pr\`{e}s de leur diagonale et quelques applications. \textit{Annales de l Institut Henri Poincare (C) Non Linear Analysis}, \textbf{7}, 461-476.

\textsc{Ji, P. and Jin, J.} (2012). UPS delivers optimal phase diagram in high-dimensional variable selection. \textit{The Annals of Statistics}, \textbf{40}, 73-103.

%\textsc{Jing, B. Y., Shao, Q. M. and Zhou, W.} (2008). Towards a universal self-normalized moderate deviation. \textit{Transactions of the American Mathematical Society}, \textbf{360}, 4263-4285.

\textsc{Kim, T. Y.} (1994). Moment bounds for non-stationary dependent sequences. \textit{Journal of Applied Probability}, \textbf{31}, 731-742.

\textsc{Petrov, V. V.} (1995). \textit{Limit theorems of probability theory: sequences of independent random variables}. Clarendon Press, London.

\textsc{Richardson, A., Wang, Z., Nicolo, A., Lu, X., Brown, M., Miron, A., Liao, X., Iglehart, J., Livingston, D. and Ganesan, S.} (2006). X chromosomal abnormalities in basal-like human breast cancer. \textit{Cancer Cell}, \textbf{9}, 121-132.

%\textsc{Shao, Q. M.} (1997). Self-normalized large deviations. \textit{The Annals of Probability}, \textbf{25}, 285-328.

\textsc{Srivastava, M., Katayama, S. and Kano, Y.} (2013). A two sample test in high dimensional data. \textit{Journal of Multivariate Analysis}, \textbf{114}, 349-358.

\textsc{Sun, Q.} (2005). Wiener's lemma for infinite matrices with polynomial off-diagonal decay. \textit{Comptes Rendus Mathematique}, \textbf{340}, 567-570.

%\textsc{Wang, Q. and Hall, P.} (2009). Relative errors in central limit theorems for Student's \tau statistic, with application. \textit{Statistical Sinica}, \textbf{19}, 343-354.

%\textsc{Wu, W.B., and Pourahmadi, M.} (2003). Nonparametric estimation of large covariance matrices of longitudinal data. \textit{Biometrika}, \textbf{90}, 831-844.

\textsc{Zhong, P., Chen, S. X. and Xu M.} (2013). Tests alternative to higher criticism for high dimensional means under sparsity and column-wise dependence. \textit{The Annals of Statistics}, \textbf{41}, 2820-2851.

\newpage

\begin{table}[H]
\begin{center}
\caption{Empirical sizes of the
proposed multi-thresholding tests with  (Mult2) and without data
transformation (Mult1),   Cai, Liu and Xia's max-norm tests with
(CLX2)  and without (CLX1) data transformation, Chen and Qin's test
(CQ) and the Oracle test for Gaussian and Gamma data. Parametric bootstrapping was conducted in Mult1* and Mult2* to calibrate the size distortion of Mult1 and Mult2.}
\begin{tabular}{crrrrrrrrc}
\hline $p$ & $(n_1, n_2)$  &\multicolumn{1}{c}{Oracle} & \multicolumn{1}{c}{CQ} &
\multicolumn{1}{c}{CLX1} &\multicolumn{1}{c}{CLX2}
&\multicolumn{1}{c}{Mult1 (Mult1*)}&\multicolumn{1}{c}{Mult2 (Mult2*)}\\
\hline \multicolumn{8}{c}{Normal}\\
\hline
$200$&$(30,40)$ & 0.068 & 0.052 &0.039  & 0.022 & 0.094 (0.057) & 0.044 (0.049)\\
              &$(60,80)$ & 0.067 & 0.065 &0.048  & 0.026 & 0.099 (0.059) & 0.033 (0.035)\\
              &$(90,120)$ & 0.066 & 0.063 &0.042  & 0.032 & 0.103 (0.064) & 0.063 (0.037)\\
$400$&$(30,40)$ & 0.059 & 0.055 &0.040  & 0.031 & 0.091 (0.063) & 0.082 (0.058)\\
              &$(60,80)$ & 0.059 & 0.064 &0.040  & 0.023 & 0.093 (0.051) & 0.046 (0.052)\\
              &$(90,120)$ & 0.062 & 0.066 &0.038  & 0.027 & 0.093 (0.071) & 0.051 (0.041)\\
$600$&$(30,40)$ & 0.058 & 0.053 &0.037  & 0.054 & 0.095 (0.057) & 0.129 (0.112)\\
              &$(60,80)$ & 0.050 & 0.049 &0.047  & 0.033 & 0.080 (0.064) & 0.061 (0.051)\\
              &$(90,120)$ & 0.054 & 0.054 &0.043  & 0.036 & 0.098 (0.066) & 0.072 (0.042)\\
\hline \multicolumn{8}{c}{Gamma}\\
\hline
$200$&$(30,40)$ & 0.068 & 0.062 &0.034  & 0.027 & 0.097 (0.064) & 0.056 (0.056)\\
              &$(60,80)$ & 0.065 & 0.063 &0.036  & 0.022 & 0.103 (0.069) & 0.031 (0.029)\\
              &$(90,120)$ & 0.061 & 0.055 &0.040  & 0.027 & 0.084 (0.057) & 0.046 (0.035)\\
$400$&$(30,40)$ & 0.065 & 0.053 &0.051  & 0.032 & 0.108 (0.050) & 0.092 (0.078)\\
              &$(60,80)$ & 0.057 & 0.055 &0.042  & 0.036 & 0.110 (0.051) & 0.064 (0.043)\\
              &$(90,120)$ & 0.073 & 0.049 &0.038  & 0.038 & 0.092 (0.047) & 0.055 (0.042)\\
$600$&$(30,40)$ & 0.068 & 0.054 &0.041  & 0.059 & 0.114 (0.054) & 0.134 (0.121)\\
              &$(60,80)$ & 0.057 & 0.056 &0.039  & 0.031 & 0.090 (0.052) & 0.061 (0.060)\\
              &$(90,120)$ & 0.059 & 0.052 &0.041  & 0.037 & 0.099 (0.059) & 0.073 (0.058)\\
\hline
\end{tabular}
\end{center}
\label{table1}
\end{table}

\clearpage

\begin{figure}[!htb]
\begin{minipage}[b]{0.48\linewidth}
\centering
\includegraphics[width=\textwidth]{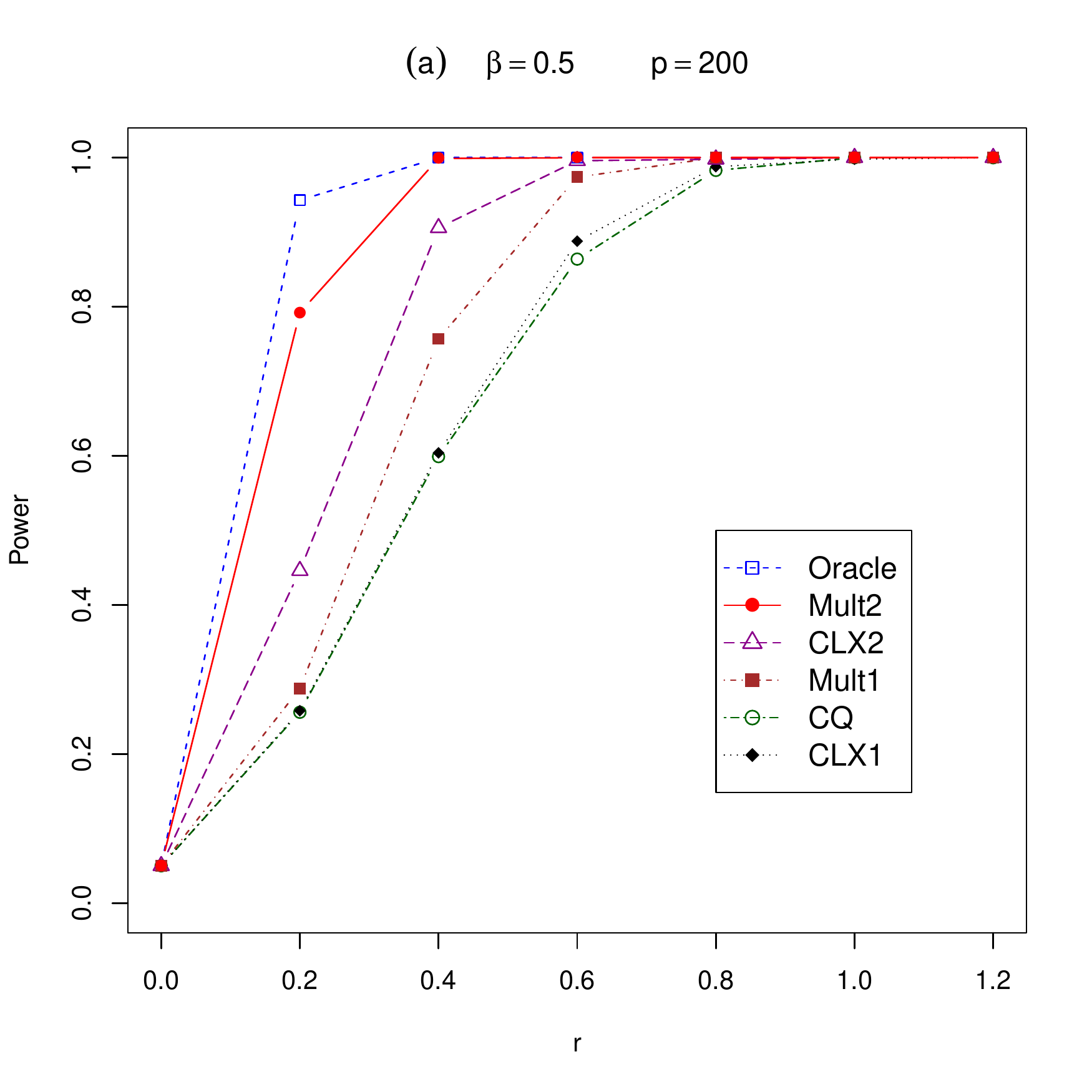}
%%\caption{default}
%%\label{fig:figure1}
\end{minipage}
\hspace{0.5cm}
\begin{minipage}[b]{0.48\linewidth}
\centering
\includegraphics[width=\textwidth]{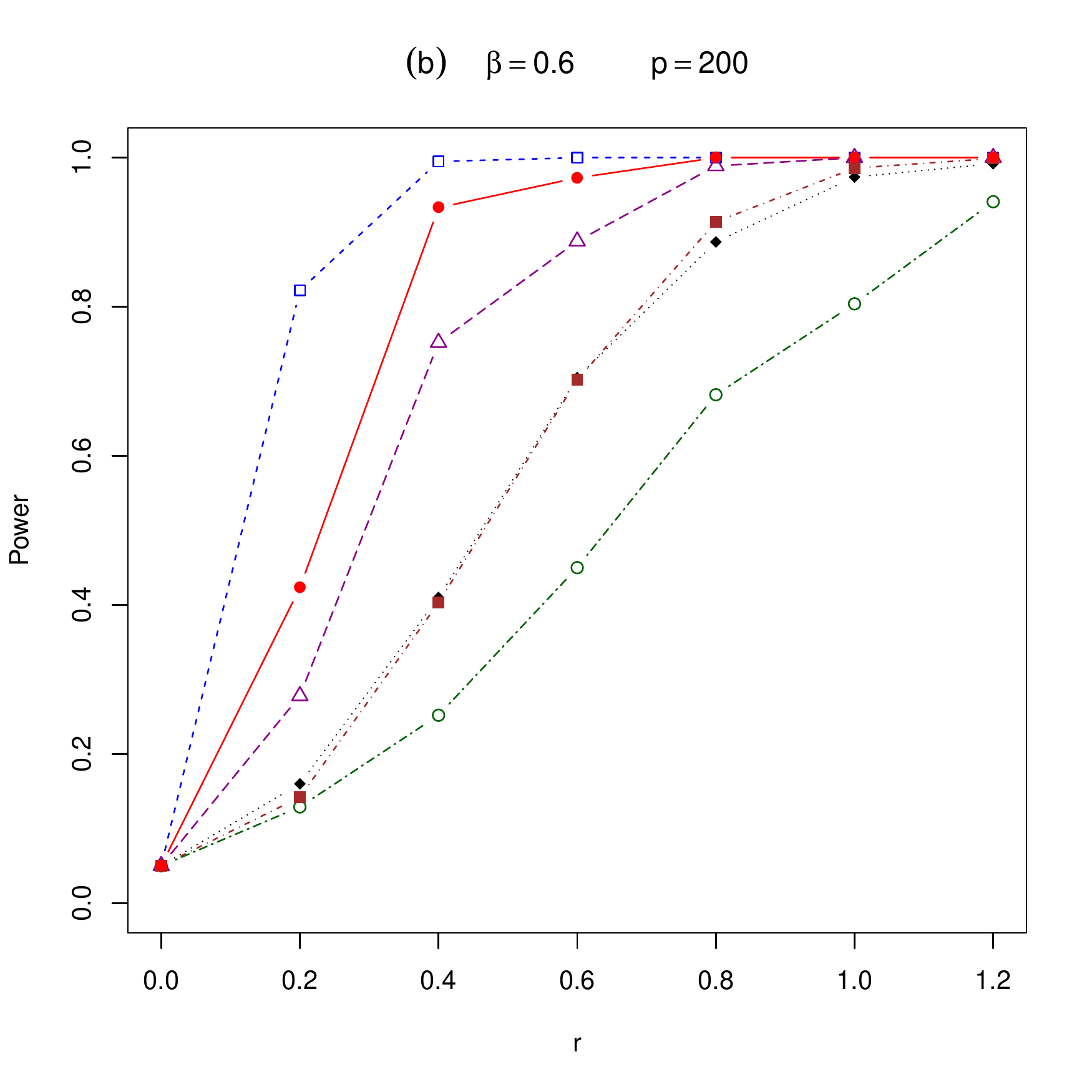}
%%\caption{default}
%%\label{fig:figure2}
\end{minipage}
\begin{minipage}[b]{0.48\linewidth}
\centering
\includegraphics[width=\textwidth]{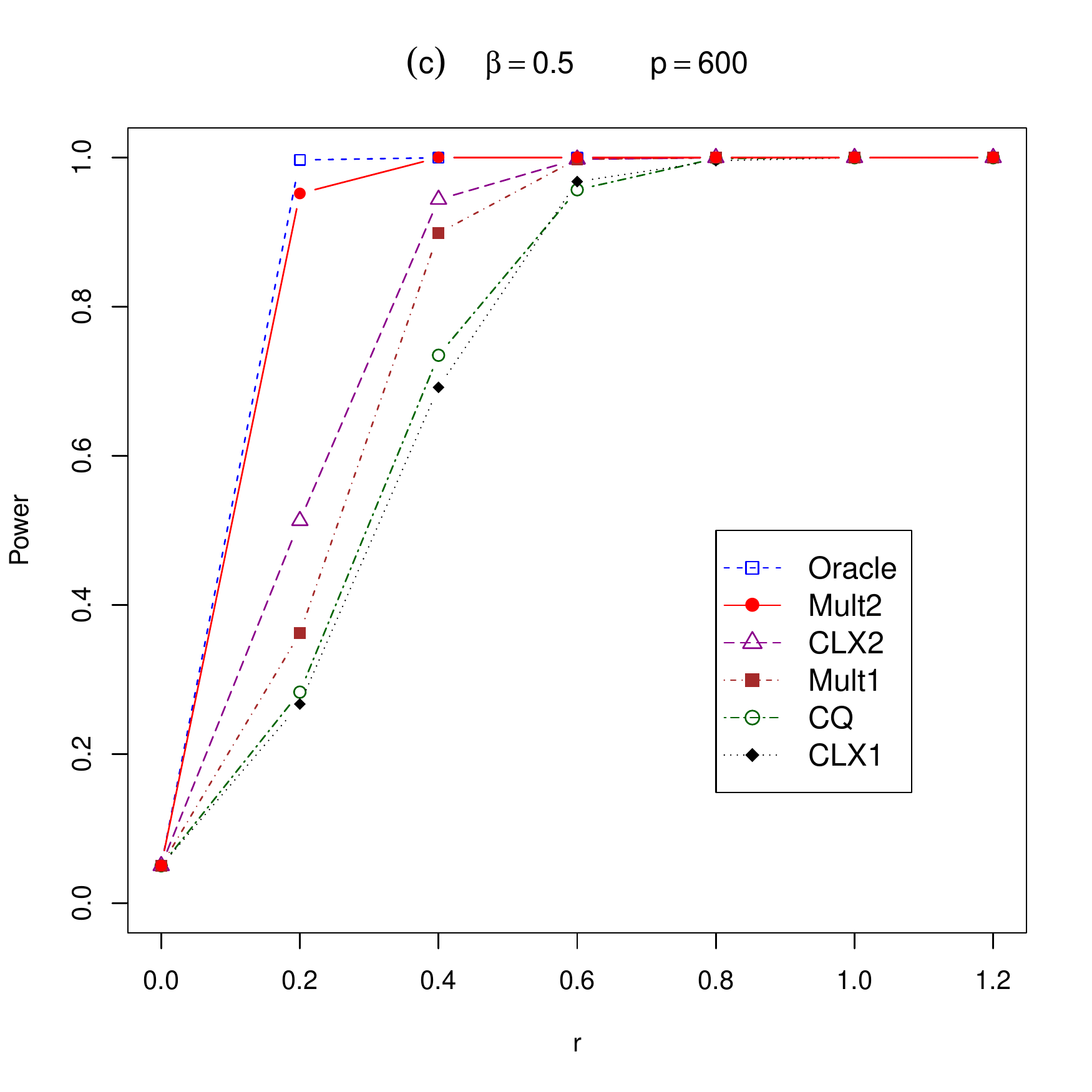}
%%\caption{default}
%%\label{fig:figure1}
\end{minipage}
\hspace{0.5cm}
\begin{minipage}[b]{0.48\linewidth}
\centering
\includegraphics[width=\textwidth]{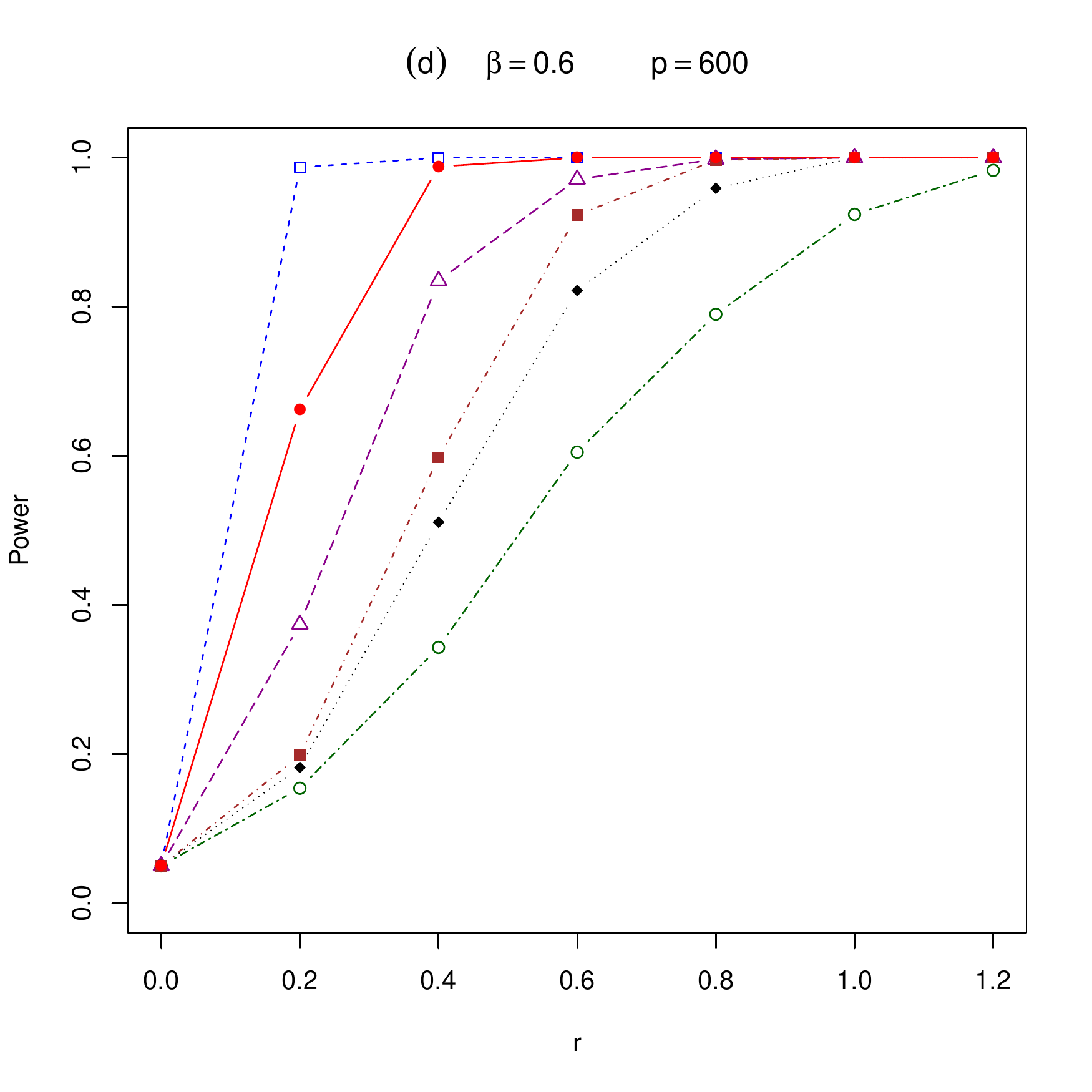}
%%\caption{default}
%%\label{fig:figure2}
\end{minipage}
\caption{Average Power with respect to the signal strength $r$ of the proposed multi-thresholding tests with  (Mult2) and without data transformation
(Mult1),   Cai, Liu and Xia's max-norm tests with (CLX2)  and without (CLX1) data transformation, Chen and Qin's test (CQ) and the Oracle test for Gaussian data with %For the first row, $p=200$, and
 $n_1=30$ and $n_2=40$.} %  For the second row, $p=600$, and $n_1=30$ and $n_2=40$.}
\end{figure}

%\begin{figure}[!htb]
%\begin{minipage}[b]{0.52\linewidth}
%\centering
%\includegraphics[width=\textwidth]{power200_gamma1.pdf}
%%%\caption{default}
%%%\label{fig:figure1}
%\end{minipage}
%\hspace{0.5cm}
%\begin{minipage}[b]{0.52\linewidth}
%\centering
%\includegraphics[width=\textwidth]{power200_gamma2.pdf}
%%%\caption{default}
%%%\label{fig:figure2}
%\end{minipage}
%\begin{minipage}[b]{0.52\linewidth}
%\centering
%\includegraphics[width=\textwidth]{power600_gamma1.pdf}
%%%\caption{default}
%%%\label{fig:figure1}
%\end{minipage}
%\hspace{0.5cm}
%\begin{minipage}[b]{0.52\linewidth}
%\centering
%\includegraphics[width=\textwidth]{power600_gamma2.pdf}
%%%\caption{default}
%%%\label{fig:figure2}
%\end{minipage}
%\caption{Average Power with respect to the signal strength $r$ of the proposed multi-thresholding tests with  (Mult2) and without data transformation
%(Mult1),   Cai, Liu and Xia's max-norm tests with (CLX2)  and without (CLX1) data transformation, Chen and Qin's test (CQ) and the Oracle test for Gamma data with %For the first row, $p=200$, and
% $n_1=30$ and $n_2=40$.}
%\end{figure}

\begin{figure}[!htb]
\begin{minipage}[b]{0.48\linewidth}
\centering
\includegraphics[width=\textwidth]{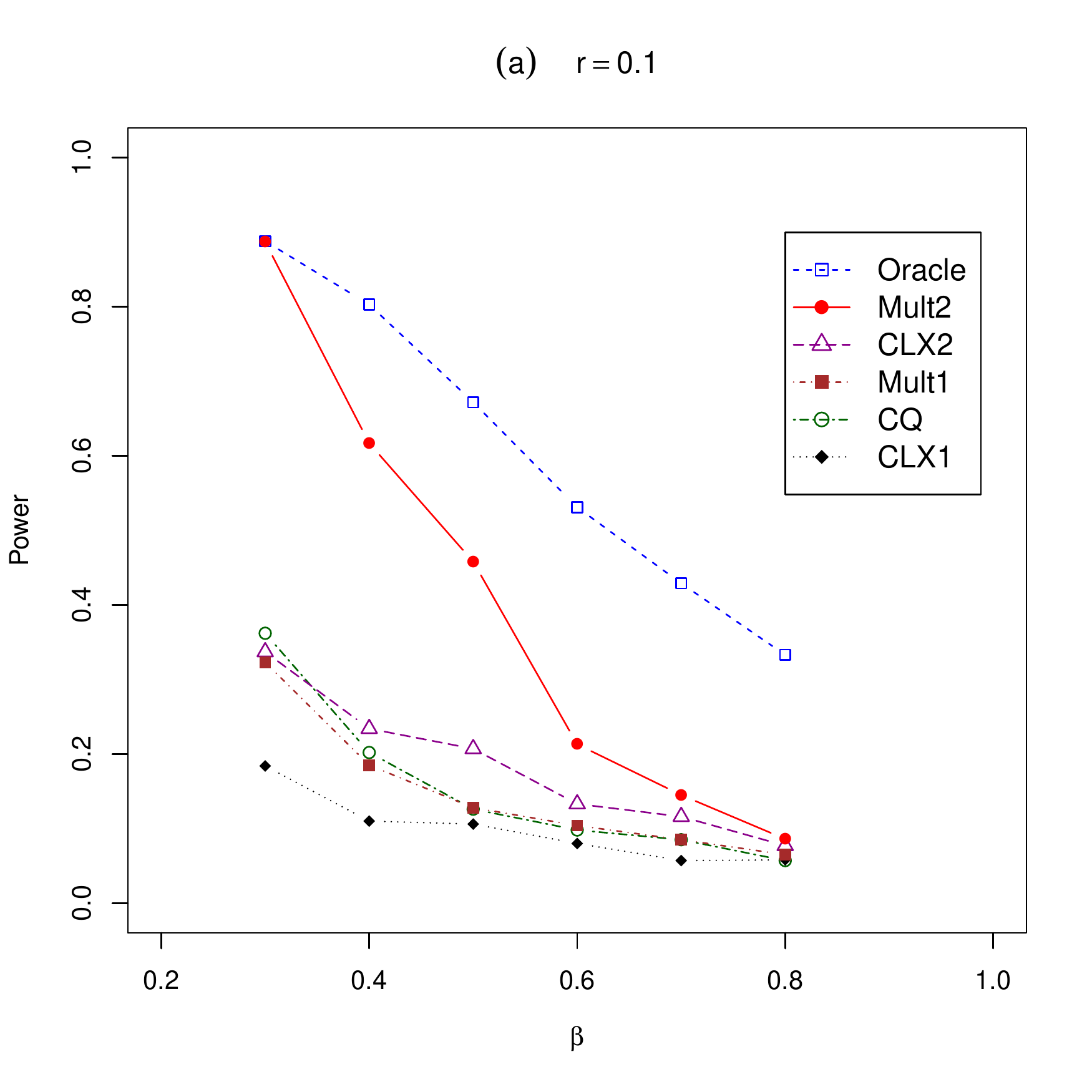}
%%\caption{default}
%%\label{fig:figure1}
\end{minipage}
\hspace{0.5cm}
\begin{minipage}[b]{0.48\linewidth}
\centering
\includegraphics[width=\textwidth]{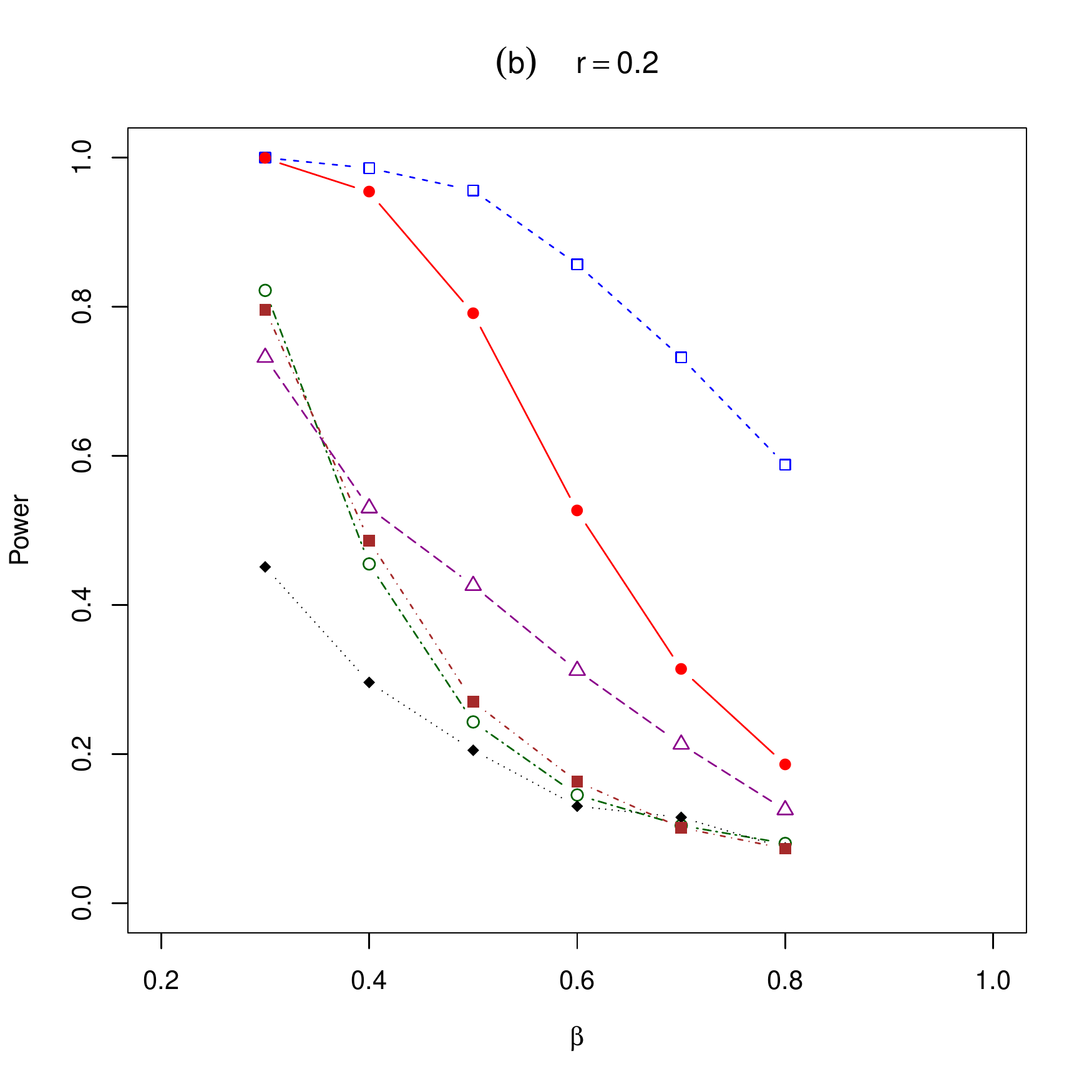}
%%\caption{default}
%%\label{fig:figure2}
\end{minipage}
\begin{minipage}[b]{0.48\linewidth}
\centering
\includegraphics[width=\textwidth]{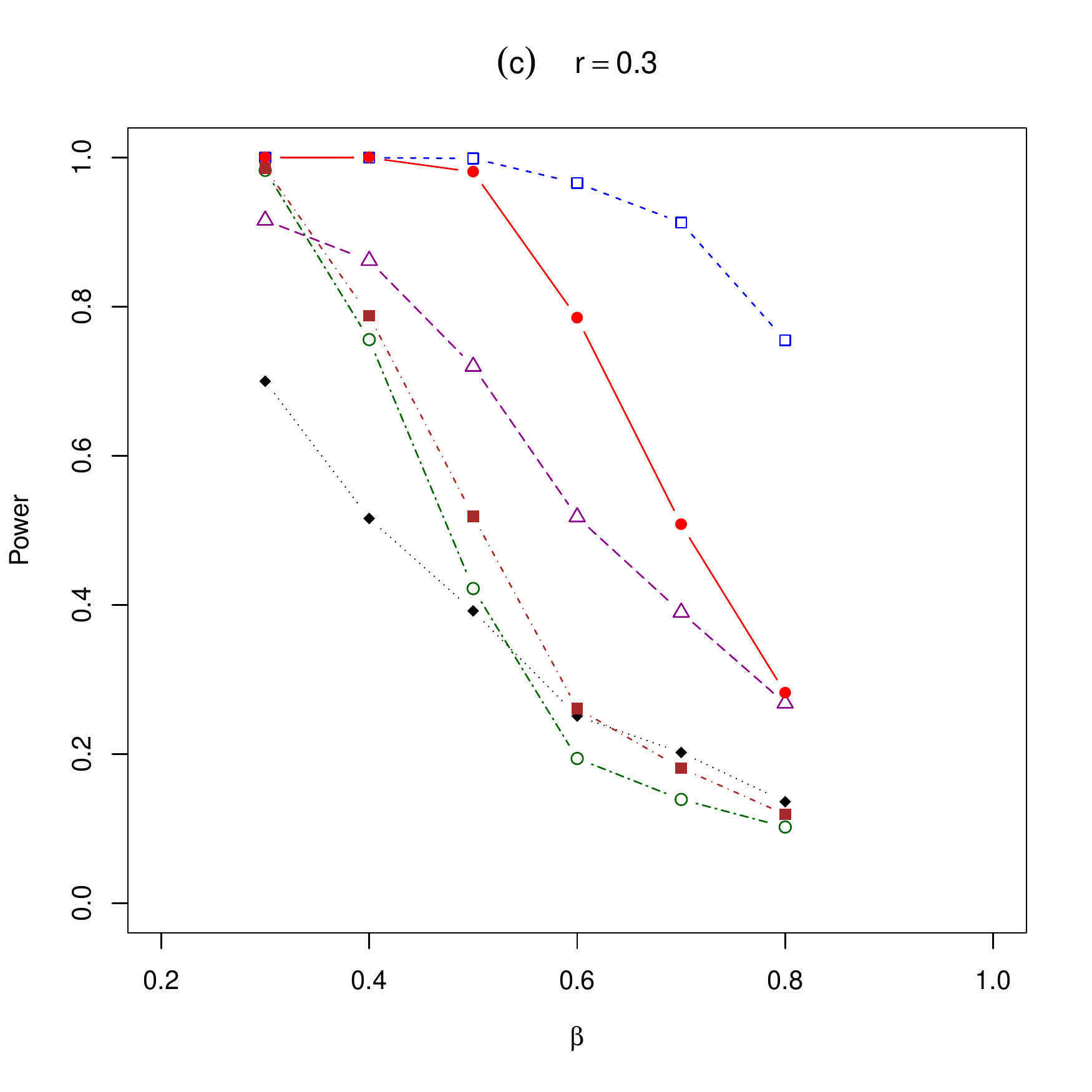}
%%\caption{default}
%%\label{fig:figure1}
\end{minipage}
\hspace{0.5cm}
\begin{minipage}[b]{0.48\linewidth}
\centering
\includegraphics[width=\textwidth]{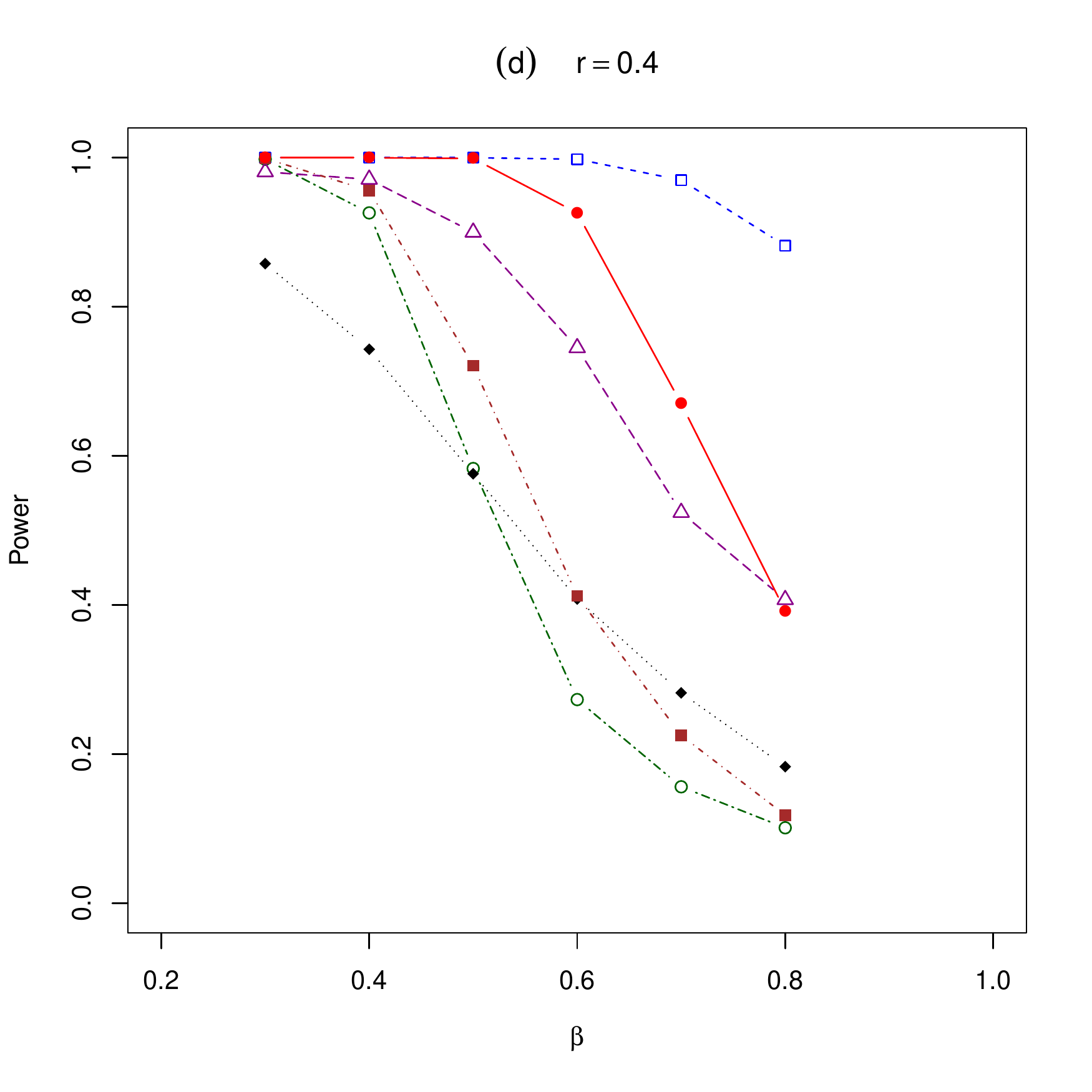}
%%\caption{default}
%%\label{fig:figure2}
\end{minipage}
\caption{Average Power with respect to the sparsity $\beta$ of the
proposed multi-thresholding tests with  (Mult2) and without data
transformation (Mult1),   Cai, Liu and Xia's max-norm tests with
(CLX2)  and without (CLX1) data transformation, Chen and Qin's test
(CQ) and the Oracle test for Gaussian data with $p=200$,
 $n_1=30$ and $n_2=40$.}
%\caption{Empirical power profiles of the proposed multi-level thresholding test  without (Mult1) and with (Mult2) data transformation,  the maximum norm tests of Cai et al. (2013) without (CLX1) and with (CLX2) data transformation, the CQ test and the Oracle %test for Gaussian data;   $p=200$, and $n_1=30$ and $n_2=40$.}
\end{figure}

\begin{figure}[!htb]
\begin{minipage}[b]{0.48\linewidth}
\centering
\includegraphics[width=\textwidth]{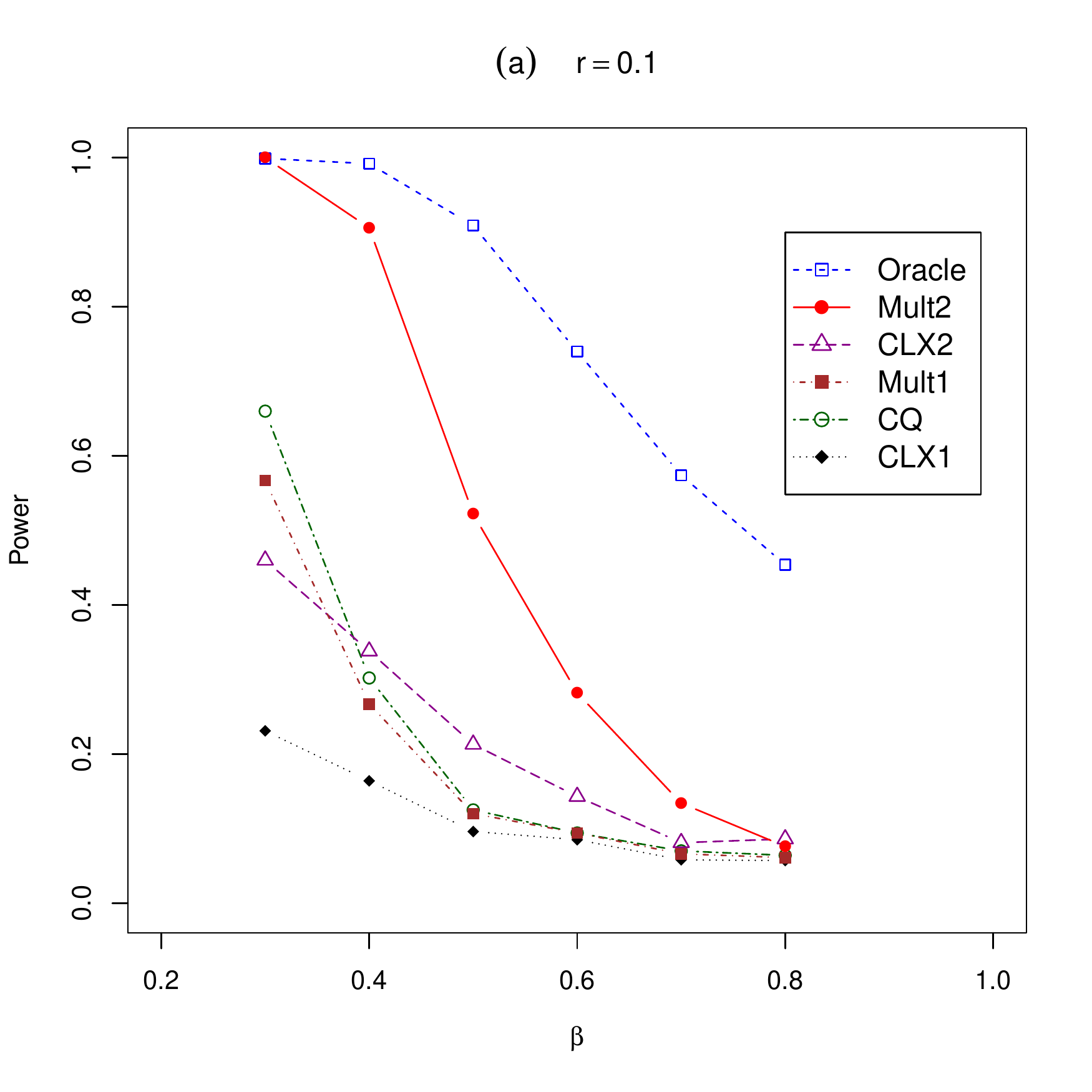}
%%\caption{default}
%%\label{fig:figure1}
\end{minipage}
\hspace{0.5cm}
\begin{minipage}[b]{0.48\linewidth}
\centering
\includegraphics[width=\textwidth]{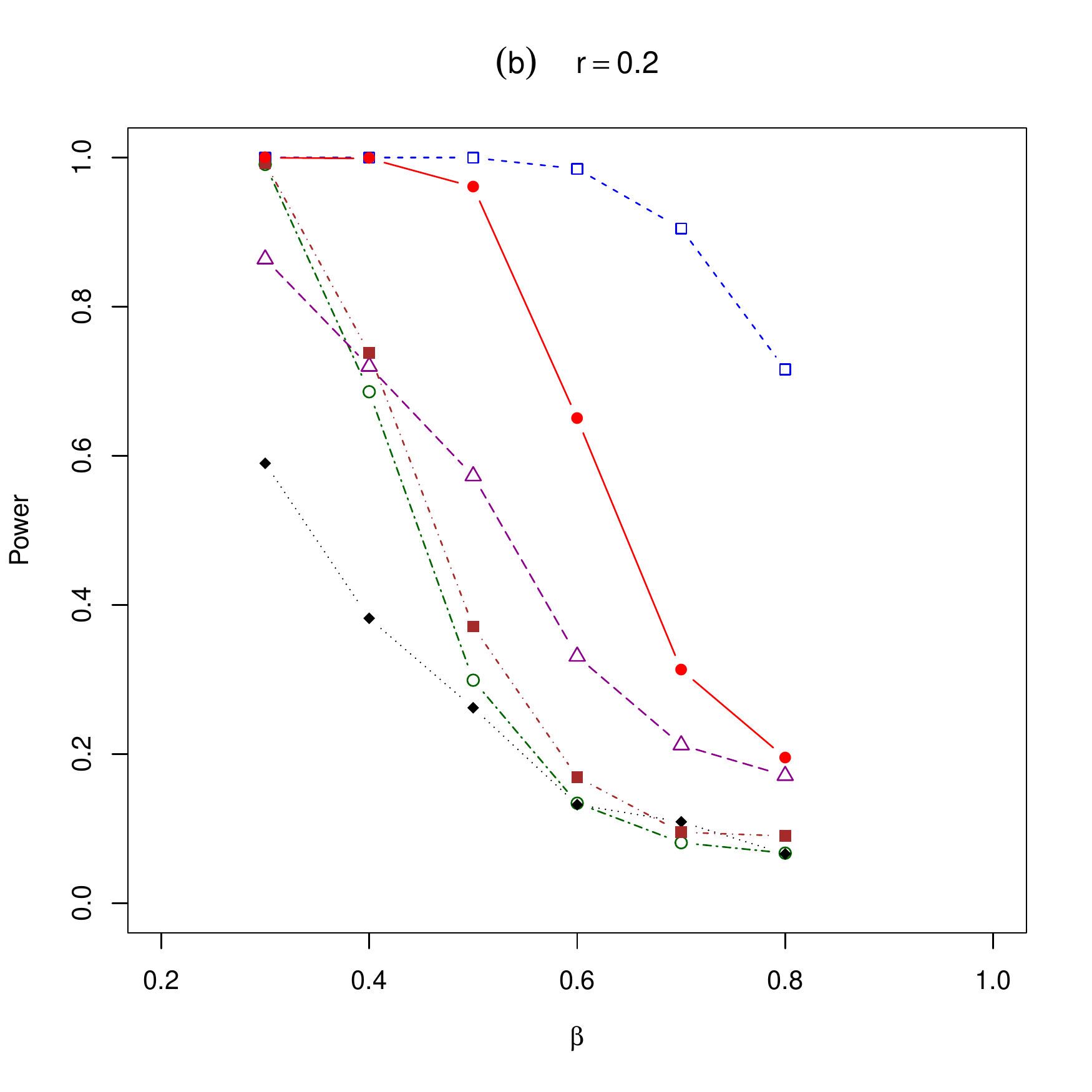}
%%\caption{default}
%%\label{fig:figure2}
\end{minipage}
\begin{minipage}[b]{0.48\linewidth}
\centering
\includegraphics[width=\textwidth]{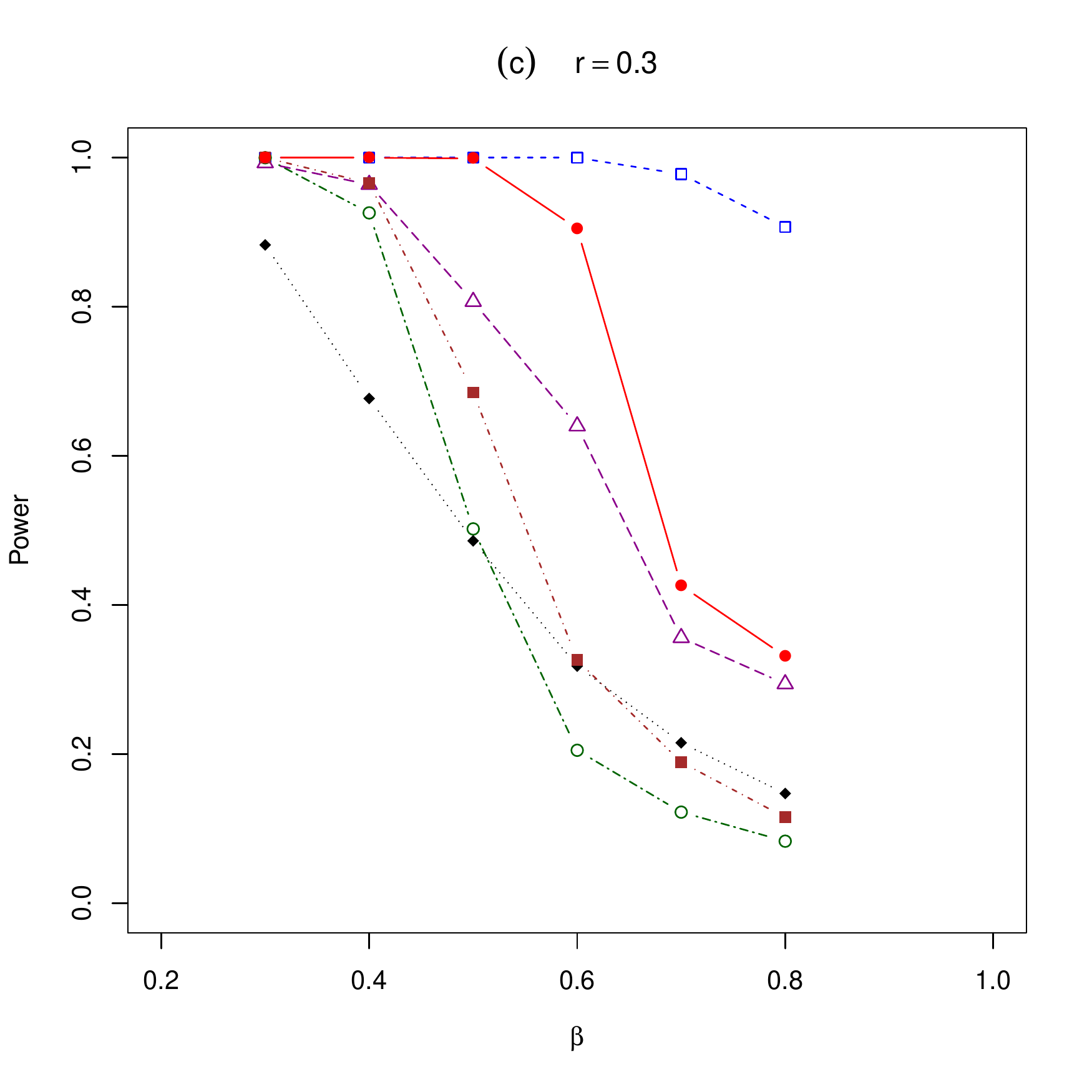}
%%\caption{default}
%%\label{fig:figure1}
\end{minipage}
\hspace{0.5cm}
\begin{minipage}[b]{0.48\linewidth}
\centering
\includegraphics[width=\textwidth]{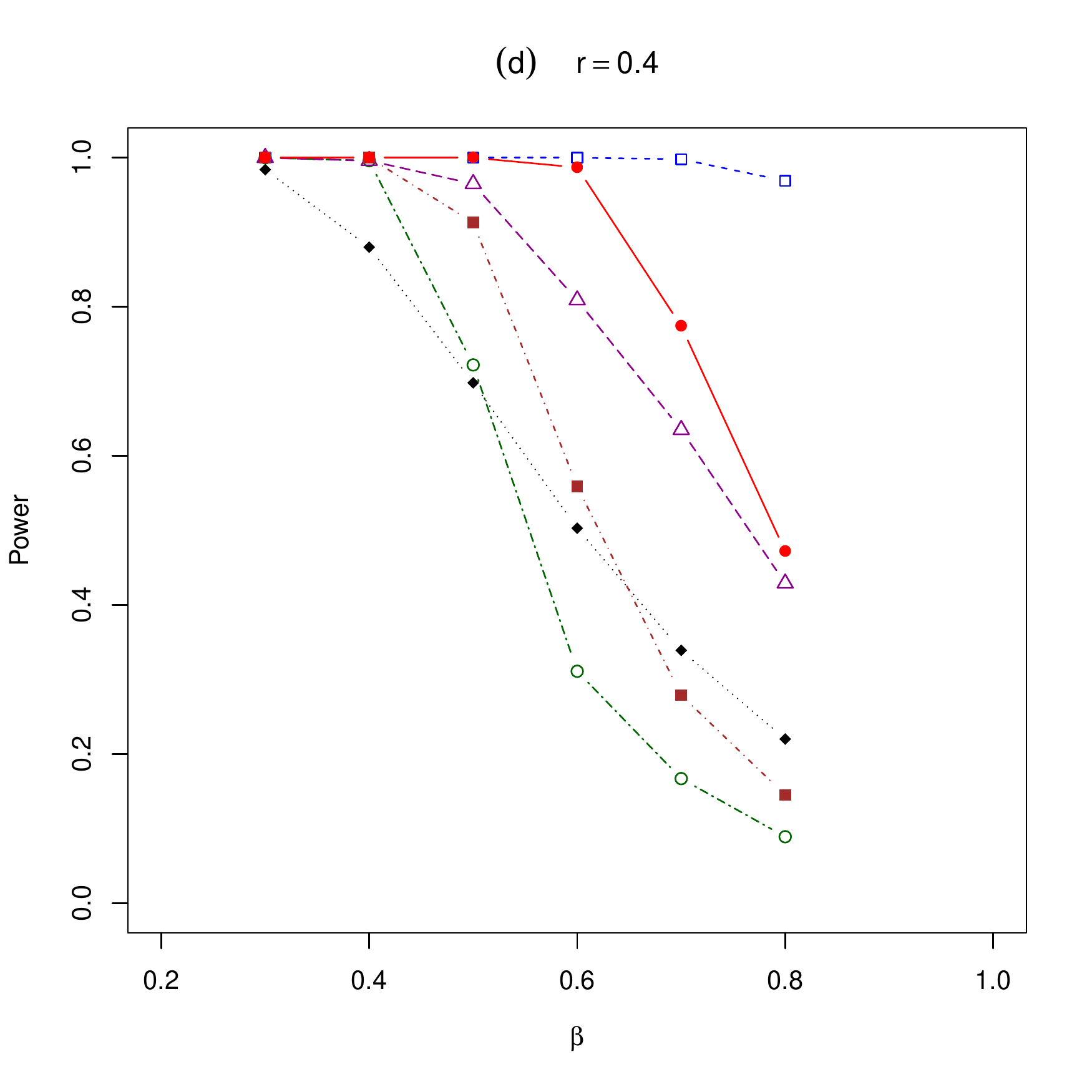}
%%\caption{default}
%%\label{fig:figure2}
\end{minipage}
\caption{Average Power with respect to the sparsity $\beta$ of the
proposed multi-thresholding tests with  (Mult2) and without data
transformation (Mult1),   Cai, Liu and Xia's max-norm tests with
(CLX2)  and without (CLX1) data transformation, Chen and Qin's test
(CQ) and the Oracle test for Gaussian data with $p=600$,
 $n_1=30$ and $n_2=40$.}
%\caption{Empirical power profiles of the proposed multi-level thresholding test  without (Mult1) and with (Mult2) data transformation,  the maximum norm tests of Cai et al. (2013) without (CLX1) and with (CLX2) data transformation, the CQ test and the Oracle %test for Gaussian data;   $p=600$, and $n_1=30$ and $n_2=40$.}
\end{figure}

\begin{table}[!htb]
\footnotesize
\caption{Top ten most significant GO terms by the multi-level
thresholding test on chromosomes $X$, 14 and 17 with false discovery rate at
0.05, where $*$ refers to GO terms  not being declared significant
by the CQ test.}
\begin{center}
\begin{tabular}{| l | c| c| c|c| r|  }
  \hline
  \multicolumn{2}{|c|}{Chromosome X}&  \multicolumn{2}{ |c|}{Chromosome 14}&  \multicolumn{2}{|c|}{Chromosome 17} \\
  \hline
 GO ID & GO term name & GO ID & GO term name &GO ID & GO term name \\
 \hline
 0005524 & ATP binding & 0005524 & ATP binding & 0005524 & ATP binding\\
 0000166 & nucleotide binding &0000166 & nucleotide binding  &0000166 & nucleotide binding\\
 0005515 & protein binding &0005515 & ${\mbox{protein binding}}^*$  &0005515 & protein binding\\
 0004872 & receptor activity &0016740& ${\mbox{transferase activity}}^*$ &0004872      &  ${\mbox{receptor activity}}^*$ \\
 0016020 & membrane &0006468 & protein amino acid &0016740 & transferase activity \\
                  &                     &                 &  phosphorylation                       &    &       \\
 0005634 & nucleus &0005576 & ${\mbox{extracellular region}}^*$ & 0006468 & protein amino acid  \\
 &   &   &   &    &phosphorylation    \\
 0003677 & DNA binding &0016020  & membrane &0005576  & extracellular region\\
 0003700 & transcription factor &0005634 & nucleus &0005887 & integral to plasma \\
                  & $\mbox{activity }^* $&     &  &  &membrane                                     \\
 0007165 & ${\mbox{signal transduction}}^*$ &0003677  & DNA binding &0016020       & membrane \\
 0046872 & ${\mbox{metal ion binding}}^*$ &0003700 & transcription factor &0005654    & ${\mbox{nucleoplasm}}^*$\\
                  &                                       &      &  activity     &      &\\
  \hline
\end{tabular}
\end{center}
\label{table2}
\end{table}

\end{document}